\documentclass[a4paper,fleqn,usenatbib]{mnras}

\usepackage{newtxtext,newtxmath}

\usepackage[T1]{fontenc}
\usepackage{ae,aecompl}

\usepackage{graphicx}	
\usepackage{amsmath}	
\usepackage{amssymb}	

\defcitealias{Labbe2013}{L13}
\defcitealias{Roberts-Borsani2016}{RB16}

\title[]{MMT/MMIRS spectroscopy of $\mathbf{z=1.3-2.4}$ extreme [OIII] emitters: Implications for galaxies in the reionization-era}

\author[M. Tang et al.]{
Mengtao Tang$^{1}$\thanks{tangmtasua@email.arizona.edu}, 
Daniel P. Stark$^{1}$,
Jacopo Chevallard$^{2}$ \&
St\'{e}phane Charlot$^{3}$ \\
\\
$^{1}$ Steward Observatory, University of Arizona, 933 N Cherry Ave, Tucson, AZ 85721 USA \\
$^{2}$ Scientific Support Office, Directorate of Science and Robotic Exploration, ESA/ESTEC, Keplerlaan 1, 2201 AZ Noordwijk, The Netherlands \\
$^{3}$ Sorbonne Universit\'{e}s, UPMC-CNRS, UMR7095, Institut d'Astrophysique de Paris, F-75014 Paris, France \\
}

\pubyear{2019}

\begin{document}
\label{firstpage}
\pagerange{\pageref{firstpage}--\pageref{lastpage}}
\maketitle

\begin{abstract}
Galaxies in the reionization-era have been shown to have prominent 
[O {\scriptsize III}]+H$\beta$ emission.  Little is known 
about the gas conditions and radiation field of this population, making it challenging to 
interpret the spectra  emerging at $z\gtrsim6$.  Motivated by this shortcoming, we have initiated a large MMT 
spectroscopic survey identifying rest-frame optical emission lines in 227 intense [O {\scriptsize III}] emitting galaxies at $1.3<z<2.4$.  
This sample complements the MOSDEF and KBSS surveys, extending 
to much lower stellar masses ($10^7-10^8 M_\odot$) and larger specific star formation rates 
($5-300$ Gyr$^{-1}$), providing a window on galaxies directly following a burst or recent upturn in star formation.  
The hydrogen ionizing production efficiency ($\xi_{\rm{ion}}$) is found to increase with the [O {\scriptsize III}] EW, in a manner similar 
to that found in local galaxies by \citet{Chevallard2018}.  We describe how this relationship 
helps explain the anomalous success rate in identifying Ly$\alpha$ emission in $z\gtrsim 7$ galaxies 
with strong [O {\scriptsize III}]+H$\beta$ emission.  We probe the impact of the intense radiation field on the ISM using O32 and Ne3O2, 
two ionization-sensitive indices.  Both are found to scale with the [O {\scriptsize III}] EW, revealing extreme ionization conditions not commonly 
seen in older and more massive galaxies.  In the most intense line emitters,  
the indices have very large average values (O32 $=9.1$, Ne3O2 $=0.5$) that have been shown to 
be linked to ionizing photon escape.  We discuss implications for the nature of galaxies 
most likely to have O32 values associated with significant LyC escape.  Finally we consider the optimal 
strategy for {\it JWST} spectroscopic investigations of galaxies at $z\gtrsim 10$ where the strongest 
rest-frame optical lines are no longer visible with NIRSpec.  
\end{abstract}





\section{Introduction} \label{sec:introduction}

Over the past decade, deep imaging surveys with the {\it Hubble Space Telescope (HST)} have 
uncovered thousands of color-selected galaxies at $z>6$ 
\citep[e.g.][]{McLure2013,Bouwens2015a,Finkelstein2015,Livermore2017,Atek2018,Oesch2018}, revealing an abundant 
population of low mass systems that likely contribute greatly 
to the reionization of intergalactic hydrogen \citep{Bouwens2015b,Robertson2015,Stanway2016,Dayal2018}. These 
galaxies have been shown to be compact with low stellar masses, blue ultraviolet continuum slopes, and large 
specific star formation rates (see \citealt{Stark2016} for a review).

The {\it James Webb Space Telescope (JWST)} will eventually build on this physical picture, providing spectroscopic 
constraints on the stellar populations and gas conditions within the reionization era.  Dedicated efforts throughout the 
last decade have delivered our first glimpse of what we are likely to see with {\it JWST}.  Ground-based spectroscopic 
surveys have revealed a rapidly declining Ly$\alpha$ emitter fraction at $z>6$ 
\citep[e.g.][]{Caruana2014,Pentericci2014,Schenker2014}, suggesting that the intergalactic medium (IGM) is likely 
partially neutral at $z\simeq 7-8$.  While prominent Ly$\alpha$ emission is very rare in early galaxies, the rest-frame 
optical emission lines are found to be very strong relative to the underlying continuum, as expected for galaxies with 
large specific star formation rates (sSFR).  The average [O {\scriptsize III}]+H$\beta$ equivalent width (EW) inferred from broadband 
flux excesses in stacked spectral energy distributions (SEDs) at $z\simeq 7-8$ is found to be $670$ \AA\ \citep[][hereafter \citetalias{Labbe2013}]{Labbe2013}, while individual objects 
have been located with combined [O {\scriptsize III}]+H$\beta$ equivalent widths reaching up to $1000-2000$ \AA\ \citep{Smit2014,Smit2015,Roberts-Borsani2016}. 
While direct spectral measurements with {\it JWST} will ultimately be required to confirm the $z\sim7$ equivalent width distribution, 
these measurements suggest large [O {\scriptsize III}] EWs ($>300$ \AA) are likely to be very common.

These results indicate that the majority of reionization-era galaxies are caught in an extreme emission line phase, implying 
a large ratio of the flux density in the ionizing UV continuum ($\lambda<912$ \AA) to that in the optical ($\lambda\simeq 4000-6000$ \AA).  
There are a variety of ways a stellar population can produce such spectra.  A likely pathway 
follows a recent upturn in the star formation rate (SFR), as would occur for systems undergoing rapidly 
rising star formation histories (which are expected at $z\gtrsim 6$; i.e. \citealt{Behroozi2013}) or punctuated bursts of star formation 
that greatly exceed  earlier activity.  In both cases, the stellar population will be dominated by recently 
formed stars, leading to an enhanced contribution from O stars relative to the longer-lived main-sequence stars 
that dominate the non-ionizing UV continuum and optical continuum.  In this framework, the galaxies with 
the largest optical line equivalent widths (i.e., those approaching EW $=1000-2000$ \AA) will be dominated by 
extremely hot massive stellar populations, resulting in an intense extreme UV (EUV) radiation field.

The interstellar gas reprocesses this radiation field, powering the observed nebular emission lines.  Uncertainties 
in the form of the ionizing spectrum thus have a significant effect on the interpretation of the nebular spectrum.   
While the implementation of new physics in population synthesis 
models (i.e., stellar multiplicity, rotation) continues to improve predictions of the EUV radiation field 
\citep[e.g.][]{Levesque2012,Eldridge2017,Gotberg2018}, the shape 
of the ionizing spectrum between 1 and 4 Ryd remains very poorly constrained for a given age and metallicity.  
This is perhaps especially true among the highest equivalent width line emitters for which the EUV spectrum is extremely 
sensitive to the nature of the hottest and most massive stars present in the galaxy.   While large spectroscopic surveys  
have begun to constrain the ionizing spectrum of typical galaxies at $z\simeq 2-3$ \citep[e.g.][]{Sanders2016,Steidel2016,Strom2017}, 
much less work has focused on the extreme emission line population. 
 
As the first nebular spectra of $z\gtrsim 6$ galaxies emerge, we are already 
beginning to see indications that the ionizing spectra of reionization-era sources differ greatly 
from typical galaxies at lower redshifts.  Perhaps most surprising has been the detection of strong nebular C {\scriptsize IV} emission 
in two of the first galaxies observed at $z\gtrsim 6$ \citep{Stark2015b,Mainali2017,Schmidt2017}, 
implying a very hard EUV ionizing spectrum capable of triply ionizing carbon in the ISM.  
The presence of C {\scriptsize IV} in a high redshift galaxy  typically indicates the power law spectrum of an active galactic nucleus (AGN).  
However, \citet{Mainali2017} have demonstrated that  the observed line ratios in one of the $z>6$ C {\scriptsize IV} emitters requires  
a strong break at the He$^+$-ionizing edge, suggesting a metal poor ($<0.1\ Z_\odot$) stellar ionizing spectrum is likely responsible 
for powering the nebular emission.   The precise form of the EUV spectrum produced by massive 
stars in this metallicity regime is very uncertain, making it difficult to reliably link the observed nebular emission to a unique 
physical picture. Similar challenges have emerged following detections of intense C {\scriptsize III}] emission 
in two galaxies at $z\gtrsim 6$ \citep{Stark2015a,Stark2017}.  Whereas some have 
argued that the UV line emission can be powered by moderately metal poor, young stellar populations \citep{Stark2017}, others suggest an AGN contribution is 
necessary to explain the large C {\scriptsize III}] equivalent width \citep{Nakajima2018}. Until more is known about the EUV radiation field of very young 
stellar populations that dominate in extreme emission line galaxies (EELGs), it will be difficult to reliably interpret the the $z\gtrsim 6$ nebular line spectra 
that will be common in the {\it JWST} era.

The uncertainties in the EUV spectra of early galaxies also plague efforts to use Ly$\alpha$ as a 
probe of reionization.  The success of the Ly$\alpha$ reionization test hinges on our ability to isolate the impact of the IGM on the 
evolving Ly$\alpha$ equivalent width distribution, requiring careful attention to variations in the 
intrinsic production of Ly$\alpha$ within the galaxy population.   The intrinsic Ly$\alpha$ luminosity of a 
given stellar population (per SFR) depends sensitively on the production efficiency of Lyman continuum (LyC) radiation \citep[e.g.][]{Wilkins2016}, 
often parameterized as $\xi_{\rm{ion}}=N(\rm{H}^0)/L_{\rm{UV}}$, the 
ratio of production rate of hydrogen ionizing photons ($N(\rm{H}^0)$) and the continuum luminosity of non-ionizing 
UV photons ($L_{\rm{UV}}$).   By definition, the extreme emission line galaxies that dominate in the reionization era should be 
extraordinarily efficient at producing hydrogen ionizing radiation.   Efforts to measure $\xi_{\rm{ion}}$ from both high 
quality spectra \citep[e.g.][]{Chevallard2018,Shivaei2018} and narrowband imaging \citep{Matthee2017} 
have ramped up in recent years.  \citet{Chevallard2018} demonstrated a correlation exists between 
$\xi_{\rm{ion}}$ and [O {\scriptsize III}] EW among ten nearby galaxies, such that the most extreme line emitters are the 
most efficient in their ionizing photon production.  The first hints at high redshift suggest a 
similar picture \citep{Matthee2017, Shivaei2018}, but the LyC production efficiency of extreme 
emission line galaxies remains poorly constrained.  As a result, it is not known the extent to which $\xi_{\rm{ion}}$ varies  over 
the full range of equivalent widths present in reionization era galaxies (EW$_{\rm{[OIII]+H}\beta}\simeq 300-2000$ \AA), making 
it challenging to accurately account for how variations in the radiation field are likely to impact Ly$\alpha$ visibility.  

The limitations of our physical picture of Ly$\alpha$ production in the reionization era has recently become apparent 
following the discovery of four massive $z\simeq 7-9$ galaxies with extremely large 
[O {\scriptsize III}]+H$\beta$ equivalent widths (EW$_{\rm{[OIII]+H}\beta}\simeq 1000-2000$ \AA; \citealt{Roberts-Borsani2016}, hereafter \citetalias{Roberts-Borsani2016}).   
Spectroscopic follow-up has revealed Ly$\alpha$ in each of the four 
galaxies from \citetalias{Roberts-Borsani2016} \citep[e.g.][]{Oesch2015,Zitrin2015,Stark2017}, corresponding to a factor of ten 
higher success rate than was found in earlier studies \citep[e.g.][]{Schenker2014,Pentericci2014}. 
The association between Ly$\alpha$ and extreme EW [O {\scriptsize III}] 
emission is also seen in other $z>7$ Ly$\alpha$ emitters with robust IRAC photometry \citep{Ono2012,Finkelstein2013}.
Remarkably these results suggest significant differences between the Ly$\alpha$ 
properties of the extreme line emitters that are typical at $z\gtrsim 7$ (EW$_{\rm{[OIII]+H}\beta}\simeq 670$ \AA) and 
massive systems with slightly larger equivalent widths (EW$_{\rm{[OIII]+H}\beta}\simeq 1000-2000$ \AA).   This could imply that the  
transmission of Ly$\alpha$ is enhanced in the higher equivalent width sources, as might be expected if these massive 
galaxies trace larger-than-average ionized patches of the IGM.  Or it could be explained if the production efficiency of 
Ly$\alpha$ is much larger in the higher EW sources.  In this case, efforts to model the evolving Ly$\alpha$ EW 
distribution would have to control for the large intrinsic variations in the Ly$\alpha$ luminosity per SFR that accompany sources 
of slightly different [O {\scriptsize III}] EW.  Unfortunately without robust measurements of how $\xi_{\rm{ion}}$ scales 
with [O {\scriptsize III}] EW at high redshift, it is impossible to know which picture is correct.

The shortcomings described above motivate the  need for a comprehensive investigation of the spectral properties of high redshift extreme 
emission line galaxies.   Prior to the launch of {\it JWST}, this is most easily accomplished at $z\simeq 2$ where both rest-UV 
and rest-frame optical nebular lines are visible from the ground.  In this paper, we present the first results from a  
near-infrared spectroscopic survey targeting rest-frame optical nebular emission lines in over 200 extreme emission line galaxies at $z=1.3-2.4$.   Our sample is 
selected using a combination of {\it HST} grism spectra \citep{Brammer2012,Skelton2014,Momcheva2016} and 
broadband photometry.   Over the past 2.5 years, we have obtained $\simeq 100$ hours of MMT and Keck spectroscopy following up 
this population in the near-infrared.  The program is ongoing, and we ultimately aim to obtain useful constraints on the strongest 
rest-frame optical lines ([O {\scriptsize II}], [Ne {\scriptsize III}], H$\beta$, [O {\scriptsize III}], H$\alpha$, [N {\scriptsize II}]) for each galaxy in our sample.  This 
program complements the KBSS \citep{Steidel2014} and MOSDEF \citep{Kriek2015} programs, including galaxies with lower masses and 
larger specific star formation rates than are common in these surveys. 

We focus on three key topics in this paper.  First we use our spectra to characterize how the production efficiency of LyC (and hence Ly$\alpha$) photons 
scales with the [O {\scriptsize III}] and H$\alpha$ EW.  Using this information, we consider how variations in Ly$\alpha$ production efficiency 
impact the visibility of Ly$\alpha$ at $z>7$.  Second, we investigate how the intense radiation field of extreme emission line galaxies 
impacts the  conditions of the ISM, characterizing how rest-frame optical line ratios that are sensitive to the ionization state of the gas vary with [O {\scriptsize III}] EW.  
We discuss the findings in the context of results showing an association between ionization-sensitive line ratios and the escape fraction of ionization radiation.  
Third we use our emission line measurements to help optimize spectroscopic observations of 
the highest redshift galaxies ($z\gtrsim 10$) that {\it JWST} will target.   At these redshifts, the strongest rest-frame optical lines ([O {\scriptsize III}], H$\alpha$) are  no 
longer visible with  NIRSpec.  We consider whether fainter rest-frame optical lines (i.e., [O {\scriptsize II}], [Ne {\scriptsize III}]) will provide viable alternatives and 
make predictions for the range of equivalent widths that are likely to be present in early galaxies.

The organization of this paper is as follows.  We describe the sample selection and  near-infrared spectroscopic 
observations in \S\ref{sec:observation}.  We introduce our photoionization modeling procedure in \S\ref{sec:modeling} before discussing the 
rest-frame optical spectroscopic properties of the EELGs in \S\ref{sec:phy_prop}. We discuss the implications in \S\ref{sec:discussion}, and summarize our results in \S\ref{sec:summary}.
We adopt a $\Lambda$-dominated, flat Universe with $\Omega_{\Lambda}=0.7$, $\Omega_{M}=0.3$ and
$\rm{H}_{0}=70$ km s$^{-1}$ Mpc$^{-1}$. 
All magnitudes in this paper are quoted in the AB system \citep{Oke1983}, and all equivalent widths are quoted in 
the rest-frame. We assume a \citet{Chabrier2003} initial mass function throughout the paper.


\section{Observations and Analysis} \label{sec:observation}

We have initiated a large spectroscopic survey of $z\simeq 2$ galaxies with 
extremely large equivalent width optical emission lines.  A 
complete description of the survey and the galaxies targeted 
will be presented in a catalog paper following the end of the survey.  Here 
we provide a detailed summary of the survey and resulting emission line sample, 
describing the pre-selection of targets  (\S\ref{sec:sample}) and our campaign to 
follow-up these sources with spectrographs on MMT and Keck (\S\ref{sec:mmirs_spec}$-$\ref{sec:rcs_spec}).  
We discuss the analysis of the spectra and our current redshift catalog in \S\ref{sec:emission_line}.   


\subsection{Pre-Selection of Extreme Emission Line Galaxies} \label{sec:sample}

The first step in our survey is to identify a robust sample of extreme emission line galaxies for spectroscopic 
follow-up.  Since we aim to study how the radiation field and gas conditions vary within the reionization 
era population, we must select galaxies at that span the full  range of 
equivalent widths expected at $z\gtrsim 6$ (EW$_{\rm{[OIII]+H}\beta}\simeq 300-2000$ \AA; e.g., \citealt{Stark2016}).   This pre-selection is most efficiently 
performed using publicly-available {\it HST} WFC3 G141 grism spectra in the five CANDELS \citep{Grogin2011,Koekemoer2011}
fields.  The G141 grism covers between $1.1\ \mu$m and $1.7\ \mu$m, enabling 
identification of H$\alpha$ at $0.7<z<1.6$ and [O {\scriptsize III}]$\lambda\lambda4959,5007$ 
(blended at the grism resolution) at $1.2<z<2.4$.    Four of the fields (AEGIS, COSMOS, GOODS-S, UDS) 
have been targeted by the 3D-HST program \citep{vanDokkum2011,Brammer2012,Skelton2014,Momcheva2016}, 
and the fifth field (GOODS-N) was observed in GO-11600 (PI:Weiner).    The latest 3D-HST public release brings together 
all the G141 data, providing photometry \citep{Skelton2014} and spectral measurements \citep{Momcheva2016} of galaxies 
in all five CANDELS fields. 

Our primary selection is focused on systems in two redshift windows 
($1.37\le z\le1.70$, $2.09\le z\le2.48$) where the full set of strong rest-frame optical lines ([O {\scriptsize II}], H$\beta$, [O {\scriptsize III}], H$\alpha$) 
can be observed with ground-based spectrographs.  
At $1.37\le z\le1.70$, we can detect H$\beta$ and [O {\scriptsize III}] in the $J$ band, and H$\alpha$ in the $H$ band.   
[O {\scriptsize II}] is situated in the $Y$ band at $1.55\le z\le1.70$ and in the optical at $z<1.55$.
At $2.09\le z\le2.48$, we can detect [O {\scriptsize II}] in the $J$ band, H$\beta$ and [O {\scriptsize III}] in the $H$ band, 
and H$\alpha$ in the $K$ band.   We also identify secondary targets at $1.70<z<1.90$ and $2.02<z<2.09$, 
where a subset of the lines are visible using ground-based spectrographs.  We will discuss the relative 
priorities we assign to galaxies when introducing the near-infrared spectroscopic follow-up in \S\ref{sec:mmirs_spec}.  

As a first pass at our  pre-selection, we use the 3D-HST v.4.1.5 grism catalogs (see \citealt{Momcheva2016}) to select 
extreme line emitters in all five CANDELS fields.  
These catalogs include extracted grism spectra for objects with WFC3 F140W magnitude $<26.0$.
We aim to select galaxies with optical line 
equivalent widths similar to those seen in the reionization era.  Stacked broadband SEDs  imply    
typical values of [O {\scriptsize III}]+H$\beta$ EW $= 670$ \AA\ at $z\simeq 8$ 
\citepalias{Labbe2013}.  To ensure our sample spans the full range of  line strengths seen at $z\simeq 8$, 
we consider all objects with [O {\scriptsize III}]+H$\beta$ EW $>340$ \AA, roughly a factor of two below 
the average value quoted above.   We thus select all galaxies in the grism catalogs with 
[O {\scriptsize III}]$\lambda\lambda4959,5007$ EW $>300$ \AA.\footnote{Adopting the theoretical flux ratio $I(5007)/I(4959)=3$, this selection 
is equivalently stated as [O {\scriptsize III}]$\lambda5007$ EW $>225$ \AA.}  Here we have assumed a 
flux ratio of [O {\scriptsize III}]$\lambda5007$/H$\beta$ $=6$, consistent with measurements 
of extreme emission line galaxies at $z\simeq 2$ \citep{Maseda2014}.  For the sake of simplicity, we 
use the same equivalent width threshold for H$\alpha$.  The precise value we choose does not impact 
our results, as we will  sample a wide range of equivalent widths in our follow-up survey.   For reference, if we crossmatch 
the MOSDEF spectroscopic sample to the 3D-HST grism catalogs (considering only galaxies in same redshift range as in our EELG 
sample), we find that the median 
[O {\scriptsize III}]$\lambda\lambda4959,5007$ EW is $154$ \AA\  for sources with S/N $>3$ grism line detections.  
Our sample thus largely probes a distinct [O {\scriptsize III}] EW regime.

The efficiency of our follow-up spectroscopy program depends sensitively on how accurately the 
equivalent widths can be measured from the grism spectra, requiring robust measurements of both the line flux and the underlying 
continuum.  To ensure the line detections 
are robust, we only include sources with significant (S/N $>3$) [O {\scriptsize III}] or H$\alpha$ flux and equivalent width measurements. 
However we find that the equivalent width S/N threshold  
excises many extreme line emitters from our sample owing to the low S/N continuum in many 
of the grism spectra.  For these fainter continuum sources, we can more reliably identify  
extreme line emitters using the line flux from the grism catalogs and the continuum from 
the {\it HST} broadband imaging data. 
We estimate the equivalent width errors for this subset by adding the uncertainties of emission line fluxes and broadband photometry in quadrature. 
If the rest-frame [O {\scriptsize III}]$\lambda\lambda4959,5007$ or H$\alpha$
equivalent width calculated in this manner is above $300$ \AA\ with S/N $>3$, we include 
the source in our catalog for follow-up spectroscopy.   

\begin{figure*}
\begin{center}
\includegraphics[width=\linewidth]{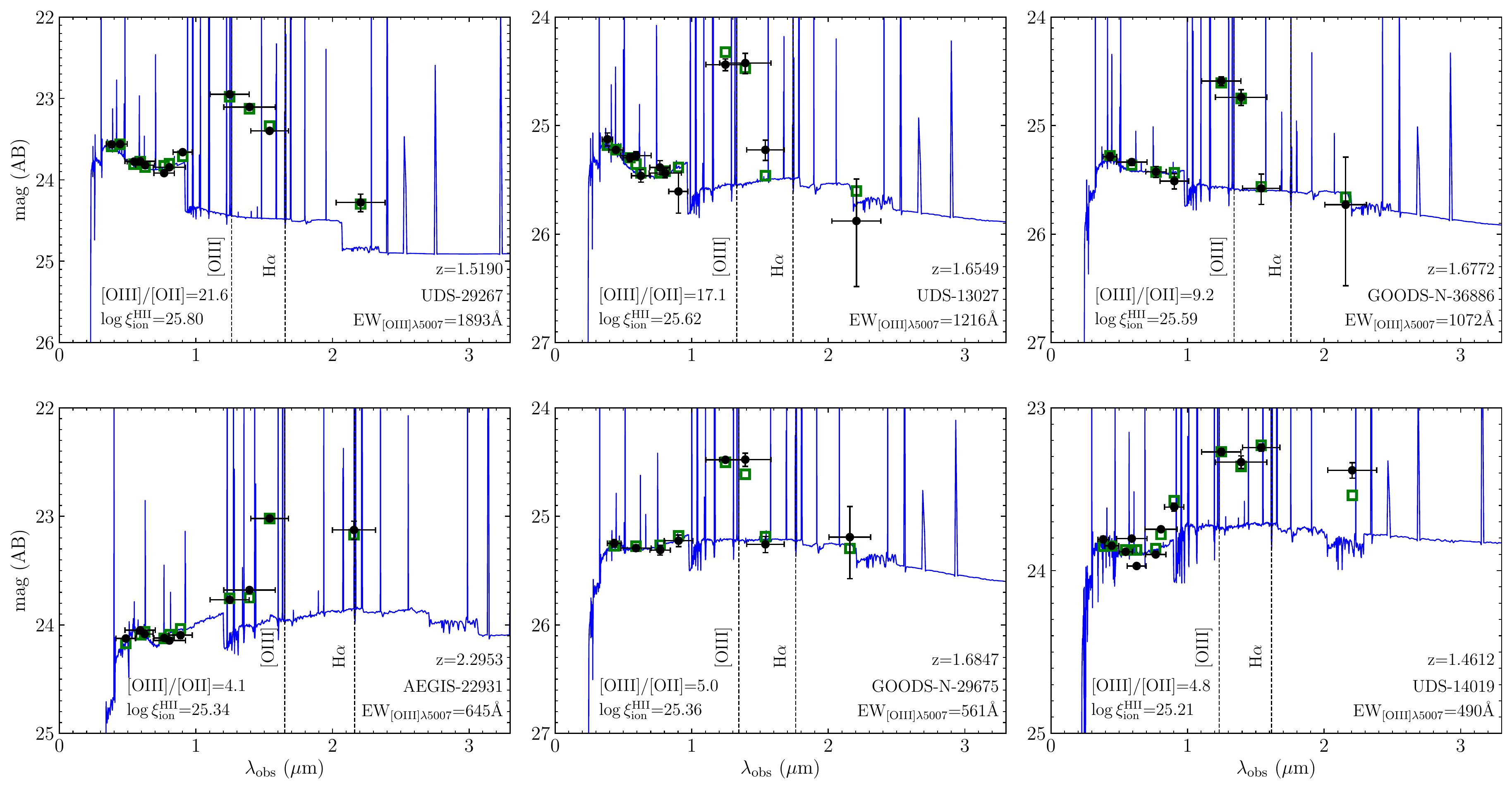}
\caption{Examples of the broadband SEDs of the extreme [O {\scriptsize III}] line emitting galaxies at $z=1.3-2.4$ in our sample. The upper panels show three of the sources with very large EWs in our 
sample ([O {\scriptsize III}]$\lambda5007$ EW $>1000$ \AA) that similar to the $z>7$ galaxies in \citetalias{Roberts-Borsani2016}. The lower panels show three 
of the sources with equivalent widths slightly above the average EW of $z>7$ galaxies ([O {\scriptsize III}]$\lambda5007$ EW $\simeq450$ \AA; \citetalias{Labbe2013}). Observed broadband 
photometry is shown as solid black circles. The best-fitting SED models (inferred from BEAGLE, see \S\ref{sec:modeling}) are plotted by solid blue lines, and synthetic photometry is shown as open 
green squares. Strong rest-frame optical emission lines, [O {\scriptsize III}]$\lambda5007$ and H$\alpha$, are highlighted by dashed black lines. 
Spectroscopic redshift, object ID, and rest-frame [O {\scriptsize III}]$\lambda5007$ EW are listed in each panel. 
The LyC photon production efficiency ($\xi^{\rm{HII}}_{\rm{ion}}$) and the ionization-sensitive line ratio [O {\scriptsize III}]/[O {\scriptsize II}] of each object are also listed.}
\label{fig:sed_blue}
\end{center}
\end{figure*}

Finally we visually inspect the grism spectra and photometry of all galaxies that satisfy these criteria, 
removing sources with grism spectra that are unreliable owing to contamination from overlapping spectra 
or those that appear to have misidentified emission lines. We are left with a sample of 1587 galaxies with grism 
measurements that imply [O {\scriptsize III}]$\lambda\lambda$4959,5007 or H$\alpha$ rest-frame EW greater than 
$300$ \AA\ across the five CANDELS fields. Examples of the broadband SEDs of the extreme emission 
line galaxies in our sample are shown in Figure \ref{fig:sed_blue}.  In all cases, the measured flux  in the specific broadband filters that sample 
the strong nebular emission lines are well above the flux of the underlying continuum.  

The grism observations cover $\sim75\%$ of the area imaged by the CANDELS survey.   While our 
spectroscopic follow-up observations are centered in these regions, often the edges of the slitmasks 
extend to areas of the CANDELS fields that lack grism spectra.  In these sub-regions, we use simple color 
cuts to identify extreme emission line galaxies based on the characteristic shape of their  broadband SEDs.  
In the redshift range $1.6<z<1.8$, galaxies with large equivalent width 
[O {\scriptsize III}]+H$\beta$ emission show a large flux excess in the $J_{\rm{125}}$-band (Figure \ref{fig:sed_blue}).  
Previous studies have successfully demonstrated that extreme [O {\scriptsize III}]+H$\beta$ emitters 
can be easily selected by identifying galaxies in which the 
$J_{\rm{125}}$-band flux is significantly greater than that in the $I$ and $H_{\rm{160}}$-bands 
 \citep{vanderWel2011,Maseda2014}.   We closely follow these selections, identifying galaxies in 
the \citet{Skelton2014} photometric catalogs that satisfy the following criteria: 
$I-J_{125} > 0.3+\sigma(I-J_{125})$ and $J_{125}-H_{160} < -0.3-\sigma(J_{125}-H_{160})$, where $\sigma$ is the color uncertainty.
The required $J$-band flux excess  is chosen to match the equivalent width threshold 
of our grism sample.  Whereas the grism selection identifies strong [O {\scriptsize III}]$\lambda\lambda$4959,5007 or 
H$\alpha$ emission, the photometric selection picks out sources based on their combined [O {\scriptsize III}]+H$\beta$ emission.  
Under the same assumptions described above, our color cuts correspond to a selection of 
galaxies at $1.6<z<1.8$ with [O {\scriptsize III}]+H$\beta$ EW  $>340$ \AA. 
We have verified (in fields where grism spectra are available) that this selection identifies the same extreme emission line galaxies selected via the grism catalogs.

Following \citet{Maseda2014}, we also select  [O {\scriptsize III}]+H$\beta$ emitter candidates at $2.0<z<2.4$ via 
an $H_{\rm{160}}$-band flux excess technique.  Because of the limited $K$-band sensitivity, the color cuts 
identify sources in which $H_{160}$ is significantly in excess of $J_{125}$.  To minimize contamination from Balmer break 
sources, \citet{Maseda2014} introduce a requirement that galaxies additionally have blue UV continuum (in $V-I$) slopes.  
We adopt similar color cuts, again making minor adjustments to the required color excess for the sake of consistency with our grism selection.  
In particular, we identify sources with $V_{606}-I < 0.1-\sigma(V_{606}-I)$ and
$J_{125}-H_{160} > 0.36+\sigma(J_{125}-H_{160})$.  We also require that the detections in  $V_{606}$, $I$,  $J_{125}$ and $H_{160}$ 
have S/N $>5$.    The two color section techniques provide an additional 612 targets, resulting in a 
total sample of 2199 galaxies in our input catalog. 

The final step is to remove galaxies that likely host AGNs from the sample.   While recent studies of EELG at $z\gtrsim 7$ 
suggests that AGN may be present in several of the sources for which we are obtaining spectra \citep[e.g.][]{Tilvi2016,Laporte2017,Mainali2018}, 
our goal is to focus first on the range of spectral properties in galaxies dominated by star formation.  
We use the deep \textit{Chandra} X-ray imaging in AEGIS \citep{Nandra2015}, COSMOS \citep{Civano2016}, 
GOODS-N \citep{Alexander2003,Xue2016}, and GOODS-S \citep{Xue2011} fields, as well as \textit{XMM} X-ray 
imaging in UDS field \citep{Ueda2008} to identify X-ray sources in the sample.  We match the coordinates of our 
EELG candidates to the X-ray source catalogs using a $1\farcs0$ search radius. There are 26 sources in our sample 
found to have X-ray counterparts within $1\farcs0$. We remove these 26 sources, leaving 2173 targets in the final grism 
spectroscopy and photometry-selected EELG candidate sample. 


\subsection{MMT/MMIRS Spectroscopy} \label{sec:mmirs_spec}

We use the catalog of extreme equivalent width line emitters described in \S\ref{sec:sample} as input for our ground-based spectroscopic 
follow-up program, the majority of which is conducted at the MMT using the MMT and Magellan Infrared Spectrograph 
(MMIRS; \citealt{McLeod2012}). MMIRS is a wide-field near-infrared imager and multi-object spectrograph (MOS) with a field of view of $4'\times6'.9$. We 
used the ``xfitmask\footnote{\url{http://hopper.si.edu/wiki/mmti/MMTI/MMIRS/ObsManual/MMIRS+Mask+Making}}'' software to design our MMIRS slit masks, 
using a slit length of $7\farcs0$ and a slit width of $1\farcs0$ for science targets.   Data were obtained over six observing runs 
between the 2015B and 2018A semesters.  Thus far, we have collected 81 hours of on-source integration with MMIRS, allowing us to 
target 313 galaxies on 17 separate masks, with 58 galaxies observed on two masks.

Our selection function is defined primarily by the redshift and equivalent width of the targets.  We first prioritize galaxies at $1.37\le z\le1.70$ 
and $2.09\le z\le2.48$, where the full set of strong rest-frame optical emission lines can be obtained from ground-based observations.
For galaxies at $1.37\le z\le1.70$, we give the highest priority to those at $1.55\le z\le1.70$, where 
all strong rest-frame optical lines can be probed by MMIRS observations.  Galaxies at $z<1.55$ will 
need additional follow-up with red-sensitive optical spectrographs to detect the [O {\scriptsize II}] emission line.  
We next adjust the target priority in each redshift interval 
based on the grism equivalent width or the flux excess implied by photometry.  Sources 
with the highest equivalent widths are extraordinarily rare, with only $15-20$ (3) galaxies per CANDELS field with 
[O {\scriptsize III}]$\lambda5007$ EW $>1000\ (2000)$ \AA.  To ensure that our sample contains an adequate number (i.e., $5-10$) of the 
largest equivalent width galaxies, we increase the priority of objects with the largest  equivalent widths in 
each redshift interval. 
This is the minimum number required for us to characterize the average line ratios in this [O {\scriptsize III}] EW range. 
Finally, we prioritize grism-selected targets over those identified by photometric excesses when 
both are available in the same region of a slitmask.

The details of the MMIRS masks obtained to date are summarized in Table \ref{tab:mmirs_obs}. 
 Spectra were taken with the \textit{J} grism + \textit{zJ} filter, \textit{H3000} grism + \textit{H} filter, and \textit{K3000} grism + \textit{Kspec} filter, 
  providing wavelength coverage of $0.95 - 1.50\ \mu$m, $1.50 - 1.79\ \mu$m and $1.95 - 2.45\ \mu$m, respectively. 
 The $1\farcs0$ slit widths result in resolving power of $R = 960$, $1200$, and $1200$ for \textit{J + zJ}, \textit{H3000 + H}, 
and \textit{K3000 + Kspec} grism and filter sets, respectively. 
The total integration time  in each filter ranges from 1 hr to 8 hr, with the specific value chosen depending on 
the predicted brightness of the set of emission lines we are targeting. 
 The average seeing  was between $0\farcs6$ and $1\farcs6$ (see Table \ref{tab:mmirs_obs}). 
 The masks all contained an isolated star to monitor the throughput and seeing during the observations and to compute the 
absolute flux calibration.  For each mask and filter combination, we also observed A0V stars at a similar airmass to derive 
the response spectrum and correct for telluric absorption. 

We reduced our MMIRS data using the publicly available data reduction pipeline\footnote{\url{https://bitbucket.org/chil\_sai/mmirs-pipeline}} developed 
by the instrument team. The MMIRS data reduction pipeline is implemented in {\footnotesize IDL} and described in \citet{Chilingarian2015}. 
Individual two-dimensional (2D) spectra are extracted from the original frames and flat-fielded after spectral tracing and optical distortion mapping. 
The wavelength calibration uses atmosphere airglow \textit{OH} lines, or the internal argon arc lamp if the \textit{OH} based computation fails. 
For sky subtraction, the pipeline creates a night sky spectrum model 
using science frames and the precise pixel-to-wavelength mapping, and subtracts the night sky emission from the 2D spectra. 
Finally, the pipeline provides telluric correction by first computing the atmosphere transmission curve 
(including spectral response function of the detector) from the observed telluric star spectra, and then correcting the transmission 
using a grid of atmospheric transmission models to account for the airmass difference between observations of the science 
target and the telluric star. The pipeline outputs the sky-subtracted and telluric-corrected 2D spectra of every slit on the mask. 

One-dimensional (1D) spectra are then extracted from the fully-reduced 2D spectra.  In cases where an emission line or 
continuum is detected with confidence, we use an optimal extraction procedure \citep{Horne1986}.  In all 
other cases we use a simple boxcar extraction, with the extraction aperture matched to the object spatial profile.   
We have verified that the extraction method does not significantly impact on the line flux measurements.   
The absolute flux calibration is then applied to the extracted 1D spectra using the slit stars. 
Slit loss corrections are performed following the similar procedures in \citet{Kriek2015}. We first 
extract a postage stamp of each galaxy from the \textit{HST} F160W image \citep{Skelton2014}, then 
smooth the postage stamp and fit the smoothed image with a 2D Gaussian profile. We compute the 
fraction of the light within the slit to that of the total Gaussian profile. Each spectrum is multiplied by the 
ratio of the in-slit light fraction measured for the slit star to that for each galaxy.


\begin{table*}
\begin{tabular}{|c|c|c|c|c|c|c|c|c|}
\hline
Mask Name & Number of Target & R.A. & Decl. & P.A. & Grism & Filter & Exposure Time & Average Seeing \\
 & & (hh:mm:ss) & (dd:mm:ss) & (deg) & & & (seconds) & ($''$) \\
(1) & (2) & (3) & (4) & (5) & (6) & (7) & (8) & (9) \\
\hline 
\hline
udsel1 & 24 & 2:17:49.000 & $-$5:13:02.00 & $-$11.00 & \textit{J} & \textit{zJ} & 14400 & 1.2 \\
udsel2 & 23 & 2:17:26.000 & $-$5:13:20.00 & $-$116.00 & \textit{J} & \textit{zJ} & 14400 & 0.8 \\
udsel3 & 24 & 2:17:26.000 & $-$5:12:13.00 & 88.00 & \textit{J} & \textit{zJ} & 28800 & 0.8 \\
udsel3 & 24 & 2:17:26.000 & $-$5:12:13.00 & 88.00 & \textit{H3000} & \textit{H} & 7200 & 0.8 \\
egsel3 & 24 & 14:19:25.973 & 52:50:08.43 & 177.00 & \textit{J} & \textit{zJ} & 21000 & 0.6 \\
egsel3 & 24 & 14:19:25.973 & 52:50:08.43 & 177.00 & \textit{H3000} & \textit{H} & 6000 & 1.0 \\
egsel4 & 23 & 14:19:58.000 & 52:55:04.01 & 179.00 & \textit{J} & \textit{zJ} & 10200 & 0.8 \\
egsel4 & 23 & 14:19:58.000 & 52:55:04.01 & 179.00 & \textit{H3000} & \textit{H} & 2400 & 1.0 \\
gdnel3 & 27 & 12:37:14.173 & 62:19:29.10 & $-$162.00 & \textit{J} & \textit{zJ} & 14400 & 0.8 \\
gdnel3 & 27 & 12:37:14.173 & 62:19:29.10 & $-$162.00 & \textit{H3000} & \textit{H} & 12000 & 1.2 \\
gdnel4 & 23 & 12:36:31.204 & 62:14:51.55 & $-$125.00 & \textit{J} & \textit{zJ} & 12000 & 0.8 \\
gdnel4 & 23 & 12:36:31.204 & 62:14:51.55 & $-$125.00 & \textit{H3000} & \textit{H} & 7200 & 0.8 \\
udsel5 & 14 & 2:17:17.088 & $-$5:14:53.13 & $-$86.00 & \textit{J} & \textit{zJ} & 7200 & 1.2 \\
udsel5 & 14 & 2:17:17.088 & $-$5:14:53.13 & $-$86.00 & \textit{H3000} & \textit{H} & 14400 & 1.2 \\
udsel6 & 21 & 2:17:49.000 & $-$5:13:02.00 & 106.00 & \textit{J} & \textit{zJ} & 7200 & 1.2 \\
udsel6 & 21 & 2:17:49.000 & $-$5:13:02.00 & 106.00 & \textit{H3000} & \textit{H} & 14400 & 1.6 \\
egsel1 & 20 & 14:19:37.358 & 52:50:22.97 & $-$179.00 & \textit{J} & \textit{zJ} & 14400 & 1.1 \\
egsel1 & 20 & 14:19:37.358 & 52:50:22.97 & $-$179.00 & \textit{K3000} & \textit{Kspec} & 3600 & 0.7 \\
egsel2 & 19 & 14:20:07.295 & 52:54:13.13 & $-$134.00 & \textit{J} & \textit{zJ} & 3600 & 0.6 \\
egsel2 & 19 & 14:20:07.295 & 52:54:13.13 & $-$134.00 & \textit{K3000} & \textit{Kspec} & 3600 & 1.6 \\
gdnel5 & 22 & 12:37:08.882 & 62:19:36.39 & $-$162.00 & \textit{H3000} & \textit{H} & 3600 & 0.8 \\
gdnel5 & 22 & 12:37:08.882 & 62:19:36.39 & $-$162.00 & \textit{K3000} & \textit{Kspec} & 3600 & 0.7 \\
gdnel6 & 17 & 12:37:22.384 & 62:18:02.04 & $-$143.00 & \textit{J} & \textit{zJ} & 7200 & 1.0 \\
egsel5 & 15 & 14:20:11.119 & 52:56:50.92 & 120.00 & \textit{J} & \textit{zJ} & 7200 & 0.8 \\
udsel7 & 27 & 2:17:15.656 & $-$5:14:04.59 & 93.00 & \textit{J} & \textit{zJ} & 10800 & 0.8 \\
udsel7 & 27 & 2:17:15.656 & $-$5:14:04.59 & 93.00 & \textit{K3000} & \textit{Kspec} & 3600 & 0.8 \\
udsel8 & 25 & 2:17:36.777 & $-$5:11:40.87 & $-$77.00 & \textit{J} & \textit{zJ} & 10800 & 0.8 \\
udsel8 & 25 & 2:17:36.777 & $-$5:11:40.87 & $-$77.00 & \textit{K3000} & \textit{Kspec} & 3600 & 0.8 \\
egsel6 & 23 & 14:19:15.000 & 52:46:29.00 & $-$146.00 & \textit{J} & \textit{zJ} & 13200 & 1.0 \\
egsel6 & 23 & 14:19:15.000 & 52:46:29.00 & $-$146.00 & \textit{H3000} & \textit{H} & 7200 & 1.0 \\
egsel6 & 23 & 14:19:15.000 & 52:46:29.00 & $-$146.00 & \textit{K3000} & \textit{Kspec} & 3600 & 1.0 \\
\hline
\end{tabular}
\caption{Summary of MMT/MMIRS observations. Column (1): mask name; Column (2): number of science 
targets on each mask, alignment stars and slit stars are not included; Column (3): right ascension of the 
mask center; Column (4): declination of the mask center; Column (5): position angle of the mask; 
Column (6): grism of the mask observed; Column (7): filter of the mask observed; Column (8): 
Total exposure time of the mask in each grism + filter set; Column (9): average seeing during the 
observation, in full width at half maximum (FWHM).}
\label{tab:mmirs_obs}
\end{table*}


\subsection{Keck/MOSFIRE Spectroscopy} \label{sec:mosfire_spec}

We also obtained spectra from the Multi-object Spectrometer for Infrared Exploration (MOSFIRE; \citealt{McLean2012}) 
on the Keck I telescope.  We placed galaxies on masks using the same selection function described in \S\ref{sec:mmirs_spec}.  
We used the MAGMA\footnote{\url{http://www2.keck.hawaii.edu/inst/mosfire/magma.html}} 
software to design the slit masks, and adopted a slit width of $0\farcs7$. Data were obtained on observing 
runs in 2015 November (one mask in the COSMOS field, one mask in the GOODS-S field, and one mask 
in the UDS field), and 2016 April (two masks in the AEGIS field). We have summarized the details of the 
Keck/MOSFIRE observations in Table \ref{tab:mosfire_obs}. Spectra were taken in the \textit{Y}, \textit{J}, 
and \textit{H} bands,  providing wavelength coverage of $0.972 - 1.125\ \mu$m, $1.153 - 1.352\ \mu$m, 
and $1.468 - 1.804\ \mu$m. The slit width of $0\farcs7$ results in resolving power of $R=3388$, $3318$, 
and $3660$ for \textit{Y}, \textit{J}, and \textit{H}, respectively.   The total integration time for each mask in 
each band ranges from 0.5 hr to 4.3 hr. The average seeing during the observations was between $0\farcs46$ and $0\farcs94$. 
The masks all contained $1-2$ isolated stars to monitor the throughput and seeing.  We use these stars 
to derive the absolute flux calibration. We also observed A0V stars in \textit{Y}, \textit{J}, and \textit{H} bands 
to derive the response spectrum and correct for telluric absorption.    In total, we observed 94 targets on 5 separate masks with MOSFIRE. 

Our MOSFIRE data were reduced using the publicly available data reduction 
pipeline\footnote{\url{https://www2.keck.hawaii.edu/inst/mosfire/drp.html}} (DRP), which is implemented 
in {\footnotesize PYTHON}. The MOSFIRE DRP first generates a flat-fielded image and traces the slit edge 
for each mask. Using a median-combined image of all science exposures, wavelength solutions are fit 
interactively for the central pixel in each slit using the night sky \textit{OH} emission lines, and propagated 
spatially along the slit.  After wavelength calibration and background subtraction, the two A$-$B and B$-$A stacks 
are shifted, combined and rectified to produce the final 2D spectra.  We derive the telluric correction spectra using 
longslit observations of a telluric standard star.  The absolute flux scaling factors are computed by comparing the 
count rates of slit star spectra with the flux densities in the 3D-HST photometric catalogs. Slit loss corrections 
were applied following the procedures in \citet{Kriek2015} as described in \S\ref{sec:mmirs_spec}. 


\begin{table*}
\begin{tabular}{|c|c|c|c|c|c|c|c|}
\hline
Mask Name & Number of Targets & R.A. & Decl. & P.A. & Band & Exposure Time & Average Seeing \\
 & & (hh:mm:ss) & (dd:mm:ss) & (deg) & & (seconds) & ($''$) \\
(1) & (2) & (3) & (4) & (5) & (6) & (7) & (8) \\
\hline 
\hline
cos\_y4 & 12 & 10:00:23.43 & 2:20:27.06 & 167.0 & \textit{Y} & 8640 & 0.74 \\
gs\_y5 & 16 & 3:32:14.63 & $-$27:44:14.42 & 158.0 & \textit{Y} & 11880 & 0.94 \\
uds\_yjh2 & 27 & 2:17:07.02 & $-$5:10:29.00 & 49.0 & \textit{Y} & 2160 & 0.52 \\
uds\_yjh2 & 27 & 2:17:07.02 & $-$5:10:29.00 & 49.0 & \textit{J} & 1680 & 0.54 \\
uds\_yjh2 & 27 & 2:17:07.02 & $-$5:10:29.00 & 49.0 & \textit{H} & 1920 & 0.46 \\
egs\_z8.6 & 20 & 14:20:02.22 & 52:54:49.08 & 292.0 & \textit{H} & 15360 & 0.72 \\
egs\_z7.5 & 19 & 14:20:28.82 & 53:00:55.53 & 107.0 & \textit{J} & 7080 & 0.90 \\
\hline
\end{tabular}
\caption{Summary of Keck/MOSFIRE observations. Column (1): mask name; Column (2): number of science targets on each mask, 
alignment stars and slit stars are not included; Column (3): right ascension of the mask center; Column (4): declination of the mask center; 
Column (5): position angle of the mask; Column (6): band of the mask observed; Column (7): Total exposure time of the mask in each 
grating; Column (8): average seeing during the observation, in full width at half maximum.}
\label{tab:mosfire_obs}
\end{table*}


\subsection{MMT/Red Channel Spectroscopy} \label{sec:rcs_spec}

For the subset of our extreme emission line sample at $z<1.55$, we require optical spectra to detect the [O {\scriptsize II}] doublet.  
As a first step toward obtaining such spectra, we observed two  galaxies at $z<1.55$ from our 
spectroscopic sample using the Red Channel Spectrograph \citep{Schmidt1989} on the MMT telescope  
on 2018 January 09 (UT).   The objects have  extremely large [O {\scriptsize III}]$\lambda\lambda4959,5007$ 
equivalent widths (EW $\simeq2500$ \AA\ for UDS-29267; EW $\simeq650$ \AA\ for UDS-14019) 
and bright continuum fluxes ($V_{606}=23.8$ for both galaxies), making them ideal targets for securing 
[O {\scriptsize II}] detections in feasible integrations with MMT Red Channel.    

The observations are summarized in Table \ref{tab:rcs_obs}.  We used the 1200 lines mm$^{-1}$ grating 
centered at $9290$ \AA\ with a UV-36 order blocking filter, providing spectral coverage of $805$ \AA. 
The observations were conducted in long slit mode, and we used a $1\farcs0$ slit width, providing a spectral 
resolution of $1.8$ \AA. We observed object UDS-29267 for 2 hours and UDS-14019 for 1.67 hours, 
with individual exposures of 20 minutes. Thin clouds were present during observations of both sources.   
The average seeing was $1\farcs1$ during the Red Channel observations of UDS-29267 and 
$1\farcs3$ for UDS-14019. 

The Red Channel long slit spectra were reduced using standard IRAF routines. Wavelength calibration was performed 
with HeAr/Ne arcs. We corrected the atmosphere transmission and instrument response by using the standard star 
spectrum. A slit star was also put on each slit, allowing us to perform an absolute flux calibration. 
Slit loss corrections were performed following the methods described in \S\ref{sec:mmirs_spec}.


\begin{table*}
\begin{tabular}{|c|c|c|c|c|c|c|c|}
\hline
Target Name & R.A. & Decl. & P.A. & $z_{\rm{spec}}$ & Configuration & Exposure Time & Average Seeing \\
 & (hh:mm:ss) & (dd:mm:ss) & (deg) & & line mm$^{-1}$ & (seconds) & ($''$) \\
 \hline 
\hline
UDS-29267 & 02:17:25.322 & $-$05:10:40.40 & $-168.99$ & $1.5190$ & 1200 & 7200 & 1.1 \\
UDS-14019 & 02:17:28.554 & $-$05:13:44.86 & $+114.81$ & $1.4616$ & 1200 & 6000 & 1.3 \\
\hline
\end{tabular}
\caption{Summary of MMT/Red Channel observation. Column (1): target name; Column (2): right ascension of the target; 
Column (3): declination of the target; Column (4): position angle of the slit; Column (5): spectroscopic redshift of the target; 
Column (6): configuration used; Column (7): Total exposure time on the target; 
Column (8): average seeing during the observation, in full width at half maximum.}
\label{tab:rcs_obs}
\end{table*}


\subsection{Emission Line Measurements} \label{sec:emission_line}

\begin{figure*}
\begin{center}
\includegraphics[width=\linewidth]{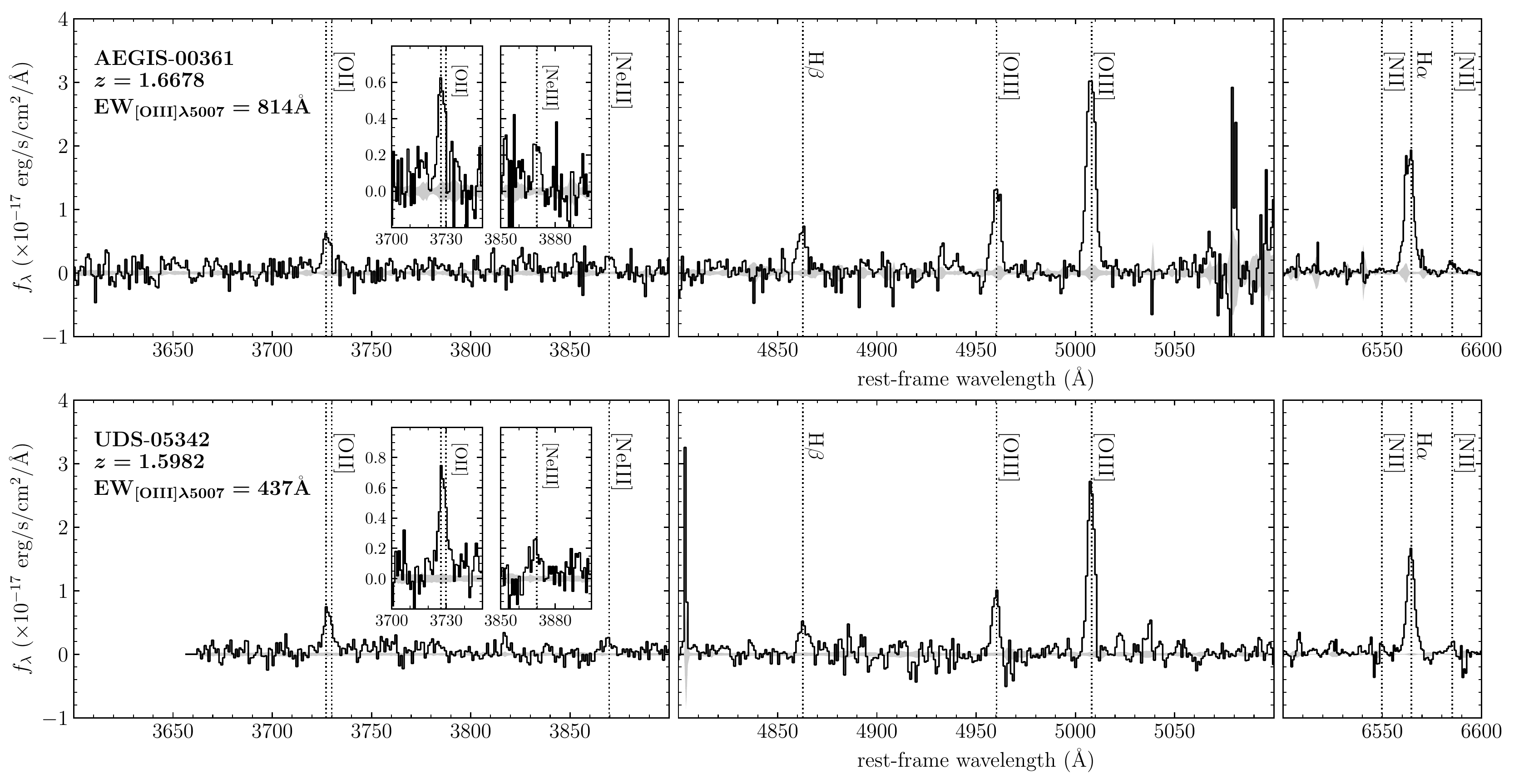}
\caption{Examples of 1D MMT and Keck spectra of two extreme [O {\scriptsize III}] emitters at $z=1.3-2.4$. 
Wavelengths are shifted to the rest-frame. The upper panels show the spectrum of a galaxy with 
similar equivalent width as the $z\sim7$ galaxies in the \citetalias{Roberts-Borsani2016} sample 
([O {\scriptsize III}]$\lambda5007$ EW $=814$ \AA). The lower panels show the spectrum 
of a galaxy with similar equivalent width as the average value of $z\sim7$ galaxies in \citetalias{Labbe2013} 
([O {\scriptsize III}]$\lambda5007$ EW $=437$ \AA). The $1\sigma$ error spectra are also 
shown as grey regions. The locations of strong rest-frame optical emission lines are highlighted by black 
dotted lines. Zoom-in spectra of relatively faint [O {\scriptsize II}] and [Ne {\scriptsize III}] emission lines 
are shown in the left panels. Emission line spectrum of the EW $=814$ \AA\ galaxy shows 
higher [O {\scriptsize III}]/[O {\scriptsize II}] and [Ne {\scriptsize III}]/[O {\scriptsize II}] ratios than that 
of the galaxy with EW $=437$ \AA.}
\label{fig:eelg_spec}
\end{center}
\end{figure*}

Examples of one-dimensional spectra are shown in Figure \ref{fig:eelg_spec}. 
Redshift confirmation is performed by visual inspection of the MMT and Keck spectra.  We require two emission line 
detections for a robust redshift measurement.  Currently we have confirmed redshifts of 240 of the galaxies 
from our grism input catalog of extreme line emitters (see \S\ref{sec:sample}).   The success rate for redshift confirmation is $92\%$ for  
those objects with complete near-infrared spectral coverage.  In the remaining 8\% of objects 
for which we fail to measure a redshift, the spectra are either very low S/N or the emission lines are 
contaminated by sky line residuals. 
The photometric flux excess selection 
contributes an additional 18 galaxies, resulting in a total sample of 258 extreme emission line 
galaxies with follow-up near-infrared spectra.  This includes 227 galaxies in the redshift range $1.3<z<2.4$; the 
remaining 31 galaxies are $z\simeq 1$ H$\alpha$ emitters.  Finally we measure redshifts for an  additional 35 galaxies
(included as filler targets) with equivalent widths below our selection threshold. 


\begin{figure}
\begin{center}
\includegraphics[width=\linewidth]{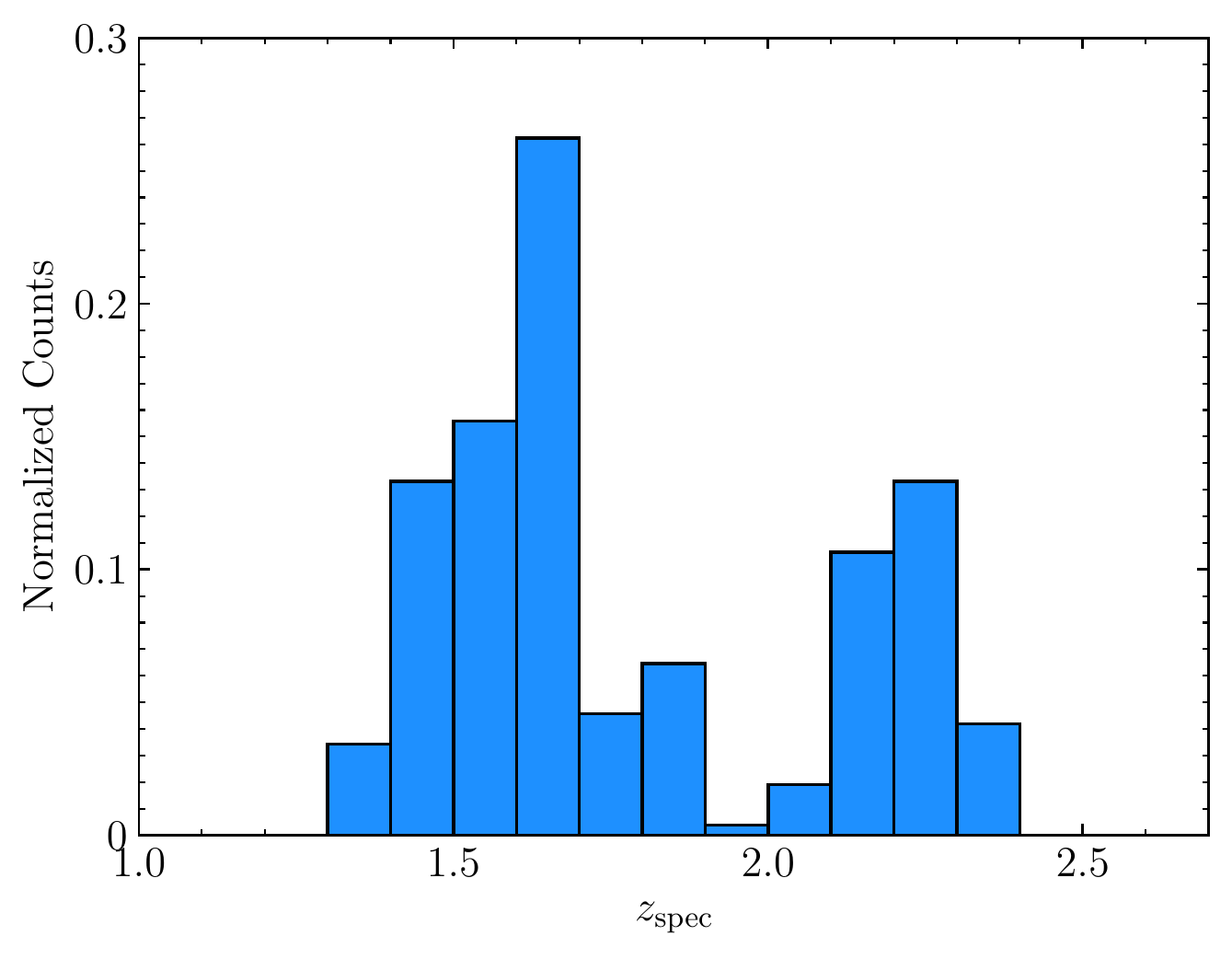}
\caption{The redshift distribution of the 227 spectroscopically-confirmed extreme emission line galaxies with rest-frame [O {\scriptsize III}]$\lambda\lambda4959,5007$ 
EW $>300$ \AA\ (or identically [O {\scriptsize III}]$\lambda5007$ EW $>225$ \AA) at $z=1.3-2.4$ in our sample.}
\label{fig:z_dist}
\end{center}
\end{figure}

The redshift distribution of the 227 galaxies at $z=1.3-2.4$  is shown in Figure \ref{fig:z_dist}.   
The distribution peaks in the $1.55\le z\le1.70$ redshift bin, with 85 of the 227 targets falling in this window. Roughly $29\%$ 
of the  sample ($66/227$) is at $z>2$, and an additional $21\%$ ($48/227$) is at $1.37<z<1.55$. 
As described above, this latter subset will require optical spectroscopic follow-up for measurement 
of [O {\scriptsize II}] and [Ne {\scriptsize III}].   
Our ultimate goal is to provide spectra coverage between [O {\scriptsize II}] and H$\alpha$ for the majority 
of extreme line emitters in this sample.  In our current sample, we have obtained complete coverage 
for 53 of the 227 galaxies.  The subset with spectral coverage between H$\beta$ and H$\alpha$ is larger (73), 
with 64 of the 73 sources having robust detections (S/N $>3$) of both H$\alpha$ and H$\beta$.
The completeness of our survey and the redshift distribution will change as we acquire more near-infrared 
and optical spectra in the future.  


Line fluxes are determined from fits to the extracted 1D spectra. 
We first fit the [O {\scriptsize III}]$\lambda5007$ emission line (or H$\alpha$ in the case that [O {\scriptsize III}] is not available) 
with a single Gaussian function.  The  central wavelength from the fit is then used to compute the redshift of the object, 
which we in turn use to identify the other emission lines.  In cases where the  lines 
are well-measured (i.e., S/N $>7$), we derive the flux using a Gaussian fit to the line profile; otherwise we derive 
the line flux using direct integration. In cases where nearby lines are partially resolved by MMIRS or MOSFIRE, 
we use multiple Gaussian functions to fit the data.   If the flux is measured with S/N $<2$, we consider the line 
undetected and derive a 2$\sigma$ upper limit.   We correct the H$\alpha$ and H$\beta$ fluxes for Balmer absorption using 
the best-fitting stellar population synthesis models (described in \S\ref{sec:modeling}).  We find that 
median correction for Balmer lines is less than $10\%$ of the measured emission line fluxes. 

The [O {\scriptsize III}]$\lambda5007$ emission line fluxes for the extreme line emitters range between $1.1\times$10$^{-17}$ 
erg cm$^{-2}$ s$^{-1}$ and $6.3\times$10$^{-16}$ erg cm$^{-2}$ s$^{-1}$  with a median of $7.5\times10^{-17}$ 
erg cm$^{-2}$ s$^{-1}$.  The range of [O {\scriptsize III}] fluxes is very similar to those of galaxies in the MOSDEF survey 
\citep{Kriek2015}; while the equivalent widths in our sample are larger, the continuum magnitudes tend to be fainter than typical MOSDEF sources. 
The H$\beta$ fluxes of our galaxies are fainter than [O {\scriptsize III}], with a median of $1.9\times$10$^{-17}$ erg cm$^{-2}$ s$^{-1}$.  
The emission line fluxes we derive from our ground-based spectra generally agree with the WFC3 grism spectra measurements.  
We find a median offset of $7\%$ and a scatter of $36\%$, in agreement with the comparison between MOSFIRE spectra 
and 3D-HST spectra reported by the MOSDEF survey \citep{Kriek2015}. 

We estimate the nebular attenuation by comparing the measured 
Balmer decrement (i.e., the observed H$\alpha$/H$\beta$ intensity ratio) with the value expected by 
Case B recombination in the case of zero dust reddening, H$\alpha$/H$\beta$ $=2.86$ (assuming 
electron temperature $T_e=10,000$ K; \citealt{Osterbrock2006}). 
To facilitate comparison with recent near-infrared spectroscopic studies of $z\sim2$ galaxies 
\citep[e.g.][]{Reddy2015,Steidel2016,Shivaei2018}, we assume the Galactic extinction 
curve in \citet{Cardelli1989} to compute the dust reddening toward H {\scriptsize II} regions.  We will 
discuss the extinction values implied by this analysis in \S\ref{sec:balmer}.

To investigate how the average spectral properties vary with optical line equivalent width, we create composite 
spectra by stacking galaxies in four bins of [O {\scriptsize III}]$\lambda5007$ EW: $0-225$ \AA, $225-450$ \AA, $450-800$ \AA, and $800-2500$ \AA.
We first shift individual spectra to the rest frame using the redshifts measured from [O {\scriptsize III}]$\lambda5007$ emission lines.  
Each spectrum is then interpolated to a common rest-frame wavelength scale of $1$ \AA\ and normalized by the 
individual H$\alpha$ luminosity (when investigating the average [Ne {\scriptsize III}]/[O {\scriptsize II}] ratio in \S\ref{sec:line_ratio}, 
we normalize the individual spectrum by its [O {\scriptsize III}] luminosity since the H$\alpha$ may not be observed in sources used for [Ne {\scriptsize III}]/[O {\scriptsize II}] analysis; 
see also Table \ref{tab:sample_size}). To minimize the contribution of sky lines from individual spectra, 
we stack spectra using  inverse-variance weighted  luminosities in each wavelength bin 
(i.e., weighted by $1/\sigma^2_i$, where $\sigma^2_i$ is the variation of the $i$th individual spectrum). 
To check that the resulting composites are not dominated by a few individual spectra, we also create 
composites using uniform weights and verify that the resulting spectra have similar line measurements as 
the weighted composite. 

\begin{figure*}
\begin{center}
\includegraphics[width=\linewidth]{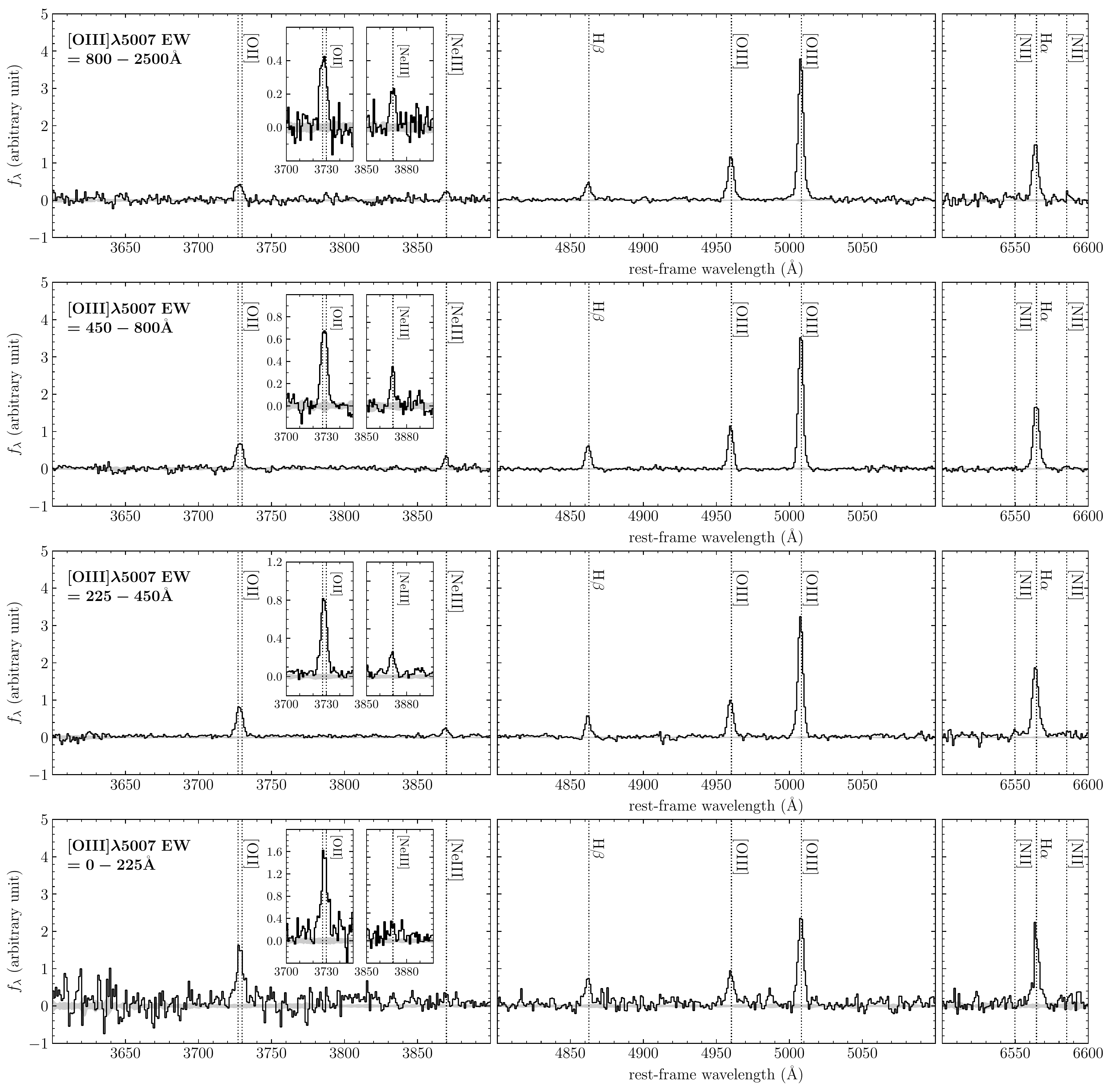}
\caption{MMT and Keck composite spectra of extreme [O {\scriptsize III}] emitters at $z=1.3-2.4$, grouped by 
sources with different [O {\scriptsize III}]$\lambda5007$ EWs. The spectra are ranked by [O {\scriptsize III}]$\lambda5007$ 
EW ranges, from the most extreme line emitters ([O {\scriptsize III}]$\lambda5007$ EW $=800-2500$ \AA) on the top to the 
objects with similar EWs as the typical galaxies at $z\sim2$ ([O {\scriptsize III}]$\lambda5007$ EW $=0-225$ \AA) in the bottom.
The $1\sigma$ error spectra are also shown as grey regions. The locations of strong rest-frame optical emission lines are 
highlighted by black dotted lines. Zoom-in spectra of relatively faint [O {\scriptsize II}] and [Ne {\scriptsize III}] emission lines 
are shown in the left panels. The composite spectra of galaxies with higher [O {\scriptsize III}]$\lambda5007$ EWs show higher 
[O {\scriptsize III}]/[O {\scriptsize II}] and [Ne {\scriptsize III}]/[O {\scriptsize II}] ratios than those of galaxies with lower EWs.}
\label{fig:comp_spec}
\end{center}
\end{figure*}

The composite spectra for all the four [O {\scriptsize III}]$\lambda5007$ EW bins are shown in Figure \ref{fig:comp_spec}. 
For these spectra, we have included the subset of our current sample with complete coverage between [O {\scriptsize II}] and H$\alpha$.   
There are 7, 26, 13, and 14 galaxies included in the $0-225$ \AA, $225-450$ \AA, $450-800$ \AA, and $800-2500$ \AA\ stacks shown in this figure, respectively.
We measure the luminosities of the strong rest-frame optical lines ([O {\scriptsize II}], [Ne {\scriptsize III}], H$\beta$, [O {\scriptsize III}], and H$\alpha$) in 
each of the four composites.   In particular, we are interested in deriving the dependence of the 
Balmer decrement, and the O32 and Ne3O2 indices on the [O {\scriptsize III}]$\lambda5007$ 
equivalent width.  Each line ratio requires slightly different spectral coverage for a robust measurement.  
For example, the Balmer decrement requires spectra that span between 
H$\beta$ and H$\alpha$, whereas the dust-corrected O32 measurement requires sources with coverage between [O {\scriptsize II}] and H$\alpha$, 
and the Ne3O2 measurement requires sources with coverage between [O {\scriptsize II}] and [O {\scriptsize III}].
Accordingly, we can include more objects in the stacks used to investigate the dependence of the Balmer decrement 
on the [O {\scriptsize III}]$\lambda5007$ EW.  We will discuss the resulting line ratios in \S\ref{sec:phy_prop}.
The detailed spectral coverage requirements and the resulting number of objects included in the composite are summarized in Table \ref{tab:sample_size}.


\begin{table*}
\begin{tabular}{|c|c|c|}
\hline
Sample & Selection criteria & $N$ \\
\hline 
\hline
Extreme [O {\scriptsize III}] emitters at $z=1.3-2.4$ & $z=1.3-2.4$ & $227$ \\
H$\alpha$/H$\beta^1$ (individual) & [O {\scriptsize III}]$\lambda5007$ $>5\sigma$, H$\beta$ and H$\alpha$ $>3\sigma$ & $64$ \\
H$\alpha$/H$\beta^1$ (composite) & [O {\scriptsize III}]$\lambda5007$ $>5\sigma$, spectral coverage between H$\beta$ and H$\alpha$ & $73$ \\
{[O {\scriptsize III}]/[O {\scriptsize II}]}$^2$ (individual) & [O {\scriptsize III}]$\lambda5007$ $>5\sigma$, [O {\scriptsize II}]$\lambda\lambda3727,3729$, H$\beta$ and H$\alpha$ $>3\sigma$ & $44$ \\
{[O {\scriptsize III}]/[O {\scriptsize II}]}$^2$ (composite) & [O {\scriptsize III}]$\lambda5007$ $>5\sigma$, spectral coverage between [O {\scriptsize II}]$\lambda\lambda3727,3729$ and H$\alpha$ & $53$ \\
{[Ne {\scriptsize III}]/[O {\scriptsize II}]}$^3$ (individual) & [O {\scriptsize III}]$\lambda5007$ $>5\sigma$, [O {\scriptsize II}]$\lambda\lambda3727,3729$ and [Ne {\scriptsize III}]$\lambda3869$ $>2\sigma$ & $26$ \\
{[Ne {\scriptsize III}]/[O {\scriptsize II}]}$^3$ (composite) & [O {\scriptsize III}]$\lambda5007$ $>5\sigma$, spectral coverage between [O {\scriptsize II}]$\lambda\lambda3727,3729$ and [O {\scriptsize III}]$\lambda5007$ & $56$ \\
\hline
\end{tabular}
\caption{Statistics for sub-samples of extreme [O {\scriptsize III}] emitters ([O {\scriptsize III}]$\lambda5007$ EW $>225$ \AA) used for line ratio analysis in \S\ref{sec:phy_prop}. 
Spectral coverage requirements and the resulting sub-sample sizes are listed. For each line ratio analysis, individual objects with robust line ratio 
(Balmer decrement H$\alpha$/H$\beta$, [O {\scriptsize III}]/[O {\scriptsize II}], or [Ne {\scriptsize III}]/[O {\scriptsize II}]) measurements have more strict requirements than 
objects used to make composite spectra. For example, the Balmer decrement requires spectra that span between H$\beta$ and H$\alpha$ (73 sources), 
while measuring robust Balmer decrement in an individual object also requires detecting H$\beta$ and H$\alpha$ with $3\sigma$ significance or higher (64 sources).}
\raggedright {\bf Notes.} \\
$^1$ Sub-sample used for Balmer decrement analysis in \S\ref{sec:balmer}. \\
$^2$ Sub-sample used for [O {\scriptsize III}]/[O {\scriptsize II}] ratio analysis in \S\ref{sec:line_ratio}. \\
$^3$ Sub-sample used for [Ne {\scriptsize III}]/[O {\scriptsize II}] ratio analysis in \S\ref{sec:line_ratio}. \\
\label{tab:sample_size}
\end{table*}


\section{Photoionization Modeling} \label{sec:modeling}

We fit the broadband fluxes of galaxies in our spectroscopic sample using the Bayesian galaxy SED modeling and 
interpreting tool BEAGLE (for BayEsian Analysis of GaLaxy sEds, version 0.10.5; \citealt{Chevallard2016}), which incorporates 
in a consistent way the production of radiation from stars and its transfer through the interstellar and intergalactic media. Broadband 
photometry is obtained from 3D-HST using the \citet{Skelton2014} catalog, and we utilize multi-wavelength data covering $0.3-2.5\ \mu$m. 
We also test the impact of adding {\it Spitzer}/IRAC constraints for the subset of galaxies that are not strongly confused. 
For each object, we remove fluxes in filters that lie blueward of Ly$\alpha$ to avoid introducing uncertain contributions from Ly$\alpha$ 
emission and Ly$\alpha$ forest absorption. We also simultaneously fit the available strong rest-frame optical emission line fluxes 
([O {\scriptsize II}]$\lambda\lambda3727,3729$, H$\beta$, [O {\scriptsize III}]$\lambda\lambda4959,5007$, and H$\alpha$). 
The version of BEAGLE used in this work adopts the recent photoionization models of star-forming galaxies of \citet{Gutkin2016}, 
which describes the emission from stars and interstellar gas based on the combination of the latest version of \citet{Bruzual2003} 
stellar population synthesis model with the photoionization code CLOUDY \citep{Ferland2013}. The main adjustable 
parameters of the photoionized gas are the interstellar metallicity, $Z_{\rm{ISM}}$, the typical ionization parameter 
of a newly ionized H {\scriptsize II} region, $U_{\rm{S}}$ (which characterizes the ratio of ionizing-photon to gas densities 
at the edge of the Str$\ddot{\rm{o}}$mgren sphere), and the dust-to-metal (mass) ratio, $\xi_{\rm{d}}$ 
(which characterizes the depletion of metals on to dust grains). We consider models with C/O abundance ratio 
equal to the standard value in nearby galaxies (C/O)$_{\odot}\approx0.44$. 
The chosen C/O ratio  does not significantly impact the results presented in this paper.  We
will explore the impact of C/O variations on the derived gas properties in a follow-up paper focused more closely on 
the photoionization model results.  To account for the effect of dust attenuation, we first assume the \citet{Calzetti2000} 
extinction curve.  We also fit galaxies assuming the Small Magellanic Cloud (SMC) extinction curve in \citet{Pei1992}. 
Finally, we adopt the prescription of \citet{Inoue2014} to include the absorption of IGM.

We assume constant star formation history for model galaxies in BEAGLE, and parameterize the 
maximum age of stars in a model galaxy in the range from 1 Myr 
to the age of the universe at the given redshift.  We fix the redshift of each object to the 
spectroscopic redshift measured from the MMIRS or MOSFIRE spectra. We adopt a standard \citet{Chabrier2003} 
initial mass function and assume that all stars in a given galaxy have the same metallicity, in the range 
$-2.2 \le \log{(Z/Z_{\odot})} \le 0.25$, where BEAGLE uses the solar metallicity value $Z_{\odot}=0.01524$ 
from \citet{Caffau2011}.  The interstellar metallicity is assumed to be the same as the stellar metallicity 
($Z_{\rm{ISM}}=Z_\star$) for each object.  The ionization parameter is allowed to freely vary in the range 
$-4.0\le\log{U_{\rm{S}}}\le-1.0$, and the dust-to-metal mass ratio is allowed to span the range $\xi_{\rm{d}} = 0.1-0.5$.  
We adopt an exponential distribution prior on the V-band dust attenuation optical depths, fixing the fraction 
of attenuation optical depth arising from dust in the diffuse ISM to $\mu=0.4$ (see \citealt{Chevallard2016}).  With the above parameterization, 
we use the BEAGLE tool to fit the broadband SEDs and available emission line constraints for the $z=1.3-2.4$ galaxies 
in our sample. Emission line fluxes and broadband fluxes are put on the same  absolute scale using the aperture 
correction procedures described in \S\ref{sec:mmirs_spec}. 
We obtain the output posterior probability distributions of the free 
parameters described above, and those of other derived physical parameters such as the ionizing photon 
production efficiency inferred from model ($\xi^\star_{\rm{ion}}$).  We use the posterior median value 
as the best-fitting value of each parameter.    In Figure \ref{fig:sed_blue}, we overlay the best-fitting 
BEAGLE models on the broadband SEDs.   It is clear that the models generally do a good job recovering the 
shape of the continuum and the large flux excesses caused by nebular emission.  

\begin{figure*}
\begin{center}
\includegraphics[width=\linewidth]{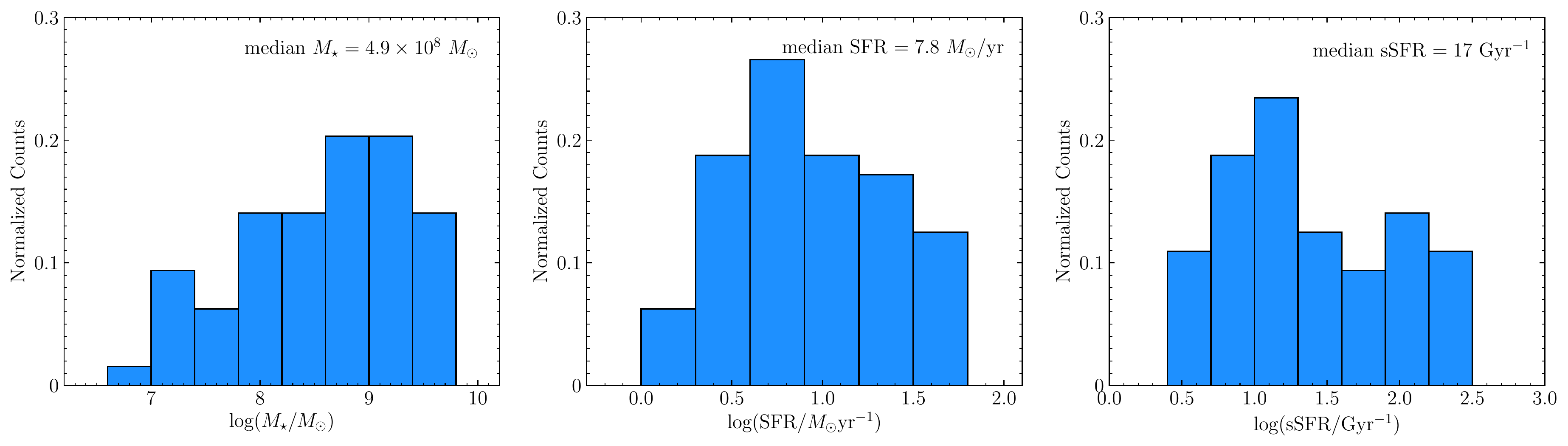}
\caption{Stellar mass, star formation rate, and specific star formation rate distributions for $z=1.3-2.4$ extreme [O {\scriptsize III}] emitters ([O {\scriptsize III}]$\lambda5007$ EW $>225$ \AA) with robust H$\alpha$ and H$\beta$ measurements (S/N $>3$) in our sample (64 of total 73 extreme [O {\scriptsize III}] emitters with spectral coverage between H$\beta$ and H$\alpha$). Stellar masses, SFRs, and sSFRs are derived from stellar population synthesis modeling using BEAGLE. The median stellar mass is $4.9\times10^8\ M_{\odot}$, the median of SFR is $7.8\ M_{\odot}$yr$^{-1}$, and the median of sSFR is $17$ Gyr$^{-1}$.}
\label{fig:star_dist}
\end{center}
\end{figure*}

\begin{figure*}
\begin{center}
\includegraphics[width=\linewidth]{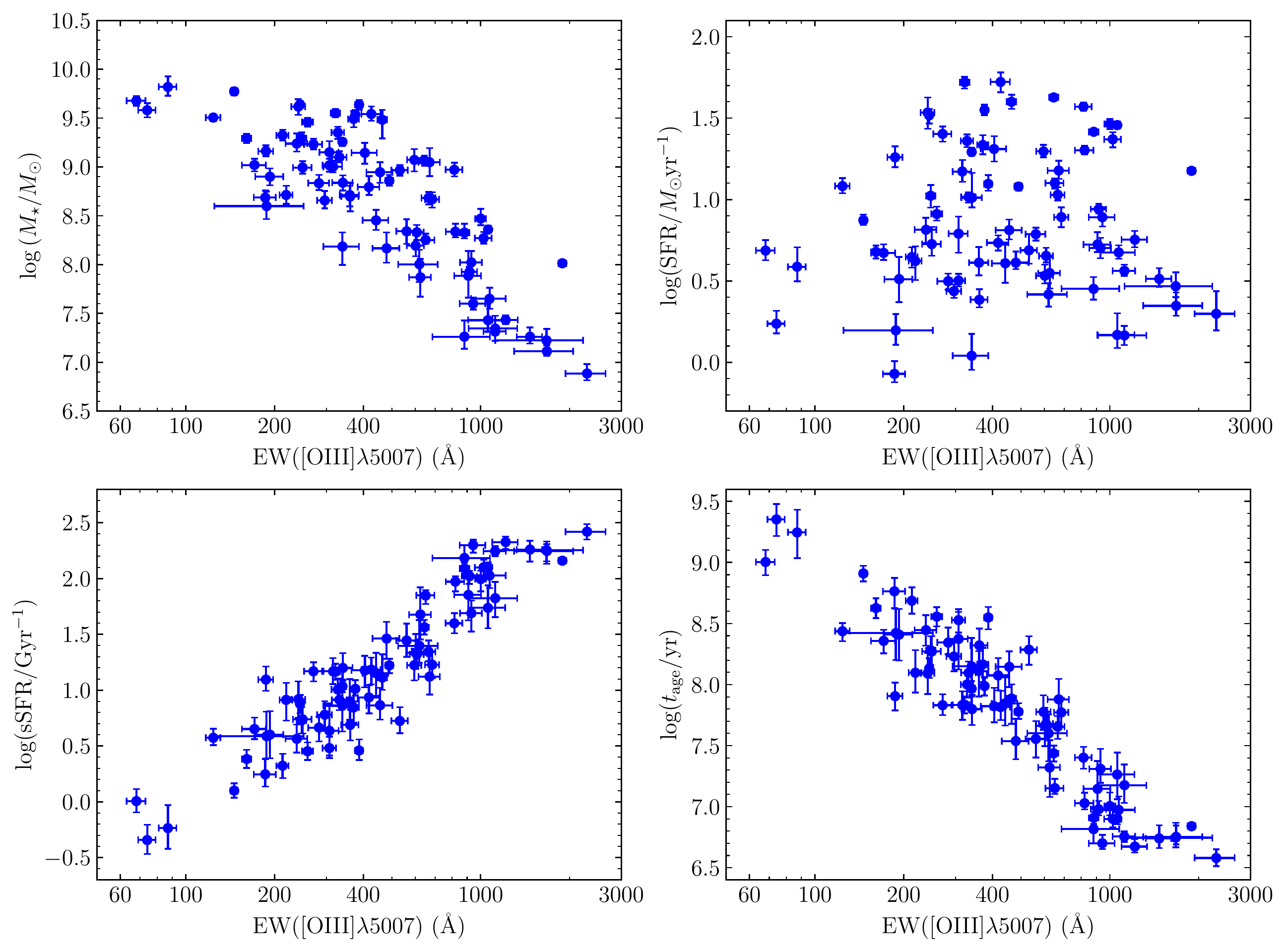}
\caption{Physical parameters (stellar mass, star formation rate, specific star formation rate, and stellar age) as functions of [O {\scriptsize III}]$\lambda5007$ EW for the 77 emission line galaxies with robust H$\alpha$ and H$\beta$ measurements (S/N $>3$) at $z=1.3-2.4$ in our sample (64 extreme [O {\scriptsize III}] emitters with EW$_{\rm{[OIII]}\lambda5007}>225$ \AA\ and 13 additional galaxies with lower EWs). All the parameters are derived from stellar population synthesis modeling using BEAGLE. The upper left, the upper right, the lower left, and the lower right panel shows the stellar mass, the star formation rate, the specific star formation rate, and the stellar age versus [O {\scriptsize III}]$\lambda5007$ EW, respectively.}
\label{fig:ew_star}
\end{center}
\end{figure*}

The distribution of stellar masses implied by the BEAGLE models is shown in the left panel of Figure \ref{fig:star_dist}.  
Here we include the 64 spectroscopically-confirmed extreme [O {\scriptsize III}] emitters with spectral coverage between H$\beta$ and H$\alpha$, 
and with significant detections of H$\beta$ and H$\alpha$ (S/N $>3$). The median stellar mass in this sample is $4.9\times10^8\ M_\odot$, 
well below the typical stellar masses ($\simeq10^{10}\ M_\odot$) found for $z\simeq 2.3$ galaxies in the KBSS and MOSDEF surveys \citep[e.g.][]{Strom2017,Sanders2018}.  
We find that galaxies in our sample with the largest [O {\scriptsize III}] equivalent widths have the lowest stellar masses 
(see  the upper left panel of Figure \ref{fig:ew_star}).  
Among the subset of sources with 
EW$_{\rm{[OIII]}\lambda5007}$ in excess of $800$ \AA, the median stellar mass is just $4.5\times10^7\ M_\odot$.   
This increases to $4.7\times10^8\ M_\odot$ for sources with 
$450$ \AA\ $<\rm{EW}_{\rm{[OIII]}\lambda5007}<800$ \AA\ and $1.4\times10^9\ M_\odot$
for $225$ \AA\ $<\rm{EW}_{\rm{[OIII]}\lambda5007}<450$ \AA.  This variation is to be expected.  
The near-infrared filters which are sensitive to the stellar 
mass are heavily contaminated by nebular emission (lines and continuum) 
in the highest equivalent width systems, implying a 
smaller contribution from stellar continuum for a given near-infrared magnitude.  This effect is further amplified 
by the dependence of the stellar mass to light ratio on the age of the stellar population.   
We note that our results do not change significantly when we add {\it Spitzer}/IRAC constraints. 
In particular, we find that the derived stellar masses are always within $0.1$ dex of the values derived without IRAC constraints.


In the middle and right panels of Figure \ref{fig:star_dist}, we  show the best-fitting model SFR and sSFR 
of the same 64 line emitters described above.  The median of the SFR distribution is 
$7.8\ M_{\odot}$yr$^{-1}$.  In contrast to the stellar mass, we do not find strong variations 
in the SFR with [O {\scriptsize III}] EW.  The median sSFR is $17$ Gyr$^{-1}$, well above the average 
values ($\simeq 2$ Gyr$^{-1}$) for galaxies in the KBSS and MOSDEF surveys 
\citep[e.g.][]{Strom2017,Sanders2018}.   As expected, the sSFR increases with [O {\scriptsize III}] EW within 
our sample.  The median sSFR ranges from $7.9$ Gyr$^{-1}$ ($225$ \AA\ $<\rm{EW}_{\rm{[OIII]}\lambda5007}<450$ \AA) 
to $21$ Gyr$^{-1}$ ($450$ \AA\ $<\rm{EW}_{\rm{[OIII]}\lambda5007}<800$ \AA) to 
$125$ Gyr$^{-1}$ ($800$ \AA\ $<\rm{EW}_{\rm{[OIII]}\lambda5007}<2500$ \AA).   
The strong variation in sSFR seen in Figure \ref{fig:ew_star} can equivalently be described as a trend in the luminosity 
weighted age of the stellar population.  Under our assumed constant star formation history, the 
BEAGLE models predict that the median of the maximum stellar age parameter is 130, 50, and 8 Myr for the three 
[O {\scriptsize III}]$\lambda5007$ equivalent width bins described above.  These very young values only refer to 
the  burst that is currently dominating the observed SED and do not negate the presence of 
faint older stars from earlier activity. 


We re-calculate  equivalent widths using the newly-obtained near-infrared spectra 
together with the underlying continuum predicted by BEAGLE.  The  models provide an improved 
determination of the continuum, accounting for the emission line contamination in the near-infrared 
broadband filters.  The ground-based spectra provide higher S/N detections of the fainter lines, 
while also allowing us to separate doublets that are blended in the grism spectra.  We calculate 
the rest-frame equivalent widths of [O {\scriptsize II}], H$\beta$, [O {\scriptsize III}]$\lambda5007$, and 
H$\alpha$.  We will use these measurements when investigating trends with rest-frame optical equivalent widths. 
In Table \ref{tab:eelg_sum}, we report the [O {\scriptsize III}]$\lambda5007$ and H$\alpha$ equivalent widths of 
the 64 extreme line emitters (EW$_{\rm{[OIII]}\lambda5007}>225$ \AA) at $z=1.3-2.4$ with robust (S/N $>$3) H$\alpha$ 
and H$\beta$  detections.  We will release the full optical line properties of our sample in 
a catalog paper after the survey is completed.


\begin{table*}
\begin{tabular}{|c|c|c|c|c|c|c|}
\hline
Object ID & R.A. & Decl. & $z_{\rm{spec}}$ & $m_{\rm{F606W}}$ & EW$_{\rm{[OIII]}\lambda5007}$ & EW$_{\rm{H}\alpha}$ \\
 & (hh:mm:ss) & (dd:mm:ss) & & (AB) & (\AA) & (\AA) \\
\hline 
\hline
AEGIS-29446 & 14:19:31.989 & $+$52:53:22.789 & $1.5506$ & $26.13\pm0.10$ & $1059\pm152$ & $ 662\pm111$ \\
AEGIS-26095 & 14:19:33.549 & $+$52:52:52.797 & $1.5625$ & $26.19\pm0.11$ & $ 340\pm 47$ & $ 164\pm 26$ \\
AEGIS-22931 & 14:19:33.787 & $+$52:52:07.560 & $2.2953$ & $24.05\pm0.02$ & $ 645\pm 23$ & $ 790\pm 51$ \\
AEGIS-23885 & 14:19:31.483 & $+$52:51:57.961 & $1.5721$ & $24.64\pm0.03$ & $ 608\pm 33$ & $ 578\pm 38$ \\
AEGIS-21396 & 14:19:32.014 & $+$52:51:27.694 & $2.3206$ & $25.35\pm0.05$ & $ 241\pm 13$ & $ 458\pm173$ \\
AEGIS-31255 & 14:19:17.457 & $+$52:51:09.896 & $1.5970$ & $24.42\pm0.03$ & $ 532\pm 34$ & $ 463\pm 36$ \\
AEGIS-19479 & 14:19:31.179 & $+$52:50:52.304 & $1.5969$ & $24.79\pm0.03$ & $ 296\pm 17$ & $ 366\pm 27$ \\
AEGIS-20493 & 14:19:24.174 & $+$52:49:48.020 & $1.6725$ & $23.73\pm0.01$ & $ 376\pm  7$ & $ 498\pm 13$ \\
AEGIS-11745 & 14:19:33.812 & $+$52:49:28.451 & $1.6069$ & $26.15\pm0.10$ & $ 620\pm 95$ & $ 728\pm165$ \\
AEGIS-11452 & 14:19:26.785 & $+$52:48:04.364 & $1.6720$ & $23.33\pm0.02$ & $ 464\pm 17$ & $ 324\pm 14$ \\
AEGIS-03127 & 14:20:07.698 & $+$52:53:11.858 & $2.3089$ & $24.51\pm0.03$ & $ 671\pm 56$ & $ 686\pm 90$ \\
AEGIS-14156 & 14:19:50.376 & $+$52:52:59.100 & $1.6738$ & $24.86\pm0.03$ & $ 249\pm 18$ & $ 249\pm 38$ \\
AEGIS-00361 & 14:20:21.976 & $+$52:54:46.588 & $1.6678$ & $23.78\pm0.01$ & $ 814\pm 53$ & $1028\pm 57$ \\
AEGIS-15032 & 14:19:26.935 & $+$52:49:02.770 & $1.6132$ & $26.21\pm0.10$ & $1121\pm210$ & $ 716\pm170$ \\
AEGIS-04337 & 14:19:35.878 & $+$52:47:54.834 & $1.3985$ & $26.80\pm0.16$ & $2296\pm356$ & $1876\pm285$ \\
AEGIS-04711 & 14:19:34.958 & $+$52:47:50.219 & $2.1839$ & $24.06\pm0.01$ & $1060\pm 25$ & $ 671\pm 33$ \\
AEGIS-16513 & 14:19:19.204 & $+$52:48:00.052 & $1.4633$ & $24.78\pm0.03$ & $ 361\pm 24$ & $ 351\pm 28$ \\
AEGIS-24361 & 14:19:08.525 & $+$52:47:59.956 & $1.4714$ & $25.86\pm0.07$ & $1471\pm145$ & $1192\pm156$ \\
AEGIS-08869 & 14:19:20.544 & $+$52:46:21.299 & $1.5792$ & $25.59\pm0.06$ & $ 909\pm 95$ & $1017\pm 88$ \\
AEGIS-15778 & 14:19:11.210 & $+$52:46:23.414 & $2.1716$ & $25.24\pm0.04$ & $1001\pm 42$ & $ 891\pm 47$ \\
AEGIS-01387 & 14:19:22.694 & $+$52:44:43.040 & $2.1700$ & $24.77\pm0.03$ & $ 426\pm 32$ & $ 461\pm 37$ \\
AEGIS-06264 & 14:19:17.443 & $+$52:45:05.370 & $2.1879$ & $24.72\pm0.03$ & $ 651\pm 46$ & $ 456\pm 46$ \\
AEGIS-07028 & 14:19:16.128 & $+$52:45:03.763 & $2.2925$ & $24.91\pm0.04$ & $1023\pm 67$ & $ 783\pm 62$ \\
AEGIS-20217 & 14:19:02.413 & $+$52:45:56.017 & $1.6306$ & $25.24\pm0.05$ & $ 341\pm 27$ & $ 601\pm 55$ \\
AEGIS-17118 & 14:19:03.204 & $+$52:45:16.919 & $1.6055$ & $24.02\pm0.02$ & $ 332\pm 19$ & $ 376\pm 21$ \\
GOODS-N-38085 & 12:37:18.165 & $+$62:22:29.258 & $1.5214$ & $24.80\pm0.02$ & $ 930\pm 79$ & $ 913\pm 80$ \\
GOODS-N-37906 & 12:37:16.752 & $+$62:22:04.152 & $1.6847$ & $24.65\pm0.02$ & $ 316\pm 23$ & $ 379\pm 30$ \\
GOODS-N-37876 & 12:37:12.919 & $+$62:21:59.569 & $1.6148$ & $25.53\pm0.04$ & $ 283\pm 23$ & $ 357\pm 32$ \\
GOODS-N-37878 & 12:37:06.160 & $+$62:22:00.005 & $1.5508$ & $25.86\pm0.05$ & $ 456\pm 46$ & $ 305\pm 32$ \\
GOODS-N-37296 & 12:37:24.594 & $+$62:21:00.900 & $1.5136$ & $24.47\pm0.01$ & $ 685\pm 38$ & $ 741\pm 44$ \\
GOODS-N-36886 & 12:37:31.259 & $+$62:20:36.596 & $1.6772$ & $25.34\pm0.03$ & $1072\pm144$ & $1114\pm151$ \\
GOODS-N-36583 & 12:37:33.171 & $+$62:20:23.489 & $1.5495$ & $25.27\pm0.03$ & $ 480\pm 50$ & $ 699\pm 70$ \\
GOODS-N-36852 & 12:37:23.247 & $+$62:20:34.465 & $1.5970$ & $25.94\pm0.06$ & $1678\pm380$ & $1209\pm274$ \\
GOODS-N-36684 & 12:37:22.580 & $+$62:20:26.984 & $1.5965$ & $25.44\pm0.04$ & $ 442\pm 44$ & $ 392\pm 40$ \\
GOODS-N-36273 & 12:37:01.427 & $+$62:20:10.543 & $2.2422$ & $26.05\pm0.06$ & $ 405\pm 40$ & $ 541\pm119$ \\
GOODS-N-29675 & 12:37:07.081 & $+$62:17:18.971 & $1.6847$ & $25.29\pm0.03$ & $ 561\pm 43$ & $ 938\pm 70$ \\
GOODS-N-35204 & 12:37:06.813 & $+$62:19:35.508 & $2.0976$ & $23.89\pm0.01$ & $ 322\pm 12$ & $ 403\pm 29$ \\
UDS-06377 & 02:17:42.856 & $-$05:15:19.134 & $1.6642$ & $25.31\pm0.05$ & $ 943\pm 94$ & $1169\pm135$ \\
UDS-12539 & 02:17:53.733 & $-$05:14:03.196 & $1.6211$ & $24.36\pm0.02$ & $ 882\pm 33$ & $ 803\pm 31$ \\
UDS-24003 & 02:17:54.729 & $-$05:11:44.020 & $1.6205$ & $24.55\pm0.03$ & $ 259\pm 11$ & $ 266\pm 17$ \\
UDS-07447 & 02:17:18.162 & $-$05:15:06.275 & $1.5972$ & $24.54\pm0.02$ & $ 666\pm 36$ & $ 529\pm 26$ \\
UDS-13027 & 02:17:16.355 & $-$05:13:56.240 & $1.6549$ & $25.28\pm0.04$ & $1216\pm121$ & $ 878\pm145$ \\
UDS-21873 & 02:17:17.096 & $-$05:12:09.652 & $1.6552$ & $24.59\pm0.02$ & $ 271\pm 19$ & $ 387\pm 35$ \\
UDS-15658 & 02:17:24.262 & $-$05:13:25.612 & $1.6869$ & $24.76\pm0.04$ & $ 307\pm 23$ & $ 334\pm 34$ \\
UDS-29267 & 02:17:25.322 & $-$05:10:40.397 & $1.5190$ & $23.77\pm0.01$ & $1893\pm 59$ & $1691\pm 73$ \\
UDS-26182 & 02:17:28.134 & $-$05:11:17.272 & $1.4867$ & $25.45\pm0.04$ & $1120\pm 98$ & $1175\pm137$ \\
UDS-28931 & 02:17:31.478 & $-$05:10:45.926 & $1.6000$ & $23.33\pm0.01$ & $ 244\pm  5$ & $ 308\pm 10$ \\
UDS-30015 & 02:17:36.517 & $-$05:10:31.256 & $1.6649$ & $24.62\pm0.02$ & $ 917\pm 55$ & $ 607\pm 48$ \\
UDS-19818 & 02:17:02.612 & $-$05:12:34.628 & $1.6241$ & $25.16\pm0.04$ & $ 418\pm 35$ & $ 470\pm 47$ \\
UDS-29624 & 02:17:00.684 & $-$05:10:34.903 & $1.6632$ & $26.26\pm0.11$ & $ 881\pm195$ & $ 779\pm208$ \\
UDS-36954 & 02:17:14.900 & $-$05:09:06.174 & $1.6578$ & $25.41\pm0.04$ & $ 238\pm 20$ & $ 347\pm 68$ \\
UDS-14019 & 02:17:28.554 & $-$05:13:44.857 & $1.4612$ & $23.80\pm0.01$ & $ 490\pm 15$ & $ 614\pm 17$ \\
UDS-05122 & 02:17:22.253 & $-$05:15:33.613 & $1.4628$ & $24.67\pm0.03$ & $ 307\pm 16$ & $ 291\pm 14$ \\
UDS-05342 & 02:17:09.216 & $-$05:15:31.990 & $1.5981$ & $23.58\pm0.01$ & $ 437\pm 12$ & $ 363\pm  8$ \\
UDS-11693 & 02:17:03.893 & $-$05:14:13.664 & $2.1854$ & $24.52\pm0.02$ & $ 596\pm 33$ & $ 592\pm 73$ \\
UDS-15128 & 02:17:38.209 & $-$05:13:32.092 & $2.3113$ & $24.25\pm0.02$ & $ 371\pm 16$ & $ 276\pm 97$ \\
UDS-12435 & 02:17:38.609 & $-$05:14:05.366 & $1.6112$ & $24.06\pm0.01$ & $ 247\pm  9$ & $ 173\pm  6$ \\
UDS-11387 & 02:17:39.228 & $-$05:14:17.689 & $1.4048$ & $24.64\pm0.02$ & $ 602\pm 40$ & $ 469\pm 34$ \\
UDS-17713 & 02:17:42.450 & $-$05:13:02.615 & $2.2484$ & $24.09\pm0.02$ & $ 328\pm 16$ & $ 249\pm 50$ \\
\hline
\end{tabular}
\caption{Summary of the 64 extreme [O {\scriptsize III}] emitters (EW$_{\rm{[OIII]}\lambda5007}>225$ \AA) with robust Balmer line detections (S/N $>3$) at  $z=1.3-2.4$. Column (1): galaxy ID in 3D-HST v4 catalogs; Column (2): right ascension of the galaxy; Column (3): declination of the galaxy; Column (4): redshift measured from MMT or Keck spectra; Column (5): {\it HST} F606W magnitude (in AB magnitude); Column (7): rest-frame [O {\scriptsize III}]$\lambda5007$ equivalent width; Column (8): rest-frame H$\alpha$ equivalent width.}
\label{tab:eelg_sum}
\end{table*}

\begin{table*}
\setcounter{table}{4}
\begin{tabular}{|c|c|c|c|c|c|c|}
\hline
Object ID & R.A. & Decl. & $z_{\rm{spec}}$ & $m_{\rm{F606W}}$ & EW$_{\rm{[OIII]}\lambda5007}$ & EW$_{\rm{H}\alpha}$ \\
 & (hh:mm:ss) & (dd:mm:ss) & & (AB) & (\AA) & (\AA) \\
\hline 
\hline
UDS-28064 & 02:17:47.396 & $-$05:10:56.197 & $1.4600$ & $24.54\pm0.03$ & $ 360\pm 27$ & $ 203\pm 16$ \\
UDS-24183 & 02:17:47.395 & $-$05:11:41.453 & $2.2448$ & $24.97\pm0.04$ & $ 821\pm 59$ & $ 756\pm132$ \\
UDS-22532 & 02:17:52.799 & $-$05:12:03.013 & $1.4207$ & $24.62\pm0.03$ & $ 387\pm 12$ & $ 197\pm  7$ \\
UDS-17891 & 02:17:45.950 & $-$05:12:57.416 & $1.6714$ & $25.99\pm0.07$ & $1673\pm552$ & $1300\pm435$ \\
UDS-12980 & 02:17:55.822 & $-$05:13:56.748 & $1.4102$ & $25.50\pm0.05$ & $ 625\pm 53$ & $ 725\pm 96$ \\
\hline
\end{tabular}
\caption{Continued.}
\label{tab:eelg_sum}
\end{table*}

The BEAGLE models allow us to characterize the production efficiency of 
hydrogen ionizing photons in the EELGs.  There are various definitions of this quantity 
in the literature.   We follow the nomenclature used in our earlier work \citep{Chevallard2018} 
which we briefly review below.  First, we define $\xi^\star_{\rm{ion}}$  
as the hydrogen ionizing photon production rate ($N(\rm{H}^0)$) per unit UV stellar continuum luminosity 
($L^{\star}_{\rm{UV}}$).  The quantity $L^{\star}_{\rm{UV}}$ is the luminosity produced by the 
stellar population prior to its transmission through the gas and dust in the galaxy, and as such it does not 
include nebular continuum emission or absorption from the ISM.   As is commonplace, we evaluate the 
UV luminosity at a rest-frame wavelength of $1500$ \AA.  BEAGLE includes a determination of 
$\xi^\star_{\rm{ion}}$ in its output parameter file for each source.  Secondly, we define 
$\xi_{\rm{ion}}$,  as the hydrogen ionizing photon production rate per 
unit {\it observed} UV luminosity ($L_{\rm{UV}}$), 
again evaluated at $1500$ \AA. $L_{\rm{UV}}$ includes emission from the stellar population and nebular 
continuum and is not corrected for the attenuation provided by the ISM.  Finally we define 
 $\xi^{\rm{HII}}_{\rm{ion}}$ as the hydrogen ionizing photon production rate per unit 
 $L^{\rm{HII}}_{\rm{UV}}$, the observed UV luminosity at $1500$ \AA\ 
(including nebular and stellar continuum) corrected for dust attenuation from the diffuse ISM. 
 This is the most commonly used definition of the hydrogen ionizing production efficiency 
in the literature, and we will focus primarily on this quantity in the following section.  

In deriving $\xi^{\rm{HII}}_{\rm{ion}}$, we follow a similar procedure adopted in several recent 
studies \citep{Matthee2017,Shivaei2018}.   We compute the H$\alpha$ luminosity 
from our spectra and apply a correction for dust attenuation using the measured Balmer 
decrement (see \S\ref{sec:emission_line}).   We then calculate the ionizing photon production rate from the 
H$\alpha$ luminosity \citep{Osterbrock2006}:
\begin{eqnarray}
L(\textrm{H}\alpha)\ [\textrm{erg s}^{-1}] = 1.36\times10^{-12}\ N(\textrm{H}^0)\ [\textrm{s}^{-1}].
\end{eqnarray}
This assumes radiation-bounded nebula, with negligible escape of ionizing radiation.
The dust correction through the Balmer decrement traces dust only outside the H {\scriptsize II} regions, 
and does not account for the absorption of ionizing photons before they ionize hydrogen \citep{Petrosian1972,Mathis1986,Charlot2000}. 
Thus, the $N(\rm{H}^0)$ computed in Eq. (1) is somewhat lower than the true production rate of ionizing photons emitted by stars. 
However, in \S\ref{sec:balmer} we demonstrate that the extreme [O {\scriptsize III}] emitters in our sample have very little dust, 
indicating that the effect of absorption of ionizing photons inside the H {\scriptsize II} regions should be very small. 
We infer the observed UV continuum luminosity from the best-fitting 
BEAGLE model using a flat $100$ \AA\ filter centered at $1500$ \AA.  We then apply a dust 
correction using the reddening inferred from BEAGLE assuming first a Calzetti and then a SMC extinction law.  
Finally, the ionizing photon production efficiency $\xi^{\rm{HII}}_{\rm{ion}}$ is computed as follows:
\begin{eqnarray}
\xi^{\textrm{HII}}_{\textrm{ion}}\ [\textrm{erg}^{-1}\ \textrm{Hz}] = \frac{N(\textrm{H}^0)\ [\textrm{s}^{-1}]}{L^{\textrm{HII}}_{\textrm{UV}}\ [\textrm{erg s}^{-1}\ \textrm{Hz}^{-1}]}.
\end{eqnarray} 

In the following section, we will investigate how the ionizing production efficiency varies with the nebular line 
equivalent widths.   Here we attempt to build some basic physical intuition about this relationship.  
Since the [O {\scriptsize III}] and H$\alpha$ EW are directly linked to the luminosity-weighted 
age of the stellar population\footnote{The  H$\alpha$ and [O {\scriptsize III}] EWs are additionally regulated by the stellar 
metallicity, and the [O {\scriptsize III}] EW will also be affected by the gas properties.  
Here we consider only the impact of age on the H$\alpha$ EW at fixed metallicity with the goal of 
building a basic understanding of the dependence of  $\xi^\star_{\rm{ion}}$ on the optical  line equivalent widths.}, 
the variation of $\xi^\star_{\rm{ion}}$ will mirror the time evolution of 
$N(\rm{H}^0)$ and $L^{\star}_{\rm{UV}}$, the former powered by O stars and the latter by O to early B stars.  
For the very young stellar populations probed by the most extreme line 
emitting galaxies in our sample (EW$_{\rm{H}\alpha}>1000$ \AA), the ratio of O to B stars 
will be maximized, resulting in very efficient ionizing photon production.   Among 
more moderate line emitters (EW$_{\rm{H}\alpha}=400-600$ \AA), a  
larger population of B stars will have emerged, boosting $L_{\rm{UV}}$ relative to $N(\rm{H}^0)$.  The 
ionizing production efficiency will thus be reduced for these systems.  Finally, at yet lower 
equivalent width  (EW$_{\rm{H}\alpha}=50-200$ \AA), the O and B star populations will be 
closer to equilibrium, with the number of newly-formed stars nearly balanced by those 
exiting the main sequence.  As a result, both $N(\rm{H}^0)$ and $L_{\rm{UV}}$ will not vary 
significantly with age, and the production efficiency of ionizing photons should 
thus begin to plateau to a near-constant value for galaxies with EW$_{\rm{H}\alpha}<200$ \AA.   
While this basic physical picture guides our expectations, the precise dependence of 
$\xi^\star_{\rm{ion}}$ on the [O {\scriptsize III}] and H$\alpha$ equivalent width depends on unknown 
physics (i.e., binary stars, rotation) governing the formation of hot stars.  Empirical 
constraints on $\xi^\star_{\rm{ion}}$ are thus critical for assessing the ionizing 
output of early galaxies.  


\section{rest-frame optical Spectroscopic Properties of Extreme [O {\scriptsize III}] Emitters at $\lowercase{z}=1.3-2.4$} \label{sec:phy_prop}

We now investigate what the rest-frame optical spectra reveal about the radiation field 
and gas conditions of galaxies with large equivalent width rest-frame optical nebular line emission.    
We will focus primarily on empirical trends that can be extracted from line ratios, leaving a 
more detailed analysis of the photoionization models to a later paper.  
In \S\ref{sec:balmer}, we use measurements of the Balmer decrements to 
constrain the nebular attenuation.  We then characterize the radiation field and 
gas conditions through measurement of the ionizing photon production efficiency (\S\ref{sec:xi_ion})  and 
standard ionization-sensitive emission line ratios  (\S\ref{sec:line_ratio}).  


\subsection{Balmer Decrement Measurements} \label{sec:balmer}
 
The physical interpretation of our spectra requires robust determination of 
the nebular attenuation between [O {\scriptsize II}] and H$\alpha$. 
As introduced in \S\ref{sec:emission_line}, this is most commonly done through comparison of the Balmer decrement, 
$I(\rm{H}\alpha)/I(\rm{H}\beta)$, to the ratio expected in absence of dust.  As noted 
in \S\ref{sec:modeling}, the Balmer decrement corrects for dust outside the H {\scriptsize II} regions but does not account for 
the possible absorption of LyC photons by dust before they ionize hydrogen. The first 
statistical measurements of the Balmer decrement distribution at high redshift 
have begun to emerge from the KBSS and MOSDEF surveys \citep[e.g.][]{Reddy2015,Steidel2016}.  These 
investigations provide an important baseline for comparison 
to the extreme emission line galaxies in our sample, so we briefly review the results below.  
To enable comparison with investigations of nebular attenuation in the KBSS and MOSDEF galaxies, 
here we adopt a \citet{Cardelli1989} extinction curve.
 
The average nebular properties of the KBSS survey are well described by the KBSS-LM1 
composite spectrum, a weighted average of galaxies at $z\simeq 2.11-2.57$ with median stellar 
mass of $6\times$10$^{9}\ M_\odot$ and specific star formation rate of $3.5$ Gyr$^{-1}$ \citep{Steidel2016}
The Balmer decrement derived from the composite  is $I(\rm{H}\alpha)/I(\rm{H}\beta)=3.61\pm0.07$ \citep{Steidel2016}, 
similar to that found in individual KBSS galaxies \citep{Strom2017}.  
For the MOSDEF survey, the average Balmer decrement measured 
from a composite of 213 galaxies with significant H$\beta$ detections is 
$I(\rm{H}\alpha)/I(\rm{H}\beta)=4.1$ \citep{Reddy2015}. For a \citet{Cardelli1989} Galactic extinction 
curve and an intrinsic Balmer decrement, $I(\rm{H}\alpha)/I(\rm{H}\beta)=2.86$, the KBSS 
and MOSDEF composites imply typical color excesses of $E(B-V)_{\rm{gas}} = 0.22$ and 
$0.36$, respectively.  We note that \citet{Steidel2016} adopt a slightly larger intrinsic Balmer decrement,  
$I(\rm{H}\alpha)/I(\rm{H}\beta)=2.89$, that is consistent with their photoionization model fits.  
As a result, they report a slightly different color excess, $E(B-V)_{\rm{gas}} = 0.21$, 
for the KBSS-LM1 composite.  

\begin{figure}
\begin{center}
\includegraphics[width=\linewidth]{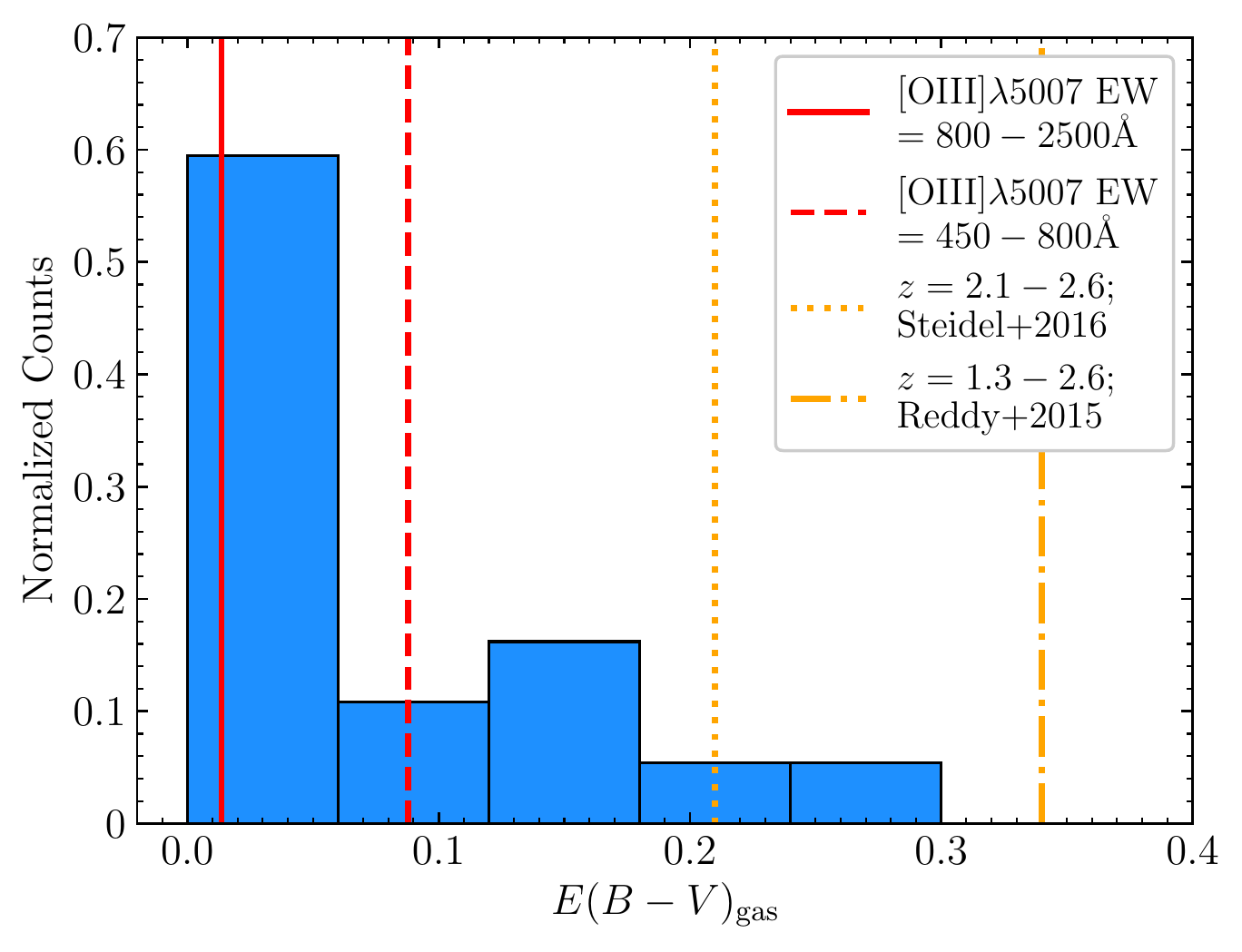}
\caption{Probability distribution of nebular dust reddening for extreme [O {\scriptsize III}] emitters with EW$_{\rm{[OIII]}\lambda5007}>450$ \AA\ at $z=1.3-2.4$. The extinction $E(B-V)_{\rm{gas}}$ is inferred from Balmer decrement, assuming the \citet{Cardelli1989} extinction curve and intrinsic H$\alpha$/H$\beta$ intensity ratio $=2.86$. The extinction values inferred from composite spectra of extreme [O {\scriptsize III}] emitters with EW$_{\rm{[OIII]}\lambda5007}=800-2500$ \AA\ and EW$_{\rm{[OIII]}\lambda5007}=450-800$ \AA\ are plotted by red solid and dashed lines. For comparison, we also show the extinction values inferred from more massive, older systems at $z\sim2$ from KBSS \citep{Steidel2016} as the orange dotted line and MOSDEF survey \citep{Reddy2015} as the orange dashed-dotted line.}
\label{fig:hab}
\end{center}
\end{figure}

In our current sample, we can measure Balmer decrements in 64 extreme emission line galaxies 
(EW$_{\rm{[OIII]}\lambda5007}>225$ \AA) with significant detections (S/N $>3$) of 
H$\alpha$ and H$\beta$ (Table \ref{tab:sample_size}).  The measurements suggest that extreme line emitters suffer   
less nebular attenuation than most galaxies in the KBSS and MOSDEF surveys.  
In Figure \ref{fig:hab}, we show the distribution of Balmer decrements 
for EELGs in our sample with EW$_{\rm{[OIII]}\lambda5007}>450$ \AA. 
This threshold is chosen to correspond roughly to the average [O {\scriptsize III}]+H$\beta$ 
equivalent width at $z\simeq 7-8$ \citepalias{Labbe2013}, assuming an [O {\scriptsize III}]/H$\beta$ 
ratio that is characteristic of EELGs (see \S\ref{sec:sample}).  The average Balmer decrement 
in this histogram is $I(\rm{H}\alpha)/I(\rm{H}\beta)=3.09$, implying a color 
excess of just  $E(B-V)_{\rm{gas}}=0.08$ for the \citet{Cardelli1989} extinction curve.  For 
reference we also measure $\beta$, the UV continuum slope, for this same subset of galaxies. 
We compute the UV slope from broadband fluxes by fitting a power-law ($f_{\lambda} \propto \lambda^{\beta}$) 
at rest-frame wavelengths $1250-2600$ \AA\ (the same wavelength range used in 
\citealt{Calzetti1994}; filters covering Ly$\alpha$ emission line are not included). 
The resulting UV slopes are very blue, with a median of $\beta=-2.04$ for galaxies with 
EW$_{\rm{[OIII]}\lambda5007}>450$ \AA, implying that the stars are also minimally reddened by dust.   


The level of nebular attenuation varies with the [O {\scriptsize III}] EW, with the most extreme line emitters 
having the least dust.  In our sample, the median Balmer decrement for the 16 
galaxies with EW$_{\rm{[OIII]}\lambda5007}=450-800$ \AA\ is $I(\rm{H}\alpha)/I(\rm{H}\beta)=3.12$.  When we 
consider the 21 more extreme line emitters with EW$_{\rm{[OIII]}\lambda5007}=800-2500$ \AA, 
we find a Balmer decrement of just $I(\rm{H}\alpha)/I(\rm{H}\beta)=2.95$.  This implies color excesses of 
$E(B-V)_{\rm{gas}}=0.09$ and $0.03$ for the two respective EW bins.  The composite spectra 
reveal a similar picture.  The measured Balmer decrement in the stacks decreases from 
$I(\rm{H}\alpha)/I(\rm{H}\beta)=3.12\pm0.04$ (EW$_{\rm{[OIII]}\lambda5007}=450-800$ \AA) to 
$I(\rm{H}\alpha)/I(\rm{H}\beta)=2.90\pm0.05$  
(EW$_{\rm{[OIII]}\lambda5007}=800-2500$ \AA).   This indicates that among the most extreme line 
emitters, the Balmer decrements are very close to the intrinsic case B recombination value 
(for $T_e=10^4$ K gas), suggesting that the emission lines in these galaxies face little to no  
attenuation from dust.  

The results described above clearly demonstrate that the EELG population has a  
different distribution of Balmer decrements than the typical KBSS and MOSDEF 
galaxies.  This result is not surprising given trends between the Balmer decrement and galaxy 
properties found previously \citep{Reddy2015}, likely reflecting both the low stellar mass (and 
hence moderately low metallicities) and large specific star formation rates (and hence young 
stellar populations) of the galaxies in our sample (Figure \ref{fig:ew_star}).  Importantly this implies that very 
small adjustments are required to correct the observed fluxes for reddening.  

An important consequence  is that uncertainties in the high redshift attenuation law 
should not significantly affect our interpretation of the nebular line spectra of EELGs.  Among 
more massive star forming galaxies at $z\simeq 2$, this is not always the case. \citet{Shivaei2018} 
demonstrated that the $\xi^{\rm{HII}}_{\rm{ion}}$ inferred assuming a Calzetti attenuation 
law is systematically 0.3 dex lower than that derived for an SMC attenuation law, 
making it difficult to robustly determine $\xi^{\rm{HII}}_{\rm{ion}}$.  While our measurements of  
$\xi^{\rm{HII}}_{\rm{ion}}$ also depend on the assumed dust law, the variation is typically 
only 0.1 dex when considering the SMC and Calzetti attenuation laws.  For the most extreme 
line emitting sources (EW$_{\rm{[OIII]}\lambda5007}>800$ \AA), the dust content is low enough 
that the average offset in $\xi^{\rm{HII}}_{\rm{ion}}$ is just 0.05 dex.
 

\subsection{The Ionizing Photon Production Efficiency} \label{sec:xi_ion}

As a first step in our investigation of the radiation field of extreme emission line galaxies, 
we characterize the production efficiency of hydrogen ionizing photons in our spectroscopic sample.   
Recent efforts have quantified $\xi^{\rm{HII}}_{\rm{ion}}$ in  more 
massive star forming galaxies at $z\simeq 2$ \citep[e.g][]{Matthee2017,Shivaei2018}, 
revealing typical values of $\log{[\xi^{\rm{HII}}_{\rm{ion}}\ (\rm{erg}^{-1}\ \rm{Hz})]}=25.06$ 
for a Calzetti UV attenuation law \citep{Shivaei2018}.   These systems tend to 
have [O {\scriptsize III}]$\lambda5007$ equivalent widths of $115$ \AA\ (see \S\ref{sec:sample}), well below those considered in this paper, 
as expected from older stellar populations. It has been shown that 
$\xi^\star_{\rm{ion}}$ scales  with the [O {\scriptsize III}] equivalent width among local 
star forming galaxies \citep{Chevallard2018}, suggesting that  the higher equivalent width systems 
which become common in the reionization era may be more efficient ionizing agents.  Here we seek to investigate 
whether a similar relationship between the ionizing production efficiency and optical lines holds at $z\simeq 1-2$.  In the following, we will first describe the distribution of 
$\xi^{\rm{HII}}_{\rm{ion}}$ values, using the 
dust corrections discussed in \S\ref{sec:modeling}.  Then we will compare  our measurements of 
$\xi^\star_{\rm{ion}}$ to the relation found locally, investigating whether there is any evidence for 
strong redshift evolution in $\xi^\star_{\rm{ion}}$ values at fixed [O {\scriptsize III}] EW.
   
\begin{figure*}
\begin{center}
\includegraphics[width=\linewidth]{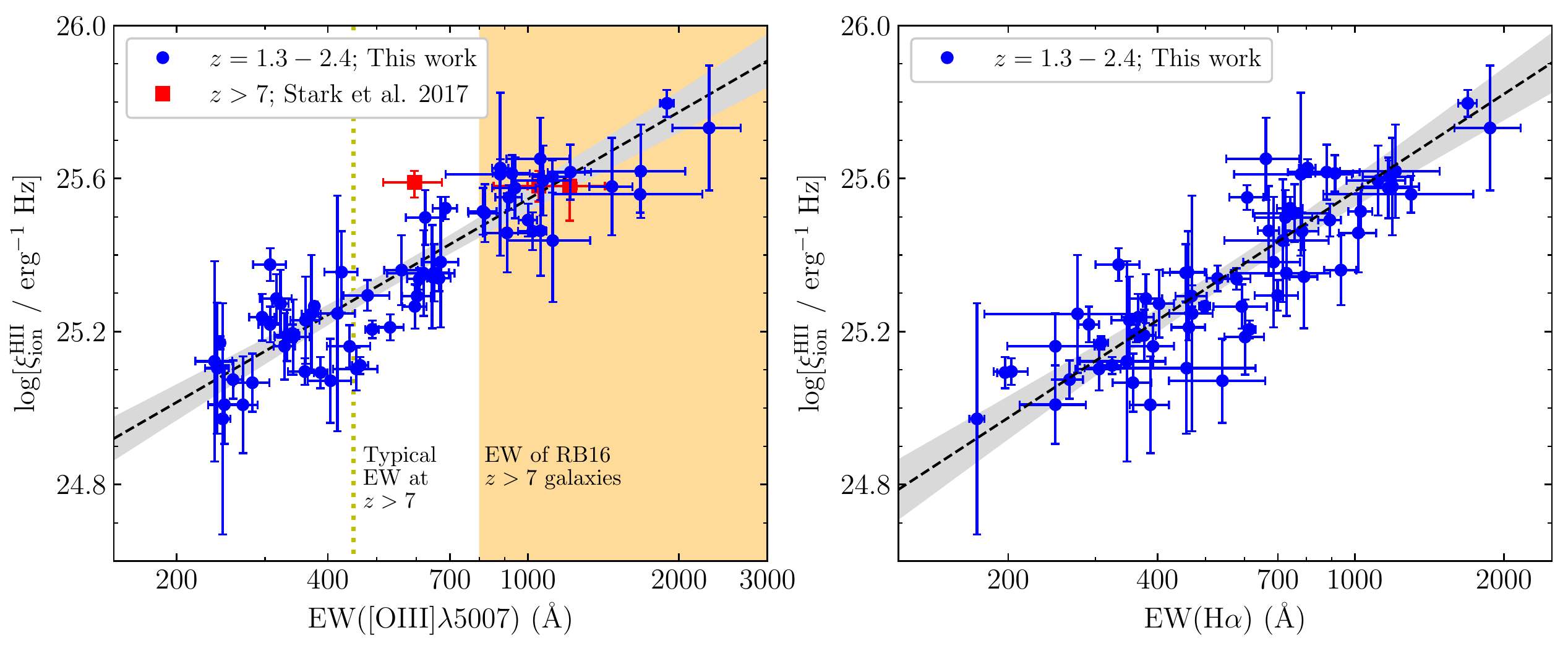}
\caption{\textit{Left Panel}: Relation between the ionizing photon production efficiency $\xi^{\rm{HII}}_{\rm{ion}}$ and [O {\scriptsize III}]$\lambda5007$ EW for the $z=1.3-2.4$ galaxies in our sample (blue circles). The $\xi^{\rm{HII}}_{\rm{ion}}$ is computed based on the spectroscopic measurements of H$\alpha$ and H$\beta$. A linear relation is fitted between $\xi^{\rm{HII}}_{\rm{ion}}$ and $\log{\rm{EW}_{\rm{[OIII]}\lambda5007}}$ (black dashed line, with grey area shows the 95 percent credible region). The average EW of $z\simeq7-8$ galaxies (EW$_{\rm{[OIII]}\lambda5007}\simeq450$ \AA; \citetalias{Labbe2013}) is highlighted by yellow dashed line. The orange area shows EWs of $z>7$ galaxies in \citetalias{Roberts-Borsani2016}, and red squares show the individual sources in \citetalias{Roberts-Borsani2016}, in which the $\xi^{\rm{HII}}_{\rm{ion}}$ are inferred from photoionization models \citep{Stark2017}. \textit{Right Panel}: Relation between $\xi^{\rm{HII}}_{\rm{ion}}$ and H$\alpha$ EW for the $z\sim2$ galaxies in our sample (blue circles). A linear relation is fitted between $\xi^{\rm{HII}}_{\rm{ion}}$ and $\log{\rm{EW}_{\rm{H}\alpha}}$ (black dashed line, with grey area shows the 95 percent credible region).}
\label{fig:xi_ion}
\end{center}
\end{figure*}

Our current sample contains 64 large equivalent width [O {\scriptsize III}] emitters (EW$_{\rm{[OIII]}\lambda5007}>225$ \AA)  
with the spectroscopic measurements of H$\alpha$ and H$\beta$ necessary to infer $\xi^{\rm{HII}}_{\rm{ion}}$.
In Figure \ref{fig:xi_ion}, we show the implied $\xi^{\rm{HII}}_{\rm{ion}}$ values as a function of the 
[O {\scriptsize III}]$\lambda5007$ and H$\alpha$ equivalent widths.  It is immediately clear that 
$\xi^{\rm{HII}}_{\rm{ion}}$ scales with both optical lines at EW$_{\rm{[OIII]}\lambda5007}>225$ \AA.  
We first consider galaxies with EW$_{\rm{[OIII]}\lambda5007}\simeq450$ \AA.  For the [O {\scriptsize III}]/H$\beta$ line ratios exhibited 
by extreme emission line galaxies (see \S\ref{sec:sample}), this [O {\scriptsize III}]$\lambda5007$ EW is comparable to that 
implied by IRAC flux excesses in composite SEDs of $z\simeq 7-8$ galaxies \citepalias{Labbe2013}. 
To estimate the typical ionizing efficiency for galaxies with this [O {\scriptsize III}] EW, we group those systems 
in our sample with EW$_{\rm{[OIII]}\lambda5007}=300-600$ \AA.  Assuming a Calzetti attenuation law for the UV, 
the median ionizing production efficiency of this sub-sample is $\log{[\xi^{\rm{HII}}_{\rm{ion}}\ (\rm{erg}^{-1}\ \rm{Hz})]}=
25.22$, $\sim 1.5\times$ greater than found in more typical galaxies at $z\simeq 2$ \citep{Shivaei2018}.  
For an SMC extinction law for the UV, the ionizing production efficiency is 
$\log{[\xi^{\rm{HII}}_{\rm{ion}}\ (\rm{erg}^{-1}\ \rm{Hz})]}=25.32$, 
slightly larger than what we found above for the Calzetti law.  This suggests that galaxies which are 
common at $z\gtrsim 6$ should  make a greater contribution to reionization than 
was originally thought, as has been noted by several recent investigations of IRAC excesses and 
UV metal lines at $z\gtrsim 5$ \citep{Rasappu2016,Stark2015b,Stark2017}.


Attention is now focused on galaxies at $z\gtrsim 7$ with even larger [O {\scriptsize III}] equivalent widths 
(EW$_{\rm{[OIII]}\lambda5007}=800-2500$ \AA).  While such sources may not be the 
norm at $z\gtrsim 7$, they are readily detectable in the CANDELS fields \citep[e.g.][]{Roberts-Borsani2016} 
and appear to have an enhanced Ly$\alpha$ visibility at redshifts where the IGM is thought to be significantly 
neutral \citep[e.g.][]{Oesch2015,Zitrin2015,Stark2017}.   One possible 
explanation is that these systems are situated in the largest ionized bubbles, thereby allowing 
enhanced transmission of their Ly$\alpha$ radiation.  Alternatively, in addition to sitting in ionized bubbles, 
the galaxies may be producing more Ly$\alpha$ than other galaxies of similar far-UV luminosities.  
This would follow if the production efficiency of Lyman-alpha is larger in systems with the most prominent [O {\scriptsize III}] 
emission. 

Our results provide some insight into the situation.  Since Ly$\alpha$ is produced 
by reprocessed hydrogen ionizing photons, the Ly$\alpha$ production efficiency 
($L_{\rm{Ly}\alpha}/L^{\rm{HII}}_{\rm{UV}}$) should be directly correlated with 
$\xi^{\rm{HII}}_{\rm{ion}}$.   Sources that are efficient at producing ionizing 
radiation should also be efficient in powering Ly$\alpha$ emission.  In Figure \ref{fig:xi_ion}, 
we see that the production efficiency of ionizing radiation steadily 
increases between EW$_{\rm{[OIII]}\lambda5007}=450$ \AA\ and $2500$ \AA.  Among 
those sources with EW$_{\rm{[OIII]}\lambda5007}>800$ \AA, the  production efficiency 
has a median of $\log{[\xi^{\rm{HII}}_{\rm{ion}}\ (\rm{erg}^{-1}\ \rm{Hz})]}=
25.58$ (Calzetti) and $25.62$ (SMC).  This value is very similar to estimates of  $\xi^{\rm{HII}}_{\rm{ion}}$ in the 
\citetalias{Roberts-Borsani2016} sample of $z\gtrsim 7$ galaxies with similarly strong [O {\scriptsize III}] (see red squares in Figure \ref{fig:xi_ion}).   
Our results suggest that galaxies with [O {\scriptsize III}] EWs similar to the \citetalias{Roberts-Borsani2016} galaxies may produce two times more ionizing 
photons than typical reionization-era sources (the latter of which have EW$_{\rm{[OIII]}\lambda5007}\simeq450$ \AA) with similar non-ionizing UV luminosities.  The  larger-than-average 
Ly$\alpha$ equivalent widths that we are now seeing in the \citetalias{Roberts-Borsani2016} population should thus partially reflect an enhanced production 
of ionizing radiation.  In \S\ref{sec:lya}, we will explore in more detail whether variations in internal galaxy properties 
can explain the anomalous visibility of Ly$\alpha$ that is now being found in strong [O {\scriptsize III}] emitters at $z\gtrsim 7$.

The relationship between the production efficiency of ionizing photons and the [O {\scriptsize III}]$\lambda5007$ EW 
shown in Figure \ref{fig:xi_ion} can be described by a simple scaling law.  We fit a linear relation between the two quantities for those galaxies 
in our sample with $225$ \AA\ $<\rm{EW}_{\rm{[OIII]}\lambda5007}<2500$ \AA.  The best-fit relation 
for the Calzetti UV attenuation law is
\begin{eqnarray}\nonumber
\log{\xi^{\textrm{HII}}_{\textrm{ion}}} &=& (0.76\pm0.05)\times\log{(\textrm{EW [O {\scriptsize III}]}\lambda5007)} + \\ && (23.27\pm0.15).
\end{eqnarray}
For an SMC attenuation law, we find a very similar relationship:
\begin{eqnarray}\nonumber
\log{\xi^{\textrm{HII}}_{\textrm{ion}}} &=& (0.73\pm0.08)\times\log{(\textrm{EW [O {\scriptsize III}]}\lambda5007)} + \\ && (23.45\pm0.23).
\end{eqnarray}
The derived fitting function is overlaid on the data in Figure \ref{fig:xi_ion}.  We emphasize 
that the scaling laws is only valid for the large equivalent widths stated above.  Indeed, as  we motivated  
at the end of \S\ref{sec:modeling}, we expect $\xi^{\rm{HII}}_{\rm{ion}}$ may deviate from the relationship at lower equivalent widths.  
But for sources with EW$_{\rm{[OIII]}\lambda5007}>225$ \AA, these fitting 
functions can be used to predict the distribution of $\xi^{\rm{HII}}_{\rm{ion}}$ given an 
observed distribution of [O {\scriptsize III}] equivalent widths. As suggested in \citet{Chevallard2018}, 
and provided that this relationship does not evolve strongly with redshift (as we will show below for $0<z<2$),  
this can be used in conjunction with IRAC flux excesses to estimate the production efficiency 
of ionizing photons in the reionization era.

We can also derive a relationship between H$\alpha$ and $\xi^{\rm{HII}}_{\rm{ion}}$ using the 
results in Figure \ref{fig:xi_ion}. Because the H$\alpha$ strength is less sensitive to the gas physical conditions 
than [O {\scriptsize III}], we expect it to have a more universal relationship with the ionizing production efficiency. 
Applying a similar fitting procedure and assuming a Calzetti attenuation law for the UV, 
we derive the following relationship:
\begin{eqnarray}\nonumber
\log{\xi^{\textrm{HII}}_{\textrm{ion}}} &=& (0.85\pm0.06)\times\log{(\textrm{EW H}\alpha)} + \\ && (23.03\pm0.15).
\end{eqnarray}
For the SMC law, we again find a very similar relationship:
\begin{eqnarray}\nonumber
\log{\xi^{\textrm{HII}}_{\textrm{ion}}} &=& (0.87\pm0.07)\times\log{(\textrm{EW H}\alpha)} + \\ && (23.06\pm0.19).
\end{eqnarray}

To assess the magnitude of the redshift evolution in the relationship between the production efficiency 
and the rest-frame optical line equivalent widths, we compare our results to those derived for local 
galaxies in \citet{Chevallard2018}.  The local galaxy relation is derived using $\xi^\star_{\rm{ion}}$, 
the ratio of the intrinsic production rate of ionizing photons to the stellar continuum luminosity in 
the non-ionizing far-UV, with both quantities derived from BEAGLE model fits to the data (see definitions in \S\ref{sec:modeling}). 
In cases where the nebular continuum contributes 
significantly to the observed UV luminosity, 
we expect $\xi^\star_{\rm{ion}}$ to deviate from $\xi^{\rm{HII}}_{\rm{ion}}$, 
providing a more realistic description of the efficiency of a stellar population at producing ionizing radiation. 
We find that $\xi^\star_{\rm{ion}}$ is indeed systematically larger than 
$\xi^{\rm{HII}}_{\rm{ion}}$ in our sample, but the differences are relatively small.  Among galaxies with 
[O {\scriptsize III}]$\lambda5007$ equivalent widths between $450$ and $800$ \AA\ ($800$ and $2500$ \AA), the 
difference is  0.04 (0.08) dex.   This suggests that the relations between $\xi^{\rm{HII}}_{\rm{ion}}$ 
and optical line equivalent widths that we derived above should closely approximate those 
inferred using $\xi^\star_{\rm{ion}}$.  Nevertheless for the sake of consistency, we consider trends with 
$\xi^\star_{\rm{ion}}$ when comparing the production efficiencies derived at $z=1.3-2.4$ with 
those found locally.  

\begin{figure*}
\begin{center}
\includegraphics[width=\linewidth]{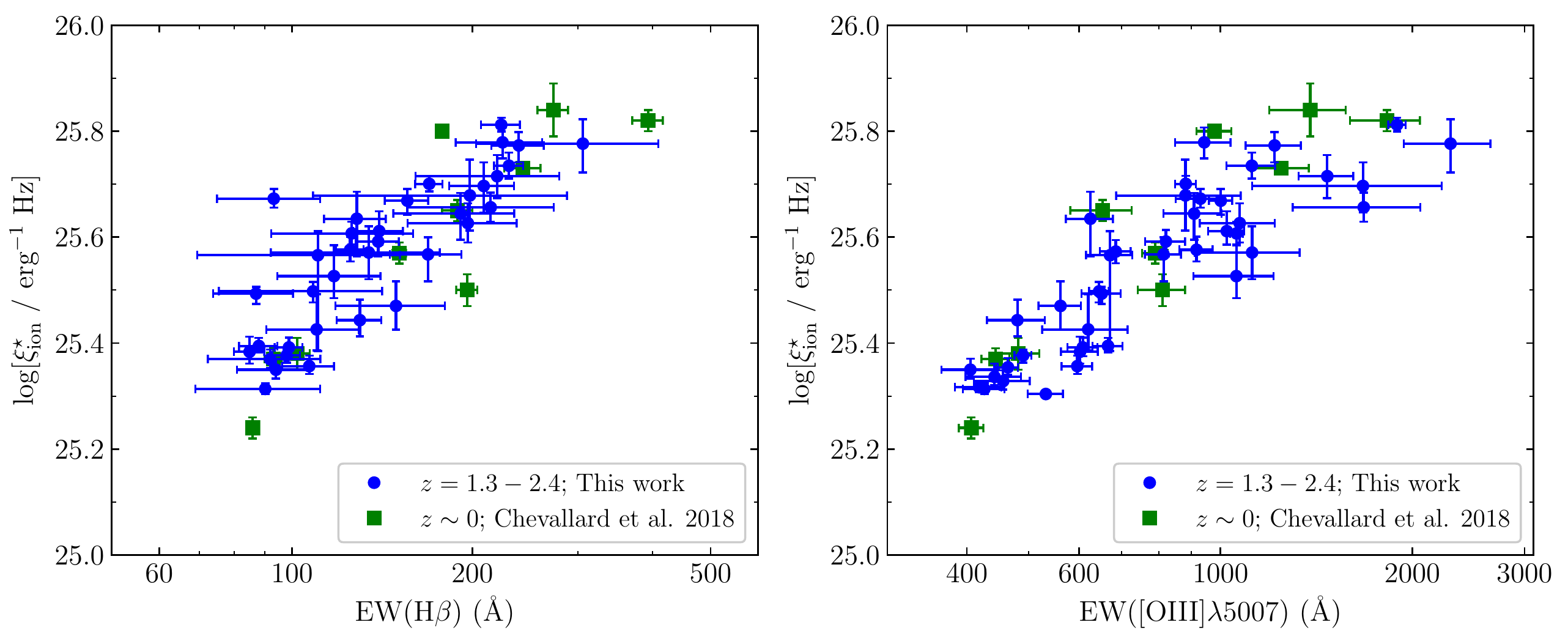}
\caption{BEAGLE model-derived $\xi^{\star}_{\rm{ion}}$ as a function of H$\beta$ EW (left panel) and [O {\scriptsize III}]$\lambda5007$ EW (right panel) at $z=1.3-2.4$ (blue circles; this work) and $z\sim0$ (green squares; \citealt{Chevallard2018}). The $\xi^{\star}_{\rm{ion}}$ is computed based on the intrinsic UV stellar luminosity (assuming a SMC extinction curve in order to be consistent with \citealt{Chevallard2018}), and only objects with similar EWs in the two samples (H$\beta$ EW $=80-400$ \AA\ in the left panel; [O {\scriptsize III}]$\lambda5007$ EW $=400-2500$ \AA\ in the right panel) are shown.}
\label{fig:xi_ion_model}
\end{center}
\end{figure*}

In Figure \ref{fig:xi_ion_model}, we show the dependence of $\xi^\star_{\rm{ion}}$ on the H$\beta$ and [O {\scriptsize III}] EW for our 
high redshift sample (blue circles) and the nearby galaxies (green squares) investigated 
in \citet{Chevallard2018}.  The local sample is comprised of ten systems with deep UV 
and optical spectra \citep{Senchyna2017}, each selected to have extreme radiation 
fields based on the presence of He {\scriptsize II} emission.  The relationship between 
$\xi^\star_{\rm{ion}}$ and H$\beta$ EW appears largely similar in the two redshift samples.  
We can quantify this by comparing the linear fit to the $\log{\xi^\star_{\rm{ion}}}-\log{\rm{EW}_{\rm{H}\beta}}$ 
relationship in the range EW$_{\rm{H}\beta}=80-400$ \AA. The best-fit slope and intercept 
for the $z\simeq0$ sample ($1.23\pm0.25$ and $22.89\pm0.56$) are both consistent (within $1\sigma$) 
with those derived for the $z\simeq2$ galaxies ($1.09\pm0.07$  and $23.23\pm0.16$).  Given the 
dependence of [O {\scriptsize III}] on ionized gas conditions (i.e., metallicity, ionization parameter), we 
expect the relationship between [O {\scriptsize III}] EW and $\xi^\star_{\rm{ion}}$ may potentially show more variation 
with redshift.  However, the current data reveal little evidence to this effect.  The best fit 
intercept and slope at $z\simeq0$ ($1.14\pm0.16$ and $22.28\pm0.44$) remain broadly  
similar (within $2\sigma$) to those derived at $z\simeq2$ ($0.88\pm0.06$ and $22.98\pm0.16$).  
While larger samples may eventually reveal some mild differences, our results suggest that the redshift 
evolution in the relationship between $\xi^\star_{\rm{ion}}$ and the H$\beta$ and [O {\scriptsize III}] EW 
is not likely to be strong.


\subsection{The Physical Conditions of the Nebular Gas} \label{sec:line_ratio}

The results of \S\ref{sec:xi_ion} show that extreme emission line galaxies are considerably more efficient at 
producing ionizing radiation than the more massive star-forming galaxies which are typical at $z\sim 2$.  
The intense radiation field of these system may  impact the 
ionization state of the ISM, potentially aiding the escape of Ly$\alpha$ and LyC radiation.   
In this subsection, we investigate  the ISM of  extreme emission line galaxies 
by quantifying the dependence of [O {\scriptsize III}]/[O {\scriptsize II}] (O32) and [Ne {\scriptsize III}]/[O {\scriptsize II}] (Ne3O2), two 
ionization-sensitive emission line ratios, on the [O {\scriptsize III}] and H$\alpha$ EW.  

The O32 index is one of the most commonly used probes of nebular gas ionization state 
and is often employed as an empirical proxy for the ionization parameter (the ratio of the 
number density of incident hydrogen-ionizing photons to the number density of hydrogen atoms in the H {\scriptsize II} region; \citealt{Penston1990}).   
The average O32 ratios of the massive star forming galaxies in the KBSS (O32 $=2.0$; \citealt{Steidel2016}) 
and MOSDEF (median O32 $=1.3$; \citealt{Sanders2016}) surveys are $\sim4-7\times$ higher than 
those of $z\sim0$ galaxies with similar stellar masses (median O32 $=0.3$; \citealt{Abazajian2009,Sanders2016}).
While the origin of the higher ionization parameters remains a matter of some debate, it has been 
argued to reflect a shift toward lower metallicities at fixed mass in high redshift galaxies \citep{Sanders2016}. 

\begin{figure}
\begin{center}
\includegraphics[width=\linewidth]{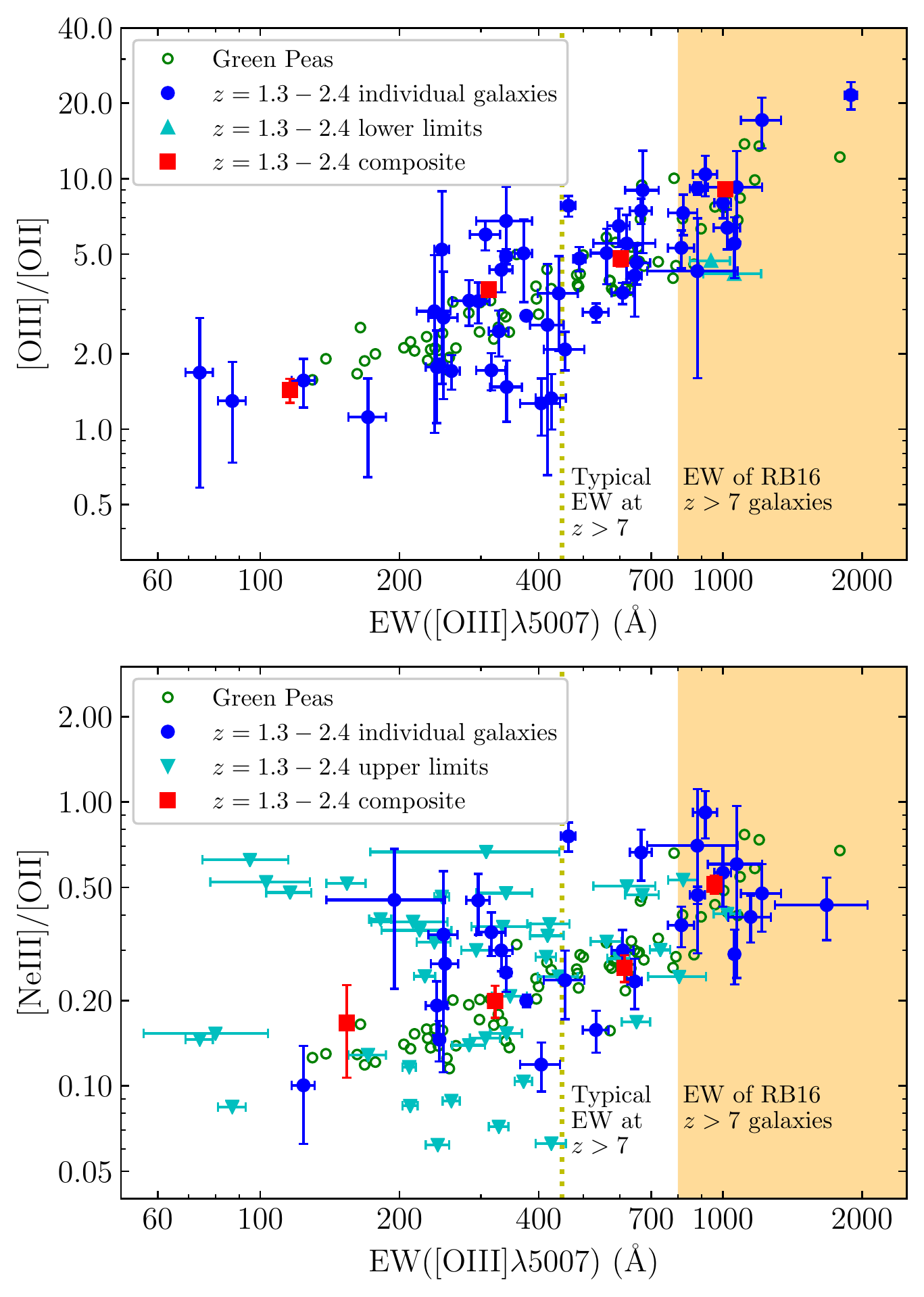}
\caption{Relation between [O {\scriptsize III}]$\lambda5007$ EW and ionization-sensitive line ratios ([O {\scriptsize III}]/[O {\scriptsize II}], upper panel; [Ne {\scriptsize III}]/[O {\scriptsize II}], lower panel) of the $z=1.3-2.4$ galaxies in our sample (blue circles). Line ratios of objects without robust [O {\scriptsize II}] (cyan triangles in the upper panel) or [Ne {\scriptsize III}] detections (cyan upside-down triangles in the lower panel) are shown as $2\sigma$ lower or upper limits. The average line ratios derived from composite spectra of galaxies with different EWs are shown by red squares. The average EW of $z\simeq7-8$ galaxies \citepalias{Labbe2013} is highlighted by yellow dashed line, and the orange area shows EWs of $z>7$ galaxies in the \citetalias{Roberts-Borsani2016} sample. For comparison, [O {\scriptsize III}]$\lambda5007$ EWs and line ratios of Green Peas \citep{Cardamone2009,Hawley2012} are shown by green open circles.}
\label{fig:line_ratio}
\end{center}
\end{figure}

The extreme emission line galaxies targeted in our survey have lower stellar masses and 
higher specific star formation rates than most galaxies from the KBSS and MOSDEF 
surveys (see \S\ref{sec:modeling}).  Both factors could lead to different ionization conditions.  Our current sample contains O32 measurements with dust corrections 
via the Balmer decrement for 44 galaxies (Table \ref{tab:sample_size}) with large equivalent width [O {\scriptsize III}] emission (EW$_{\rm{[OIII]}\lambda5007}>225$ \AA).  
In the top panel of Figure \ref{fig:line_ratio}, we show the dependence of O32 on the [O {\scriptsize III}]$\lambda5007$ EW for individual 
galaxies (blue circles) and for our composite spectra (red squares).   The results 
clearly demonstrate that O32 increases with the [O {\scriptsize III}]$\lambda5007$ EW, indicating more extreme 
ionizing conditions in the high EW galaxies.   The composite spectra reveal [O {\scriptsize III}]/[O {\scriptsize II}] 
ratios of O32 $=1.4\pm0.2$, $3.6\pm0.2$, $4.8\pm0.3$, and $9.1\pm0.5$ for stacks including galaxies with  
EW$_{\rm{[OIII]}\lambda5007}=0-225$ \AA, $225-450$ \AA, $450-800$ \AA, and $800-2500$ \AA, respectively.   


We see a comparable O32 trend in the individual galaxy measurements.  For galaxies with 
EW$_{\rm{[OIII]}\lambda5007}=300-600$ \AA\ (similar to the average EW expected in the reionization era), 
we find a median [O {\scriptsize III}]/[O {\scriptsize II}] ratio of O32 $=3.5$.  Among the galaxies in our sample with 
[O {\scriptsize III}] equivalent widths comparable to the \citetalias{Roberts-Borsani2016} $z\gtrsim 7$ Ly$\alpha$ emitters 
(EW$_{\rm{[OIII]}\lambda5007}=800-2500$ \AA), we find evidence for even more extreme ISM conditions, 
with 8 of the 11 sources with robust [O {\scriptsize III}]/[O {\scriptsize II}] measurements having O32 $>6$.   
Not surprisingly given these trends, the two highest [O {\scriptsize III}]/[O {\scriptsize II}] 
ratios observed in our survey (O32 $=17\pm4$ and $22\pm3$) are found in  two of the most extreme line 
emitters (EW$_{\rm{[OIII]}\lambda5007}=1216\pm121$ \AA\ and $1893\pm59$ \AA, respectively). 
The highest O32 values appear to become commonplace in the extremely young systems ($\lesssim 10$ Myr 
for constant star formation, \S\ref{sec:modeling}) that power intense optical line emission.  

The Ne3O2 index is another proxy for the ionization parameter  \citep[e.g.][]{Levesque2014}, 
providing an independent probe of the ionization state of the ISM.   Owing to the short wavelength 
baseline between [Ne {\scriptsize III}] and [O {\scriptsize II}], the flux ratio is largely insensitive to the effect of reddening.
At the highest redshifts probed by {\it JWST} ($z\gtrsim 10$), [Ne {\scriptsize III}] and [O {\scriptsize II}] will be among 
the brightest lines visible in the NIRSpec bandpass (see \S\ref{sec:jwst} for more discussion), motivating considerable interest in the 
effectiveness of Ne3O2 in proving nebular gas conditions.   Among the more massive star forming galaxies probed by 
the KBSS-LM1 composite spectrum, [Ne {\scriptsize III}] tends to be much fainter than [O {\scriptsize II}], with Ne3O2 $=0.15$ \citep{Steidel2016}.  
Based on the dependence of O32 on [O {\scriptsize III}] EW described above, we expect to see larger Ne3O2 values among 
the extreme emission line population. 

In our spectroscopic sample, we have obtained measurements of Ne3O2 for 26 galaxies (Table \ref{tab:sample_size}) with large 
equivalent width [O {\scriptsize III}] emission (EW$_{\rm{[OIII]}\lambda5007}>225$ \AA).  
In the bottom panel of Figure \ref{fig:line_ratio}, we plot the dependence of Ne3O2 on the [O {\scriptsize III}]$\lambda5007$ EW, 
showing  individual galaxies (blue circles) and measurements from the composite spectra (red squares).  
The results show that the Ne3O2 index increases with the [O {\scriptsize III}] EW (albeit with scatter), reaching 
values that are much greater than found among more massive galaxies in the KBSS survey.   In our 
composite spectra, we measure Ne3O2 $=0.17\pm0.06$, $0.20\pm0.02$, $0.26\pm0.03$, 
and $0.51\pm0.04$ for stacks including galaxies with EW$_{\rm{[OIII]}\lambda5007}=0-225$ \AA, 
$225-450$ \AA, $450-800$ \AA, and $800-2500$ \AA.   Among the largest equivalent width line 
emitters in our sample, the [Ne {\scriptsize III}] line is just as strong  as the individual components [O {\scriptsize II}] doublet.  
In \S\ref{sec:jwst}, we will consider the feasibility of detecting both features at $z\gtrsim 10$ with {\it JWST}.  

Both the O32 and Ne3O2 measurements suggest a picture in which the ISM of  galaxies 
with prominent optical line emission is characterized by extreme ionization conditions.  
 A large number of factors can modulate O32 \citep[e.g.][]{Nakajima2014}. 
 We will discuss some of these in \S\ref{sec:ism}, but a detailed investigation of the physical origin of the trend 
between O32 and [O {\scriptsize III}] EW will be considered in a follow-up paper focused on the photoionization 
models described in \S\ref{sec:modeling}.  


\section{Discussion} \label{sec:discussion}

In the previous section, we showed that the ionizing photon production efficiency and the ionization state of 
the nebular gas scales with optical line equivalent width.  Here we consider implications for the escape of 
ionizing radiation (and its potential association with large O32), the Ly$\alpha$ visibility test, and observability of 
extreme line emitting galaxies with {\it JWST} at $z\gtrsim 10$.  


\subsection{Implications for the ISM Conditions and the Escape of Ionizing Radiation in Reionization-Era Galaxies} \label{sec:ism}

In \S\ref{sec:line_ratio}, we demonstrated that [O {\scriptsize III}]/[O {\scriptsize II}] ratio increases with [O {\scriptsize III}] EW, 
reaching very large value (O32 $>10$) for sources with EW$_{\rm{[OIII]}\lambda5007}>1000$ \AA.  
The physical factors regulating O32 have been discussed in detail elsewhere \citep[e.g.][]{Kewley2013,Nakajima2014,Sanders2016}.  
These studies show that the [O {\scriptsize III}]/[O {\scriptsize II}] ratio depends strongly on the ionization parameter of the gas with 
a weaker dependence on its metallicity.  Assuming ionization balance, the ionization parameter in turn 
depends on the ionization photon production rate, the gas density, and the volume filling factor of 
ionized gas.   The hardness of the ionizing spectrum also impacts the [O {\scriptsize III}]/[O {\scriptsize II}] ratio, increasing 
O32 at fixed ionization parameter \citep{Sanders2016}.  While a comprehensive investigation of the 
trend seen in Figure \ref{fig:line_ratio} is beyond the scope of this paper, the boosted O32 values 
likely reflect the combined effect of a large ionization parameter and hard ionizing 
spectrum, the latter expected in the very young stellar populations present in the largest [O {\scriptsize III}] EW galaxies. 

Extreme optical line emitters are not the only sources at high redshift known to have  
large O32 ratios.  Spectroscopic investigations have shown large O32 values in high redshift 
galaxies with large EW Ly$\alpha$ emission.  \citet{Nakajima2016} measure [O {\scriptsize III}]/[O {\scriptsize II}] 
ratios in the range O32 $=6.0-11.5$ for four Ly$\alpha$ emitting galaxies (EW$_{\rm{Ly}\alpha}=26-61$ \AA) 
at $z\simeq 3.1$, comparable to the galaxies with EW$_{\rm{[OIII]}\lambda5007}>800$ \AA\ in our 
sample.  Similarly \citet{Erb2016} report [O {\scriptsize III}]/[O {\scriptsize II}] ratios of 2.4 to 12.2 (median O32 $=4.9$) 
for 11 Ly$\alpha$ emitters at $z\simeq 2$ with EW$_{\rm{Ly}\alpha}>20$ \AA.  The large EW 
Ly$\alpha$ emitters with low masses likely probe a very similar population.    
Strong [O {\scriptsize III}] emitters at $z\simeq 1$ have previously been shown to have [O {\scriptsize III}]/[O {\scriptsize II}] 
ratios (O32 $\gtrsim 3-100$) similar to those we report for our sample \citep{Kakazu2007}.  
Among local galaxy samples, the Green Peas also exhibit large [O {\scriptsize III}]/[O {\scriptsize II}] ratios and large [O {\scriptsize III}] 
equivalent widths \citep{Cardamone2009}.  In Figure \ref{fig:line_ratio}, we overlay the O32 and Ne3O2 values of the Green 
Peas on our high redshift relationships (green open circles).  The scaling between the ionization-sensitive ratios 
and the [O {\scriptsize III}] EW appears very similar to what we have found at high redshift.

Over the past few years, attention has focused on a possible connection between large 
values of O32 and the escape of ionizing radiation \citep[e.g.][]{Izotov2016,Izotov2017,Izotov2018,Naidu2018,Fletcher2018}. 
It has been argued  that large [O {\scriptsize III}]/[O {\scriptsize II}] ratios may arise in density-bounded nebula in which ionizing 
radiation is able to escape \citep{Guseva2004,Jaskot2013,Nakajima2014}, although as we note below this is not a unique 
interpretation.   An investigation of five compact $z\simeq 0.3$ Green Peas with relatively large [O {\scriptsize III}]/[O {\scriptsize II}] ratios 
(O32 $=6.4-8.9$ with a median of O32 $=6.5$) revealed substantial ionizing photon escape fractions 
($f_{\rm{esc}}=0.06-0.13$) in all five systems \citep{Izotov2016}.  More recent work has shown 
even larger escape fractions ($f_{\rm{esc}}>0.3$) in several galaxies with O32 $>10$, 
while also demonstrating that there is considerable scatter in LyC escape fractions at large O32 
\citep[e.g.][]{Izotov2018}.  Large O32 may be a necessary but not sufficient condition for the 
detection of LyC emission in low redshift samples \citep{Izotov2017}.  Work at higher redshift indicates a similar picture, with 
the small number of known LyC leakers typically exhibiting large [O {\scriptsize III}]/[O {\scriptsize II}] ratios 
\citep[e.g.][]{Vanzella2016,deBarros2016,Fletcher2018}.    High redshift galaxies with moderate [O {\scriptsize III}]/[O {\scriptsize II}] ratios 
(O32 $>3$) tend not to show LyC detections \citep{Naidu2018}, implying relatively low 
escape fractions ($f_{\rm{esc}}<0.1$).  This is consistent with the local results indicating 
the largest escape fractions are generally found in galaxies with the largest O32 values.  
We emphasize that the relationship between O32 and LyC leakage does not uniquely point to 
density-bounded nebula, as stellar ionizing spectra are capable of producing the majority of O32 
values seen in these studies \citep[e.g.][]{Stasinska2015}.  But it does suggest that the ISM of galaxies 
with large O32 ($\gtrsim 6.5$) is conducive to the escape of ionizing radiation.  

While we do not have direct constraints on the escape fraction of the galaxies in our sample, 
the results in \S\ref{sec:phy_prop} show that not all extreme emission line galaxies power the O32 
ratios that appear to be associated with large escape fractions at lower redshifts.  
The extreme [O {\scriptsize III}]/[O {\scriptsize II}] ratios (O32 $>10$) associated with many LyC leakers 
($f_{\rm{esc}}\gtrsim 0.3$) become the norm for galaxies with 
extremely strong [O {\scriptsize III}] emission (EW$_{\rm{[OIII]}\lambda5007}>1100$ \AA),  
implying very large specific star formation rates ($\gtrsim 100$ Gyr$^{-1}$) and  
young stellar populations ($\sim 3-10$ Myr for constant star formation; see \S\ref{sec:modeling}).    The O32 
values that appear to be associated with more moderate LyC escape fractions (O32 $\simeq 6.5$, 
$f_{\rm{esc}}\simeq 0.1$) map to galaxies with EW$_{\rm{[OIII]}\lambda5007}\simeq760$ \AA,  
corresponding to (constant star formation) ages of $10-30$ Myr.  There are several important caveats to this picture.  
First the relation between O32 and $f_{\rm{esc}}$ remains poorly sampled and needs additional 
constraints at low and high redshift.  Second, significant LyC leakage has recently been reported by 
\citet{Shapley2016} in a high redshift galaxy that appears to be very different from what we have described 
above, with a comparatively old age ($\simeq 1$ Gyr for constant star formation) and relatively weak [O {\scriptsize III}]+H$\beta$ emission 
(EW$_{\rm{[OIII]+H}\beta}<256$ \AA).  Clearly young stellar populations are not the only route toward LyC escape.  
Finally, the absolute ages we quote above are subject to standard systematic uncertainties associated 
with our assumed star formation history and population synthesis model (see \S\ref{sec:modeling}) and may be somewhat 
larger if binary stars contribute significantly to ionizing production.   Nevertheless the data suggest a relative trend 
whereby the conditions for LyC escape appear to be optimized in galaxies with 
extremely young stellar populations and intense optical line emission.  This is consistent with a picture 
whereby LyC escape can reach large values for a short window following a burst of star formation (and after the 
birth cloud has been cleared), as has been predicted in many simulations \citep[e.g.][]{Kimm2014,Wise2014,Ma2015,Paardekooper2015}.

The potential relationship between LyC leakage and rest-frame optical line properties (in particular, O32 and [O {\scriptsize III}] EW) 
has  implications for our understanding of ionizing photon escape in the reionization era.  
Over the past few years, attention has focused on the intense [O {\scriptsize III}] emission  that is thought to 
become common at $z\simeq 7-8$ \citep{Labbe2013,Smit2014,Smit2015}.  While extreme 
emission line galaxies may be the norm at these redshifts, the average equivalent width 
(EW$_{\rm{[OIII]}\lambda5007}\simeq450$ \AA; see \S\ref{sec:balmer}) is considerably lower than 
that typically seen in LyC leakers.  If the mapping between [O {\scriptsize III}] EW and O32 (valid at $0.3 \lesssim z\lesssim 2.4$; 
Figure \ref{fig:line_ratio}) is similar at $z\gtrsim 6$, it would imply typical [O {\scriptsize III}]/[O {\scriptsize II}] ratios of O32 $\simeq 3.6$.  
As noted above, high redshift galaxies with similar optical line properties tend to have relatively 
low escape fractions \citep{Naidu2018}.   It is of course possible that the galaxies at $z\gtrsim 6$ 
have different ISM conditions at fixed [O {\scriptsize III}] EW.   Spectroscopic measurements with {\it JWST} will 
soon clarify the [O {\scriptsize III}] EW and O32 distribution at $z\gtrsim 6$, allowing constraints to be placed on the fraction 
of the population with  rest-frame optical line properties that are correlated with significant LyC escape.  


 \subsection{Implications for the Ly$\alpha$ visibility test} \label{sec:lya}

The Ly$\alpha$ visibility test provides one of  the few probes of the IGM ionization state 
at $7\lesssim z\lesssim 8$.  The rapid decline in the population at $z\gtrsim 7$ is consistent 
with a substantially neutral IGM at $7\lesssim z\lesssim 8$ \citep[e.g.][]{Mason2018a,Weinberger2018}.  
The   $100\%$ success rate in identifying Ly$\alpha$ in the 
\citetalias{Roberts-Borsani2016} sample of [O {\scriptsize III}] emitters \citep{Oesch2015,Zitrin2015,Stark2017} stands in 
sharp contrast to the weak Ly$\alpha$ that is typical of the reionization era, likely 
indicating that Ly$\alpha$ visibility is 
enhanced in massive galaxies with intense optical line emission.  Specifically, the fraction of these 
sources with Ly$\alpha$ EW $>25$ \AA\  (x$_{\rm{Ly}\alpha}=0.50\pm0.29$) is a factor 
of five greater than the general population \citep{Stark2017}.   
It has been suggested that these systems may trace overdense regions that produce extremely large 
ionized bubbles, boosting transmission of Ly$\alpha$ in a mostly neutral IGM. 
While the galaxies must  lie in ionized regions for Ly$\alpha$ to escape, we suggest that 
their atypical visibility may instead be related to properties internal to the galaxies.  
The trend between $\xi^{\rm{HII}}_{\rm{ion}}$ and [O {\scriptsize III}] EW (Figure 8) 
suggests the \citetalias{Roberts-Borsani2016} galaxies (EW$_{\rm{[OIII]}\lambda5007}>800$ \AA) are likely to be twice 
as efficient at producing Ly$\alpha$ as typical galaxies (EW$_{\rm{[OIII]}\lambda5007}\simeq450$ \AA) 
in the reionization era.  In addition to enhanced production, the transmission through the galaxy is 
likely to be increased due to the highly ionized state of the ISM in the most extreme line emitters 
(see Figure \ref{fig:line_ratio}).  

To test the impact of $\xi^{\rm{HII}}_{\rm{ion}}$ and ISM variations on the Ly$\alpha$ EW 
distribution, we consider the Ly$\alpha$ properties of lower redshift samples with 
intense optical line emission.  \citet{Stark2017} characterized the fraction of large 
EW ($>25$ \AA) Ly$\alpha$ emission in a sample galaxies at $3.8<z<5.0$ with 
intense H$\alpha$ emission.  The H$\alpha$ EWs are determined using {\it Spitzer}/IRAC color 
excesses.  For the sake of comparison with the results at $z\gtrsim 7$, \citet{Stark2017} 
isolate luminous galaxies ($M_{\rm{UV}}<-21$) with blue UV slopes ($\beta<-1.8$) and an 
H$\alpha$ threshold (EW$_{\rm{H\alpha+[NII]+[SII]}}>600$ \AA) that is is chosen to match the 
[O {\scriptsize III}] EW cut of the \citetalias{Roberts-Borsani2016} sample.  The Ly$\alpha$ emitter fraction in this sub-sample is 
found to be x$_{\rm{Ly}\alpha}=0.53\pm0.17$, nearly a factor of five greater than that of 
the general population of $z\simeq 4-5$ galaxies of the same luminosity (x$_{\rm{Ly}\alpha}=
0.12$; \citealt{Stark2010}).  Because the IGM is highly ionized at these redshifts, the increased Ly$\alpha$ 
output that is observed should reflect factors internal to the galaxies.   

A similar test can be conducted with Green Peas, the population of EELGs 
at $z\simeq 0.3$.   The Ly$\alpha$ properties of 43 Green Peas  
have recently been reported by \citet{Yang2017}.  The [O {\scriptsize III}] emission lines in their sample are 
very strong, with a median EW of EW$_{\rm{[OIII]}\lambda5007}\simeq900$ \AA, 
comparable to the \citetalias{Roberts-Borsani2016} galaxies.  Large EW Ly$\alpha$ is shown to be common in the {\it HST}/COS spectra, with 
28 of 43 galaxies having Ly$\alpha$ EW $>25$ \AA.  This implies a Ly$\alpha$ emitter 
fraction (x$_{\rm{Ly}\alpha}=0.65$) that is similar to that measured in the extreme 
line emitters at $3.8<z<5.0$ and at $z\gtrsim 7$.  

The low redshift samples described above confirm that the conditions are conducive to strong  
Ly$\alpha$ emission in galaxies with intense optical line emission.  The factor of five enhancement in 
 x$_{\rm{Ly}\alpha}$ seen in the \citetalias{Roberts-Borsani2016} galaxies is fully consistent with that measured in EELGs at $z\simeq 4-5$, 
indicating that IGM variations (i.e., larger than average ionized bubbles) are not required to explain the large Ly$\alpha$ success rate 
in this sample of galaxies.   It has also been argued that the visibility is likely further enhanced in 
massive galaxies owing to their larger Ly$\alpha$ velocity offsets (e.g., \citealt{Stark2017,Mason2018a}).  
Taken together, these results suggest that massive galaxies with extremely large [O {\scriptsize III}] EW are the systems 
most likely to have detectable Ly$\alpha$ at very high redshifts where the IGM is still substantially neutral.  
While less massive systems will provide the most sensitive probe of the IGM in the middle and late 
stages of reionization (e.g., \citealt{Mason2018a,Weinberger2018}), their Ly$\alpha$ emission will be mostly 
extinguished at earlier epochs.  Because of their prodigious Ly$\alpha$ output, massive extreme emission line 
galaxies may offer a unique  path toward studying the IGM at the earliest stages of reionization (see \citealt{Mason2018b}).   



\begin{figure}
\begin{center}
\includegraphics[width=\linewidth]{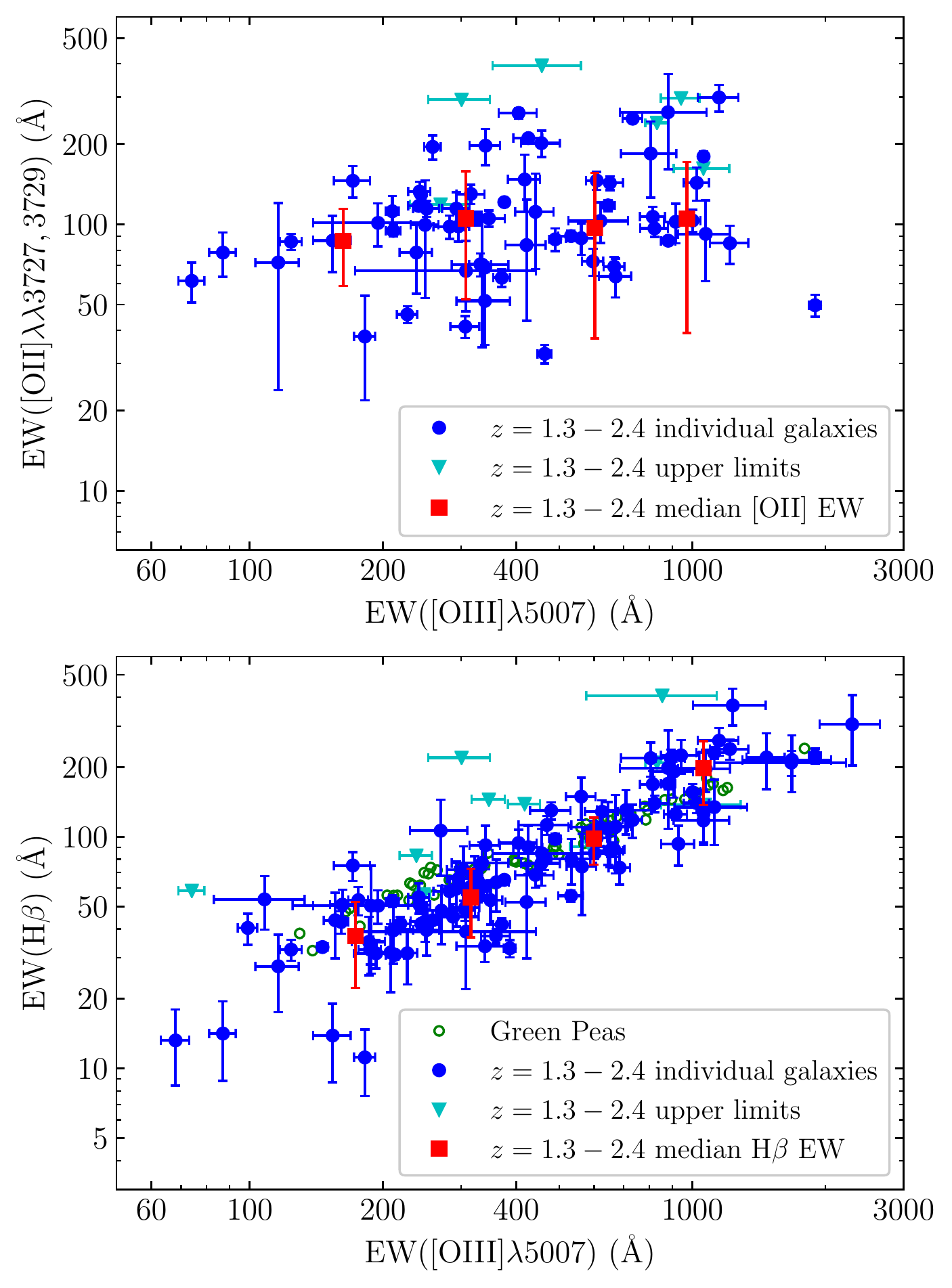}
\caption{\textit{Left Panel}: [O {\scriptsize II}]$\lambda\lambda3727,3729$ EW as a function of [O {\scriptsize III}]$\lambda5007$ EW for the $z=1.3-2.4$ galaxies in our sample. Sources with $>2\sigma$ detections of [O {\scriptsize II}] are shown by blue circles, while cyan upside-down triangles show the $2\sigma$ upper limits. Red squares present the median [O {\scriptsize II}] EWs of galaxies in different [O {\scriptsize III}]$\lambda5007$ EW bins. \textit{Right Panel}: H$\beta$ EW as a function of [O {\scriptsize III}]$\lambda5007$ EW for the $z=1.3-2.4$ galaxies in our sample. Sources with $>2\sigma$ detections of H$\beta$ are shown by blue circles, while cyan upside-down triangles show the $2\sigma$ upper limits. Red squares present the median H$\beta$ EWs of galaxies in different [O {\scriptsize III}]$\lambda5007$ EW bins. For comparison, EWs of Green Peas \citep{Cardamone2009,Hawley2012} are shown by green open circles.}
\label{fig:o2hbew}
\end{center}
\end{figure}

\subsection{Implications for {\it JWST} observations of $z\simeq 10$ galaxies} \label{sec:jwst}

The discovery that reionization era systems tend to have strong [O {\scriptsize III}] emission is important 
not only for what it tells us about the nature of early galaxies, but also for what it implies 
about the feasibility of future spectroscopic surveys with {\it JWST}.  If large equivalent line emission 
is commonplace, it will be straightforward to build large redshift samples down to faint 
continuum flux levels.   But at the highest redshifts that {\it JWST} will hope to probe ($z\gtrsim 10$), 
the prospects for spectroscopy are not so straightforward.  The [O {\scriptsize III}] line shifts out of the 
NIRSpec bandpass above $z\simeq 9$, leaving only  faint nebular lines in the rest-frame optical 
([O {\scriptsize II}], [Ne {\scriptsize III}])) and rest-UV (C {\scriptsize III}], O {\scriptsize III}], C {\scriptsize IV}).  
Over the past few years, there has been considerable efforts invested in the UV lines, 
revealing equivalent widths that approach $20$ \AA\ for $10-20\%$ of the population \citep{Mainali2018}.   
In this section, we turn our attention to the EW distribution of the fainter rest-frame optical lines in the extreme emission line 
population and consider which features are likely to provide the best route to spectroscopic confirmation at $z\gtrsim 9$.  

We first consider the feasibility of detecting [O {\scriptsize II}] emission.  Because of the large O32 values 
expected at high redshift, [O {\scriptsize II}] is often regarded as prohibitively faint. We have 
obtained [O {\scriptsize II}] measurements for 61 galaxies with EW$_{\rm{[OIII]}\lambda5007}>225$ \AA.  
In the top panel of Figure \ref{fig:o2hbew}, we show the dependence of the [O {\scriptsize II}] doublet EW on the [O {\scriptsize III}]$\lambda5007$ EW.  
For the 55 sources with robust [O {\scriptsize II}] detections, we measure a median EW of EW$_{\rm{[OII]}}=103$ \AA, 
with the sample ranging between EW$_{\rm{[OII]}}=63$ \AA\ (25th percentile) and EW$_{\rm{[OII]}}=195$ \AA\ 
(75th percentile), considerably larger than the equivalent widths of the most extreme UV metal line emitters. 
It is also evident from Figure \ref{fig:o2hbew} that 
there is no strong correlation between [O {\scriptsize II}] EWs and [O {\scriptsize III}] EWs
at EW$_{\rm{[OIII]}\lambda5007}>225$ \AA. 
Given what we know about the [O {\scriptsize III}] equivalent widths in the reionization-era \citep[e.g.][]{Labbe2013,Smit2015}, 
we can use Figure \ref{fig:o2hbew} to predict the observability of [O {\scriptsize II}] at $z\gtrsim 7$.   If we consider a $z\simeq 10$ galaxy with the 
median [O {\scriptsize II}] EW found in EELGs ($103$ \AA), we predict [O {\scriptsize II}] doublet line fluxes of 
$2.8\times10^{-18}$, $1.1\times10^{-18}$, and $4.5\times10^{-19}$ erg cm$^{-2}$ s$^{-1}$ for  a 
$z\simeq 10$ galaxy with apparent magnitudes of $H=26$, $27$, and $28$, respectively.  Here we assume a 
flat continuum in $f_\nu$ to extrapolate to the underlying continuum near [O {\scriptsize II}].  We compute the 
detectability of [O {\scriptsize II}] assuming each component of the 
doublet has a line width typical of the EELGs in our sample (FWHM $=190$ km s$^{-1}$).  For  
$R\simeq 1000$ NIRSpec MSA observations with the G395M/F290LP disperser-filter combination, 
the NIRSpec exposure time calculator predicts the doublet can be detected with S/N $=5$ in 
27 minutes (assuming underlying continuum of $H=26$), 2.7 hours ($H=27$), and 16 hrs ($H=28$).  This is considerably more efficient 
than targeting the C {\scriptsize III}] doublet with NIRSpec.  Assuming the most optimistic value for its equivalent 
width (EW$_{\rm{CIII]}}\simeq20$ \AA), likely present in only a subset of early galaxies \citep{Mainali2018}, 
we predict exposure times that are $\gtrsim 6\times$ larger than what is required for detection of [O {\scriptsize II}].  

If the ISM conditions at very high redshift are similar to what we 
have observed at $z\simeq 0-2$, we expect [Ne {\scriptsize III}] will be $0.5\times$ the flux of the combined 
[O {\scriptsize II}] doublet in the most extreme line emitters.   [Ne {\scriptsize III}] will be narrower than [O {\scriptsize II}] 
in the $R\simeq 1000$ grism, allowing it to be detected in just 1 hr ($H=26$) and 6 hrs ($H=27$) 
for sources with EW$_{\rm{[NeIII]}}=50$ \AA.   As recently pointed out in \citet{Shapley2017}, deep NIRSpec integrations 
targeting $z\simeq 10$ galaxies should easily be able to recover both lines in sources with $H\lesssim 27$, 
providing a unique path toward constraining the ISM conditions in the highest redshift galaxies {\it JWST} will probe.  

The H$\beta$ line will not be detectable by NIRSpec at $z\gtrsim 10$, as it shifts out of the bandpass at  
roughly the same redshift as [O {\scriptsize III}].   But it will serve as an important diagnostic  for reionization-era systems. 
Since H$\beta$ is likely to be among the faintest rest-frame optical lines targeted, it is valuable to know the range of 
H$\beta$ equivalent widths expected in the extreme emission line population.  In the bottom panel of Figure 
\ref{fig:o2hbew}, we plot the dependence of the H$\beta$ EW on the [O {\scriptsize III}]$\lambda5007$ EW in our spectroscopic sample.  
The H$\beta$ EW scales with the [O {\scriptsize III}] EW, with values of  EW$_{\rm{H}\beta}=53$, $98$, and $198$ \AA\ for 
[O {\scriptsize III}]$\lambda5007$ EW bins of EW$_{\rm{[OIII]}\lambda5007}=225-450$, $450-800$, and $800-2500$ \AA, respectively.   
As can be seen in Figure \ref{fig:o2hbew}, the relationship between the H$\beta$ EW and the [O {\scriptsize III}] EW is very similar for the Green Peas 
(at EW$_{\rm{[OIII]}\lambda5007}>350$ \AA), suggesting very little evolution in the trend over $0.3\lesssim z\lesssim 2.4$.
If the evolution remains slow at $z\gtrsim 3$, it indicates that H$\beta$ should be detectable in typical ($H\lesssim 27$) 
reionization-era galaxies with deep ($\lesssim 10$ hr) NIRSpec integrations.  

While much attention has recently focused on the rest-UV metal lines, we show here that the fainter rest-frame optical 
emission lines (i.e., [O {\scriptsize II}], [Ne {\scriptsize III}], H$\beta$) are likely to be provide a more efficient route toward spectroscopic 
confirmation of EELGs at $9\lesssim z\lesssim 12$.  For sources that are bright enough ($H\lesssim 26$), the UV nebular lines will 
complement the picture provided by the rest-frame optical lines, providing unique constraints on the nature of the 
ionizing sources \citep[e.g.][]{Feltre2016,Jaskot2016,Mainali2017,Senchyna2017,Byler2018} 
and the abundance of carbon in the early universe \citep[e.g.][]{Berg2016}.  
But initial {\it JWST}/NIRSpec efforts aimed at building a statistical sample of redshifts for $z\simeq 10$ galaxies are likely 
to be more successful targeting the observed wavelengths around [O {\scriptsize II}] and [Ne {\scriptsize III}].  


\section{Summary} \label{sec:summary}

We report results from a large MMT and Keck spectroscopic survey targeting 
strong rest-frame optical emission lines in extreme emission line galaxies at high redshift.  
We have confirmed redshifts of 227 galaxies at $z=1.3-2.4$ with rest-frame [O {\scriptsize III}]$\lambda5007$ equivalent 
widths spanning $225$ \AA\ $<\rm{EW}_{\rm{[OIII]}\lambda5007}<2500$ \AA. This range is chosen 
to include the average [O {\scriptsize III}] EW of reionization-era galaxies 
(EW$_{\rm{[OIII]}\lambda5007}\simeq450$ \AA; \citetalias{Labbe2013}) and the more extreme values 
(EW$_{\rm{[OIII]}\lambda5007}>800$ \AA) that characterize most of the known Ly$\alpha$ emitters at $z>7$.  
By obtaining spectra of similarly intense line emitters at slightly lower redshifts, we hope to gain 
insight into the range of stellar populations and gas conditions present in the reionization era as 
well as what distinguishes the Ly$\alpha$ emitting galaxies from the rest of the population at $z\gtrsim 7$.  

We fit the broadband continuum SED and emission line fluxes using the BEAGLE tool  
\citep{Chevallard2016}.  The stellar masses implied by the models extend down to 
$10^7\ M_\odot$ with a median of $4.9\times10^8\ M_\odot$, both well below typical 
values in the KBSS and MOSDEF surveys.  The sSFR distribution extends up to $300$ Gyr$^{-1}$ 
with a median of $17$ Gyr$^{-1}$, consistent with very young stellar populations expected after a 
substantial burst or recent upturn of star formation.  We summarize our findings below:

1.  The Balmer decrement decreases with increasing [O {\scriptsize III}] and H$\alpha$ EW, reaching a 
value close to that expected for case B recombination ($I(\rm{H}\alpha)/I(\rm{H}\beta)=2.90\pm0.05$) 
in our composite spectrum of galaxies with EW$_{\rm{[OIII]}\lambda5007}=800-2500$ \AA.  
The UV continuum slopes are uniformly blue, with median values of $\beta=-2.04$ for  
galaxies with EW$_{\rm{[OIII]}\lambda5007}>450$ \AA.  This suggests that both the hot stars 
and nebular gas are minimally obscured by dust in the most extreme line emitters.

2.  We find that the ionizing production efficiency scales with the [O {\scriptsize III}] and H$\alpha$ EW at high redshift, 
reaching the largest values ($\log{[\xi^{\rm{HII}}_{\rm{ion}}\ (\rm{erg}^{-1}\ \rm{Hz})]}\simeq 25.6$)
in the most intense emitters (EW$_{\rm{[OIII]}\lambda5007}>800$ \AA). 
We derive simple scaling laws between the ionizing production efficiency and the [O {\scriptsize III}] and H$\alpha$ 
equivalent widths.  Comparison to a recent investigation of nearby galaxies \citep{Chevallard2018}
reveals that there is not strong evolution in the link between $\xi^\star_{\rm{ion}}$ and the optical lines 
over $0\lesssim z\lesssim 2$.   The trend between the ionizing efficiency and [O {\scriptsize III}] EW indicates that 
the extreme line emitters which become common in the reionization era are likely to be very 
efficient ionizing agents.  

3.  We discuss the impact of the $\xi^\star_{\rm{ion}}-$ [O {\scriptsize III}]  EW relationship on the visibility of 
Ly$\alpha$ in the reionization era.    A large ionizing efficiency results in a larger than average 
intrinsic Ly$\alpha$ EW, boosting the visibility relative to systems with lower [O {\scriptsize III}] EW.  The 
detectability is further enhanced by the ionization conditions in the ISM and CGM and the large Ly$\alpha$ velocity 
offsets associated with massive galaxies.   We  suggest that the anomalous Ly$\alpha$ success rate in massive 
galaxies with intense [O {\scriptsize III}] is likely driven by factors internal to the galaxies and does not 
necessarily require these objects to sit in the very largest ionized bubbles at these early epochs.  Because 
of their atypical Ly$\alpha$ visibility, massive extreme emission line galaxies may provide a unique  
probe of the earliest stages of reionization (see \citealt{Mason2018b}) where substantial neutral fractions (and smaller velocity 
offsets) make lower mass galaxies very difficult to detect in Ly$\alpha$ \citep{Mason2018a,Weinberger2018}.

4.  We show that two ionization-sensitive line ratios (O32 and Ne3O2) increase with the [O {\scriptsize III}] EW, 
revealing extreme ionization conditions in the ISM of the most intense line emitters.  In a stacked 
spectrum of galaxies with EW$_{\rm{[OIII]}\lambda5007}=800-2500$ \AA, the ionization-sensitive 
ratios have very large values (O32 $=9.1$ and Ne3O2 $=0.51$) that are commonly 
associated with LyC leakers at $0\lesssim z\lesssim 2$.  We find that very similar O32 and Ne3O2 
relationships are also present in Green Pea population, indicating very little redshift evolution 
over $0.3\lesssim z\lesssim 2.4$ and suggesting a close connection between the [O {\scriptsize III}] EW and the 
ionization state of the ISM in the extreme line emitter population.

5.  We discuss the O32 $-$ [O {\scriptsize III}] EW relationship in the context of emerging evidence that 
large O32 values tend to accompany LyC leakers.  Our results demonstrate that the 
O32 values  that appear associated with moderate ($f_{\rm{esc}}\simeq 0.1$) 
escape fractions (O32 $\gtrsim 6.5$) are most commonly found in EELGs with  
EW$_{\rm{[OIII]}\lambda5007}>760$ \AA.  The extreme O32 (O32 $>10$) values 
linked with larger escape fractions ($f_{\rm{esc}}\gtrsim 0.3$) become the norm 
for galaxies with extremely strong [O {\scriptsize III}] emission (EW$_{\rm{[OIII]}\lambda5007}>1100$ \AA).
These [O {\scriptsize III}] EW values are well above the threshold commonly adopted for EELG selections, 
corresponding to extremely young stellar populations ($\simeq 3-30$ Myr for constant star formation history), 
consistent with a picture whereby LyC leakage is enhanced during a short window after a burst of
star formation. 

6.  We show that [O {\scriptsize II}] is likely to be one of the efficient spectroscopic probes at 
very high redshifts ($z\simeq 10$) where the strongest optical lines are shifted out of the 
NIRSpec bandpass.   In spite of the large O32 values in the strongest line emitters, the 
median [O {\scriptsize II}] doublet EW remains fairly large (EW$_{\rm{[OII]}}\simeq100$ \AA) 
throughout the EELG population.  We show that [O {\scriptsize II}] is likely to be detected in 
considerably less exposure time than the strongest of the UV nebular lines (i.e., C {\scriptsize III}], O {\scriptsize III}]).  
[Ne {\scriptsize III}] should also be readily detectable if Ne3O2 ratios are as large as indicated by our survey.


\section*{Acknowledgements}

We are grateful for enlightening conversations with Michael Maseda, Alice Shapley, Irene Shivaei, and David Sobral. 
DPS acknowledges support from the National Science Foundation  through the grant AST-1410155. 
Observations reported here were obtained at the MMT Observatory, a joint facility of the University of Arizona 
and the Smithsonian Institution. We thank the MMT/MMIRS queue observers, ShiAnne Kattner and Chun Ly, 
who helped a lot collect the data. This work was partially supported  by a NASA Keck PI Data Award, administered by the 
NASA Exoplanet Science Institute. Data presented herein were obtained at the W. M. Keck Observatory 
from telescope time allocated to the National Aeronautics and Space Administration through the agency's 
scientific partnership with the California Institute of Technology and the University of California. 
The Observatory was made possible by the generous financial support of the W. M. Keck Foundation.
The authors acknowledge the very significant cultural role that the
summit of Mauna Kea has always had within the indigenous Hawaiian community.
We are most fortunate to have the opportunity to conduct observations from this mountain. 



\bibliographystyle{mnras}
\bibliography{z12eelg_reion}

\begin{thebibliography}{}
\makeatletter
\relax
\def\mn@urlcharsother{\let\do\@makeother \do\$\do\&\do\#\do\^\do\_\do\%\do\~}
\def\mn@doi{\begingroup\mn@urlcharsother \@ifnextchar [ {\mn@doi@}
  {\mn@doi@[]}}
\def\mn@doi@[#1]#2{\def\@tempa{#1}\ifx\@tempa\@empty \href
  {http://dx.doi.org/#2} {doi:#2}\else \href {http://dx.doi.org/#2} {#1}\fi
  \endgroup}
\def\mn@eprint#1#2{\mn@eprint@#1:#2::\@nil}
\def\mn@eprint@arXiv#1{\href {http://arxiv.org/abs/#1} {{\tt arXiv:#1}}}
\def\mn@eprint@dblp#1{\href {http://dblp.uni-trier.de/rec/bibtex/#1.xml}
  {dblp:#1}}
\def\mn@eprint@#1:#2:#3:#4\@nil{\def\@tempa {#1}\def\@tempb {#2}\def\@tempc
  {#3}\ifx \@tempc \@empty \let \@tempc \@tempb \let \@tempb \@tempa \fi \ifx
  \@tempb \@empty \def\@tempb {arXiv}\fi \@ifundefined
  {mn@eprint@\@tempb}{\@tempb:\@tempc}{\expandafter \expandafter \csname
  mn@eprint@\@tempb\endcsname \expandafter{\@tempc}}}

\bibitem[\protect\citeauthoryear{{Abazajian} et~al.}{{Abazajian}
  et~al.}{2009}]{Abazajian2009}
{Abazajian} K.~N.,  et~al., 2009, \mn@doi [\apjs]
  {10.1088/0067-0049/182/2/543}, \href
  {https://ui.adsabs.harvard.edu/#abs/2009ApJS..182..543A} {182}

\bibitem[\protect\citeauthoryear{{Alexander} et~al.}{{Alexander}
  et~al.}{2003}]{Alexander2003}
{Alexander} D.~M.,  et~al., 2003, \mn@doi [\aj] {10.1086/376473}, \href
  {http://adsabs.harvard.edu/abs/2003AJ....126..539A} {126, 539}

\bibitem[\protect\citeauthoryear{{Atek}, {Richard}, {Kneib}  \&
  {Schaerer}}{{Atek} et~al.}{2018}]{Atek2018}
{Atek} H.,  {Richard} J.,  {Kneib} J.-P.,   {Schaerer} D.,  2018, \mn@doi
  [\mnras] {10.1093/mnras/sty1820}, \href
  {https://ui.adsabs.harvard.edu/#abs/2018MNRAS.479.5184A} {479, 5184}

\bibitem[\protect\citeauthoryear{{Behroozi}, {Wechsler}  \&
  {Conroy}}{{Behroozi} et~al.}{2013}]{Behroozi2013}
{Behroozi} P.~S.,  {Wechsler} R.~H.,   {Conroy} C.,  2013, \mn@doi [\apj]
  {10.1088/0004-637X/770/1/57}, \href
  {https://ui.adsabs.harvard.edu/#abs/2013ApJ...770...57B} {770, 57}

\bibitem[\protect\citeauthoryear{{Berg}, {Skillman}, {Henry}, {Erb}  \&
  {Carigi}}{{Berg} et~al.}{2016}]{Berg2016}
{Berg} D.~A.,  {Skillman} E.~D.,  {Henry} R.~B.~C.,  {Erb} D.~K.,   {Carigi}
  L.,  2016, \mn@doi [\apj] {10.3847/0004-637X/827/2/126}, \href
  {http://adsabs.harvard.edu/abs/2016ApJ...827..126B} {827, 126}

\bibitem[\protect\citeauthoryear{{Bouwens} et~al.}{{Bouwens}
  et~al.}{2015a}]{Bouwens2015a}
{Bouwens} R.~J.,  et~al., 2015a, \mn@doi [\apj] {10.1088/0004-637X/803/1/34},
  \href {https://ui.adsabs.harvard.edu/#abs/2015ApJ...803...34B} {803, 34}

\bibitem[\protect\citeauthoryear{{Bouwens}, {Illingworth}, {Oesch}, {Caruana},
  {Holwerda}, {Smit}  \& {Wilkins}}{{Bouwens} et~al.}{2015b}]{Bouwens2015b}
{Bouwens} R.~J.,  {Illingworth} G.~D.,  {Oesch} P.~A.,  {Caruana} J.,
  {Holwerda} B.,  {Smit} R.,   {Wilkins} S.,  2015b, \mn@doi [\apj]
  {10.1088/0004-637X/811/2/140}, \href
  {http://adsabs.harvard.edu/abs/2015ApJ...811..140B} {811, 140}

\bibitem[\protect\citeauthoryear{{Brammer} et~al.}{{Brammer}
  et~al.}{2012}]{Brammer2012}
{Brammer} G.~B.,  et~al., 2012, \mn@doi [\apjs] {10.1088/0067-0049/200/2/13},
  \href {http://adsabs.harvard.edu/abs/2012ApJS..200...13B} {200, 13}

\bibitem[\protect\citeauthoryear{{Bruzual} \& {Charlot}}{{Bruzual} \&
  {Charlot}}{2003}]{Bruzual2003}
{Bruzual} G.,  {Charlot} S.,  2003, \mn@doi [\mnras]
  {10.1046/j.1365-8711.2003.06897.x}, \href
  {http://adsabs.harvard.edu/abs/2003MNRAS.344.1000B} {344, 1000}

\bibitem[\protect\citeauthoryear{{Byler}, {Dalcanton}, {Conroy}, {Johnson},
  {Levesque}  \& {Berg}}{{Byler} et~al.}{2018}]{Byler2018}
{Byler} N.,  {Dalcanton} J.~J.,  {Conroy} C.,  {Johnson} B.~D.,  {Levesque}
  E.~M.,   {Berg} D.~A.,  2018, \mn@doi [\apj] {10.3847/1538-4357/aacd50},
  \href {https://ui.adsabs.harvard.edu/#abs/2018ApJ...863...14B} {863, 14}

\bibitem[\protect\citeauthoryear{{Caffau}, {Ludwig}, {Steffen}, {Freytag}  \&
  {Bonifacio}}{{Caffau} et~al.}{2011}]{Caffau2011}
{Caffau} E.,  {Ludwig} H.~G.,  {Steffen} M.,  {Freytag} B.,   {Bonifacio} P.,
  2011, \mn@doi [Sol. Phys.] {10.1007/s11207-010-9541-4}, \href
  {https://ui.adsabs.harvard.edu/#abs/2011SoPh..268..255C} {268, 255}

\bibitem[\protect\citeauthoryear{{Calzetti}, {Kinney}  \&
  {Storchi-Bergmann}}{{Calzetti} et~al.}{1994}]{Calzetti1994}
{Calzetti} D.,  {Kinney} A.~L.,   {Storchi-Bergmann} T.,  1994, \mn@doi [\apj]
  {10.1086/174346}, \href
  {https://ui.adsabs.harvard.edu/#abs/1994ApJ...429..582C} {429, 582}

\bibitem[\protect\citeauthoryear{{Calzetti}, {Armus}, {Bohlin}, {Kinney},
  {Koornneef}  \& {Storchi-Bergmann}}{{Calzetti} et~al.}{2000}]{Calzetti2000}
{Calzetti} D.,  {Armus} L.,  {Bohlin} R.~C.,  {Kinney} A.~L.,  {Koornneef} J.,
   {Storchi-Bergmann} T.,  2000, \mn@doi [\apj] {10.1086/308692}, \href
  {http://adsabs.harvard.edu/abs/2000ApJ...533..682C} {533, 682}

\bibitem[\protect\citeauthoryear{{Cardamone} et~al.}{{Cardamone}
  et~al.}{2009}]{Cardamone2009}
{Cardamone} C.,  et~al., 2009, \mn@doi [\mnras]
  {10.1111/j.1365-2966.2009.15383.x}, \href
  {http://adsabs.harvard.edu/abs/2009MNRAS.399.1191C} {399, 1191}

\bibitem[\protect\citeauthoryear{{Cardelli}, {Clayton}  \& {Mathis}}{{Cardelli}
  et~al.}{1989}]{Cardelli1989}
{Cardelli} J.~A.,  {Clayton} G.~C.,   {Mathis} J.~S.,  1989, \mn@doi [\apj]
  {10.1086/167900}, \href {http://adsabs.harvard.edu/abs/1989ApJ...345..245C}
  {345, 245}

\bibitem[\protect\citeauthoryear{{Caruana}, {Bunker}, {Wilkins}, {Stanway},
  {Lorenzoni}, {Jarvis}  \& {Ebert}}{{Caruana} et~al.}{2014}]{Caruana2014}
{Caruana} J.,  {Bunker} A.~J.,  {Wilkins} S.~M.,  {Stanway} E.~R.,  {Lorenzoni}
  S.,  {Jarvis} M.~J.,   {Ebert} H.,  2014, \mn@doi [\mnras]
  {10.1093/mnras/stu1341}, \href
  {https://ui.adsabs.harvard.edu/#abs/2014MNRAS.443.2831C} {443, 2831}

\bibitem[\protect\citeauthoryear{{Chabrier}}{{Chabrier}}{2003}]{Chabrier2003}
{Chabrier} G.,  2003, \mn@doi [\pasp] {10.1086/376392}, \href
  {http://adsabs.harvard.edu/abs/2003PASP..115..763C} {115, 763}

\bibitem[\protect\citeauthoryear{{Charlot} \& {Fall}}{{Charlot} \&
  {Fall}}{2000}]{Charlot2000}
{Charlot} S.,  {Fall} S.~M.,  2000, \mn@doi [\apj] {10.1086/309250}, \href
  {http://adsabs.harvard.edu/abs/2000ApJ...539..718C} {539, 718}

\bibitem[\protect\citeauthoryear{{Chevallard} \& {Charlot}}{{Chevallard} \&
  {Charlot}}{2016}]{Chevallard2016}
{Chevallard} J.,  {Charlot} S.,  2016, \mn@doi [\mnras]
  {10.1093/mnras/stw1756}, \href
  {http://adsabs.harvard.edu/abs/2016MNRAS.462.1415C} {462, 1415}

\bibitem[\protect\citeauthoryear{{Chevallard} et~al.,}{{Chevallard}
  et~al.}{2018}]{Chevallard2018}
{Chevallard} J.,  et~al., 2018, \mn@doi [\mnras] {10.1093/mnras/sty1461}, \href
  {https://ui.adsabs.harvard.edu/#abs/2018MNRAS.479.3264C} {479, 3264}

\bibitem[\protect\citeauthoryear{{Chilingarian}, {Beletsky}, {Moran}, {Brown},
  {McLeod}  \& {Fabricant}}{{Chilingarian} et~al.}{2015}]{Chilingarian2015}
{Chilingarian} I.,  {Beletsky} Y.,  {Moran} S.,  {Brown} W.,  {McLeod} B.,
  {Fabricant} D.,  2015, \mn@doi [\pasp] {10.1086/680598}, \href
  {http://adsabs.harvard.edu/abs/2015PASP..127..406C} {127, 406}

\bibitem[\protect\citeauthoryear{{Civano} et~al.}{{Civano}
  et~al.}{2016}]{Civano2016}
{Civano} F.,  et~al., 2016, \mn@doi [\apj] {10.3847/0004-637X/819/1/62}, \href
  {http://adsabs.harvard.edu/abs/2016ApJ...819...62C} {819, 62}

\bibitem[\protect\citeauthoryear{{Dayal} \& {Ferrara}}{{Dayal} \&
  {Ferrara}}{2018}]{Dayal2018}
{Dayal} P.,  {Ferrara} A.,  2018, \mn@doi [\physrep]
  {10.1016/j.physrep.2018.10.002}, \href
  {https://ui.adsabs.harvard.edu/abs/2018PhR...780....1D} {780, 1}

\bibitem[\protect\citeauthoryear{{Eldridge}, {Stanway}, {Xiao}, {McClelland},
  {Taylor}, {Ng}, {Greis}  \& {Bray}}{{Eldridge} et~al.}{2017}]{Eldridge2017}
{Eldridge} J.~J.,  {Stanway} E.~R.,  {Xiao} L.,  {McClelland} L.~A.~S.,
  {Taylor} G.,  {Ng} M.,  {Greis} S.~M.~L.,   {Bray} J.~C.,  2017, \mn@doi
  [\pasa] {10.1017/pasa.2017.51}, \href
  {https://ui.adsabs.harvard.edu/#abs/2017PASA...34...58E} {34, e058}

\bibitem[\protect\citeauthoryear{{Erb}, {Pettini}, {Steidel}, {Strom}, {Rudie},
  {Trainor}, {Shapley}  \& {Reddy}}{{Erb} et~al.}{2016}]{Erb2016}
{Erb} D.~K.,  {Pettini} M.,  {Steidel} C.~C.,  {Strom} A.~L.,  {Rudie} G.~C.,
  {Trainor} R.~F.,  {Shapley} A.~E.,   {Reddy} N.~A.,  2016, \mn@doi [\apj]
  {10.3847/0004-637X/830/1/52}, \href
  {https://ui.adsabs.harvard.edu/#abs/2016ApJ...830...52E} {830, 52}

\bibitem[\protect\citeauthoryear{{Feltre}, {Charlot}  \& {Gutkin}}{{Feltre}
  et~al.}{2016}]{Feltre2016}
{Feltre} A.,  {Charlot} S.,   {Gutkin} J.,  2016, \mn@doi [\mnras]
  {10.1093/mnras/stv2794}, \href
  {https://ui.adsabs.harvard.edu/#abs/2016MNRAS.456.3354F} {456, 3354}

\bibitem[\protect\citeauthoryear{{Ferland} et~al.,}{{Ferland}
  et~al.}{2013}]{Ferland2013}
{Ferland} G.~J.,  et~al., 2013, \rmxaa, \href
  {http://adsabs.harvard.edu/abs/2013RMxAA..49..137F} {49, 137}

\bibitem[\protect\citeauthoryear{{Finkelstein} et~al.}{{Finkelstein}
  et~al.}{2013}]{Finkelstein2013}
{Finkelstein} S.~L.,  et~al., 2013, \mn@doi [\nat] {10.1038/nature12657}, \href
  {https://ui.adsabs.harvard.edu/#abs/2013Natur.502..524F} {502, 524}

\bibitem[\protect\citeauthoryear{{Finkelstein} et~al.}{{Finkelstein}
  et~al.}{2015}]{Finkelstein2015}
{Finkelstein} S.~L.,  et~al., 2015, \mn@doi [\apj]
  {10.1088/0004-637X/810/1/71}, \href
  {https://ui.adsabs.harvard.edu/#abs/2015ApJ...810...71F} {810, 71}

\bibitem[\protect\citeauthoryear{{Fletcher}, {Robertson}, {Nakajima}, {Ellis},
  {Stark}  \& {Inoue}}{{Fletcher} et~al.}{2018}]{Fletcher2018}
{Fletcher} T.~J.,  {Robertson} B.~E.,  {Nakajima} K.,  {Ellis} R.~S.,  {Stark}
  D.~P.,   {Inoue} A.,  2018, preprint, \href
  {https://ui.adsabs.harvard.edu/#abs/2018arXiv180601741F} {p.
  arXiv:1806.01741} (\mn@eprint {arXiv} {1806.01741})

\bibitem[\protect\citeauthoryear{{G{\"o}tberg}, {de Mink}, {Groh}, {Kupfer},
  {Crowther}, {Zapartas}  \& {Renzo}}{{G{\"o}tberg} et~al.}{2018}]{Gotberg2018}
{G{\"o}tberg} Y.,  {de Mink} S.~E.,  {Groh} J.~H.,  {Kupfer} T.,  {Crowther}
  P.~A.,  {Zapartas} E.,   {Renzo} M.,  2018, \mn@doi [\aap]
  {10.1051/0004-6361/201732274}, \href
  {https://ui.adsabs.harvard.edu/#abs/2018A&A...615A..78G} {615, A78}

\bibitem[\protect\citeauthoryear{{Grogin} et~al.}{{Grogin}
  et~al.}{2011}]{Grogin2011}
{Grogin} N.~A.,  et~al., 2011, \mn@doi [\apjs] {10.1088/0067-0049/197/2/35},
  \href {http://adsabs.harvard.edu/abs/2011ApJS..197...35G} {197, 35}

\bibitem[\protect\citeauthoryear{{Guseva}, {Papaderos}, {Izotov}, {Noeske}  \&
  {Fricke}}{{Guseva} et~al.}{2004}]{Guseva2004}
{Guseva} N.~G.,  {Papaderos} P.,  {Izotov} Y.~I.,  {Noeske} K.~G.,   {Fricke}
  K.~J.,  2004, \mn@doi [\aap] {10.1051/0004-6361:20035949}, \href
  {https://ui.adsabs.harvard.edu/#abs/2004A&A...421..519G} {421, 519}

\bibitem[\protect\citeauthoryear{{Gutkin}, {Charlot}  \& {Bruzual}}{{Gutkin}
  et~al.}{2016}]{Gutkin2016}
{Gutkin} J.,  {Charlot} S.,   {Bruzual} G.,  2016, \mn@doi [\mnras]
  {10.1093/mnras/stw1716}, \href
  {http://adsabs.harvard.edu/abs/2016MNRAS.462.1757G} {462, 1757}

\bibitem[\protect\citeauthoryear{{Hawley}}{{Hawley}}{2012}]{Hawley2012}
{Hawley} S.~A.,  2012, \mn@doi [\pasp] {10.1086/663866}, \href
  {https://ui.adsabs.harvard.edu/#abs/2012PASP..124...21H} {124, 21}

\bibitem[\protect\citeauthoryear{{Horne}}{{Horne}}{1986}]{Horne1986}
{Horne} K.,  1986, \mn@doi [\pasp] {10.1086/131801}, \href
  {http://adsabs.harvard.edu/abs/1986PASP...98..609H} {98, 609}

\bibitem[\protect\citeauthoryear{{Inoue}, {Shimizu}, {Iwata}  \&
  {Tanaka}}{{Inoue} et~al.}{2014}]{Inoue2014}
{Inoue} A.~K.,  {Shimizu} I.,  {Iwata} I.,   {Tanaka} M.,  2014, \mn@doi
  [\mnras] {10.1093/mnras/stu936}, \href
  {http://adsabs.harvard.edu/abs/2014MNRAS.442.1805I} {442, 1805}

\bibitem[\protect\citeauthoryear{{Izotov}, {Schaerer}, {Thuan}, {Worseck},
  {Guseva}, {Orlitov{\'a}}  \& {Verhamme}}{{Izotov} et~al.}{2016}]{Izotov2016}
{Izotov} Y.~I.,  {Schaerer} D.,  {Thuan} T.~X.,  {Worseck} G.,  {Guseva} N.~G.,
   {Orlitov{\'a}} I.,   {Verhamme} A.,  2016, \mn@doi [\mnras]
  {10.1093/mnras/stw1205}, \href
  {https://ui.adsabs.harvard.edu/#abs/2016MNRAS.461.3683I} {461, 3683}

\bibitem[\protect\citeauthoryear{{Izotov}, {Thuan}  \& {Guseva}}{{Izotov}
  et~al.}{2017}]{Izotov2017}
{Izotov} Y.~I.,  {Thuan} T.~X.,   {Guseva} N.~G.,  2017, \mn@doi [\mnras]
  {10.1093/mnras/stx1629}, \href
  {https://ui.adsabs.harvard.edu/#abs/2017MNRAS.471..548I} {471, 548}

\bibitem[\protect\citeauthoryear{{Izotov}, {Worseck}, {Schaerer}, {Guseva},
  {Thuan}, {Fricke}  \& {Orlitov{\'a}}}{{Izotov} et~al.}{2018}]{Izotov2018}
{Izotov} Y.~I.,  {Worseck} G.,  {Schaerer} D.,  {Guseva} N.~G.,  {Thuan} T.~X.,
   {Fricke} Verhamme A.,   {Orlitov{\'a}} I.,  2018, \mn@doi [\mnras]
  {10.1093/mnras/sty1378}, \href
  {https://ui.adsabs.harvard.edu/#abs/2018MNRAS.478.4851I} {478, 4851}

\bibitem[\protect\citeauthoryear{{Jaskot} \& {Oey}}{{Jaskot} \&
  {Oey}}{2013}]{Jaskot2013}
{Jaskot} A.~E.,  {Oey} M.~S.,  2013, \mn@doi [\apj]
  {10.1088/0004-637X/766/2/91}, \href
  {https://ui.adsabs.harvard.edu/#abs/2013ApJ...766...91J} {766, 91}

\bibitem[\protect\citeauthoryear{{Jaskot} \& {Ravindranath}}{{Jaskot} \&
  {Ravindranath}}{2016}]{Jaskot2016}
{Jaskot} A.~E.,  {Ravindranath} S.,  2016, \mn@doi [\apj]
  {10.3847/1538-4357/833/2/136}, \href
  {http://adsabs.harvard.edu/abs/2016ApJ...833..136J} {833, 136}

\bibitem[\protect\citeauthoryear{{Kakazu}, {Cowie}  \& {Hu}}{{Kakazu}
  et~al.}{2007}]{Kakazu2007}
{Kakazu} Y.,  {Cowie} L.~L.,   {Hu} E.~M.,  2007, \mn@doi [\apj]
  {10.1086/521333}, \href
  {https://ui.adsabs.harvard.edu/#abs/2007ApJ...668..853K} {668, 853}

\bibitem[\protect\citeauthoryear{{Kewley}, {Dopita}, {Leitherer}, {Dav{\'e}},
  {Yuan}, {Allen}, {Groves}  \& {Sutherland}}{{Kewley}
  et~al.}{2013}]{Kewley2013}
{Kewley} L.~J.,  {Dopita} M.~A.,  {Leitherer} C.,  {Dav{\'e}} R.,  {Yuan} T.,
  {Allen} M.,  {Groves} B.,   {Sutherland} R.,  2013, \mn@doi [\apj]
  {10.1088/0004-637X/774/2/100}, \href
  {https://ui.adsabs.harvard.edu/#abs/2013ApJ...774..100K} {774, 100}

\bibitem[\protect\citeauthoryear{{Kimm} \& {Cen}}{{Kimm} \&
  {Cen}}{2014}]{Kimm2014}
{Kimm} T.,  {Cen} R.,  2014, \mn@doi [\apj] {10.1088/0004-637X/788/2/121},
  \href {https://ui.adsabs.harvard.edu/#abs/2014ApJ...788..121K} {788, 121}

\bibitem[\protect\citeauthoryear{{Koekemoer} et~al.}{{Koekemoer}
  et~al.}{2011}]{Koekemoer2011}
{Koekemoer} A.~M.,  et~al., 2011, \mn@doi [\apjs] {10.1088/0067-0049/197/2/36},
  \href {https://ui.adsabs.harvard.edu/#abs/2011ApJS..197...36K} {197, 36}

\bibitem[\protect\citeauthoryear{{Kriek} et~al.}{{Kriek}
  et~al.}{2015}]{Kriek2015}
{Kriek} M.,  et~al., 2015, \mn@doi [\apjs] {10.1088/0067-0049/218/2/15}, \href
  {http://adsabs.harvard.edu/abs/2015ApJS..218...15K} {218, 15}

\bibitem[\protect\citeauthoryear{{Labb{\'e}} et~al.,}{{Labb{\'e}}
  et~al.}{2013}]{Labbe2013}
{Labb{\'e}} I.,  et~al., 2013, \mn@doi [\apjl] {10.1088/2041-8205/777/2/L19},
  \href {http://adsabs.harvard.edu/abs/2013ApJ...777L..19L} {777, L19}

\bibitem[\protect\citeauthoryear{{Laporte}, {Nakajima}, {Ellis}, {Zitrin},
  {Stark}, {Mainali}  \& {Roberts-Borsani}}{{Laporte}
  et~al.}{2017}]{Laporte2017}
{Laporte} N.,  {Nakajima} K.,  {Ellis} R.~S.,  {Zitrin} A.,  {Stark} D.~P.,
  {Mainali} R.,   {Roberts-Borsani} G.~W.,  2017, \mn@doi [\apj]
  {10.3847/1538-4357/aa96a8}, \href
  {https://ui.adsabs.harvard.edu/#abs/2017ApJ...851...40L} {851, 40}

\bibitem[\protect\citeauthoryear{{Levesque} \& {Richardson}}{{Levesque} \&
  {Richardson}}{2014}]{Levesque2014}
{Levesque} E.~M.,  {Richardson} M.~L.~A.,  2014, \mn@doi [\apj]
  {10.1088/0004-637X/780/1/100}, \href
  {http://adsabs.harvard.edu/abs/2014ApJ...780..100L} {780, 100}

\bibitem[\protect\citeauthoryear{{Levesque}, {Leitherer}, {Ekstrom}, {Meynet}
  \& {Schaerer}}{{Levesque} et~al.}{2012}]{Levesque2012}
{Levesque} E.~M.,  {Leitherer} C.,  {Ekstrom} S.,  {Meynet} G.,   {Schaerer}
  D.,  2012, \mn@doi [\apj] {10.1088/0004-637X/751/1/67}, \href
  {https://ui.adsabs.harvard.edu/#abs/2012ApJ...751...67L} {751, 67}

\bibitem[\protect\citeauthoryear{{Livermore}, {Finkelstein}  \&
  {Lotz}}{{Livermore} et~al.}{2017}]{Livermore2017}
{Livermore} R.~C.,  {Finkelstein} S.~L.,   {Lotz} J.~M.,  2017, \mn@doi [\apj]
  {10.3847/1538-4357/835/2/113}, \href
  {https://ui.adsabs.harvard.edu/#abs/2017ApJ...835..113L} {835, 113}

\bibitem[\protect\citeauthoryear{{Ma}, {Kasen}, {Hopkins},
  {Faucher-Gigu{\`e}re}, {Quataert}, {Kere{\v{s}}}  \& {Murray}}{{Ma}
  et~al.}{2015}]{Ma2015}
{Ma} X.,  {Kasen} D.,  {Hopkins} P.~F.,  {Faucher-Gigu{\`e}re} C.-A.,
  {Quataert} E.,  {Kere{\v{s}}} D.,   {Murray} N.,  2015, \mn@doi [\mnras]
  {10.1093/mnras/stv1679}, \href
  {https://ui.adsabs.harvard.edu/#abs/2015MNRAS.453..960M} {453, 960}

\bibitem[\protect\citeauthoryear{{Mainali}, {Kollmeier}, {Stark}, {Simcoe},
  {Walth}, {Newman}  \& {Miller}}{{Mainali} et~al.}{2017}]{Mainali2017}
{Mainali} R.,  {Kollmeier} J.~A.,  {Stark} D.~P.,  {Simcoe} R.~A.,  {Walth} G.,
   {Newman} A.~B.,   {Miller} D.~R.,  2017, \mn@doi [\apj]
  {10.3847/2041-8213/836/1/L14}, \href
  {https://ui.adsabs.harvard.edu/#abs/2017ApJ...836L..14M} {836, L14}

\bibitem[\protect\citeauthoryear{{Mainali} et~al.,}{{Mainali}
  et~al.}{2018}]{Mainali2018}
{Mainali} R.,  et~al., 2018, \mn@doi [\mnras] {10.1093/mnras/sty1640}, \href
  {http://adsabs.harvard.edu/abs/2018MNRAS.479.1180M} {479, 1180}

\bibitem[\protect\citeauthoryear{{Maseda} et~al.}{{Maseda}
  et~al.}{2014}]{Maseda2014}
{Maseda} M.~V.,  et~al., 2014, \mn@doi [\apj] {10.1088/0004-637X/791/1/17},
  \href {http://adsabs.harvard.edu/abs/2014ApJ...791...17M} {791, 17}

\bibitem[\protect\citeauthoryear{{Mason}, {Treu}, {Dijkstra}, {Mesinger},
  {Trenti}, {Pentericci}, {de Barros}  \& {Vanzella}}{{Mason}
  et~al.}{2018a}]{Mason2018a}
{Mason} C.~A.,  {Treu} T.,  {Dijkstra} M.,  {Mesinger} A.,  {Trenti} M.,
  {Pentericci} L.,  {de Barros} S.,   {Vanzella} E.,  2018a, \mn@doi [\apj]
  {10.3847/1538-4357/aab0a7}, \href
  {https://ui.adsabs.harvard.edu/#abs/2018ApJ...856....2M} {856, 2}

\bibitem[\protect\citeauthoryear{{Mason} et~al.,}{{Mason}
  et~al.}{2018b}]{Mason2018b}
{Mason} C.~A.,  et~al., 2018b, \mn@doi [\apj] {10.3847/2041-8213/aabbab}, \href
  {https://ui.adsabs.harvard.edu/#abs/2018ApJ...857L..11M} {857, L11}

\bibitem[\protect\citeauthoryear{{Mathis}}{{Mathis}}{1986}]{Mathis1986}
{Mathis} J.~S.,  1986, \mn@doi [\pasp] {10.1086/131859}, \href
  {https://ui.adsabs.harvard.edu/#abs/1986PASP...98..995M} {98, 995}

\bibitem[\protect\citeauthoryear{{Matthee}, {Sobral}, {Best}, {Khostovan},
  {Oteo}, {Bouwens}  \& {R{\"o}ttgering}}{{Matthee} et~al.}{2017}]{Matthee2017}
{Matthee} J.,  {Sobral} D.,  {Best} P.,  {Khostovan} A.~A.,  {Oteo} I.,
  {Bouwens} R.,   {R{\"o}ttgering} H.,  2017, \mn@doi [\mnras]
  {10.1093/mnras/stw2973}, \href
  {http://adsabs.harvard.edu/abs/2017MNRAS.465.3637M} {465, 3637}

\bibitem[\protect\citeauthoryear{{McLean} et~al.}{{McLean}
  et~al.}{2012}]{McLean2012}
{McLean} I.~S.,  et~al., 2012, in Ground-based and Airborne Instrumentation for
  Astronomy IV. p. 84460J, \mn@doi{10.1117/12.924794}

\bibitem[\protect\citeauthoryear{{McLeod} et~al.}{{McLeod}
  et~al.}{2012}]{McLeod2012}
{McLeod} B.,  et~al., 2012, \mn@doi [\pasp] {10.1086/669044}, \href
  {http://adsabs.harvard.edu/abs/2012PASP..124.1318M} {124, 1318}

\bibitem[\protect\citeauthoryear{{McLure} et~al.}{{McLure}
  et~al.}{2013}]{McLure2013}
{McLure} R.~J.,  et~al., 2013, \mn@doi [\mnras] {10.1093/mnras/stt627}, \href
  {https://ui.adsabs.harvard.edu/#abs/2013MNRAS.432.2696M} {432, 2696}

\bibitem[\protect\citeauthoryear{{Momcheva} et~al.}{{Momcheva}
  et~al.}{2016}]{Momcheva2016}
{Momcheva} I.~G.,  et~al., 2016, \mn@doi [\apjs] {10.3847/0067-0049/225/2/27},
  \href {http://adsabs.harvard.edu/abs/2016ApJS..225...27M} {225, 27}

\bibitem[\protect\citeauthoryear{{Naidu}, {Forrest}, {Oesch}, {Tran}  \&
  {Holden}}{{Naidu} et~al.}{2018}]{Naidu2018}
{Naidu} R.~P.,  {Forrest} B.,  {Oesch} P.~A.,  {Tran} K.-V.~H.,   {Holden}
  B.~P.,  2018, \mn@doi [\mnras] {10.1093/mnras/sty961}, \href
  {https://ui.adsabs.harvard.edu/#abs/2018MNRAS.478..791N} {478, 791}

\bibitem[\protect\citeauthoryear{{Nakajima} \& {Ouchi}}{{Nakajima} \&
  {Ouchi}}{2014}]{Nakajima2014}
{Nakajima} K.,  {Ouchi} M.,  2014, \mn@doi [\mnras] {10.1093/mnras/stu902},
  \href {https://ui.adsabs.harvard.edu/#abs/2014MNRAS.442..900N} {442, 900}

\bibitem[\protect\citeauthoryear{{Nakajima}, {Ellis}, {Iwata}, {Inoue},
  {Kusakabe}, {Ouchi}  \& {Robertson}}{{Nakajima} et~al.}{2016}]{Nakajima2016}
{Nakajima} K.,  {Ellis} R.~S.,  {Iwata} I.,  {Inoue} A.~K.,  {Kusakabe} H.,
  {Ouchi} M.,   {Robertson} B.~E.,  2016, \mn@doi [\apjl]
  {10.3847/2041-8205/831/1/L9}, \href
  {http://adsabs.harvard.edu/abs/2016ApJ...831L...9N} {831, L9}

\bibitem[\protect\citeauthoryear{{Nakajima} et~al.}{{Nakajima}
  et~al.}{2018}]{Nakajima2018}
{Nakajima} K.,  et~al., 2018, \mn@doi [\aap] {10.1051/0004-6361/201731935},
  \href {https://ui.adsabs.harvard.edu/#abs/2018A&A...612A..94N} {612, A94}

\bibitem[\protect\citeauthoryear{{Nandra} et~al.}{{Nandra}
  et~al.}{2015}]{Nandra2015}
{Nandra} K.,  et~al., 2015, \mn@doi [\apjs] {10.1088/0067-0049/220/1/10}, \href
  {http://adsabs.harvard.edu/abs/2015ApJS..220...10N} {220, 10}

\bibitem[\protect\citeauthoryear{{Oesch} et~al.}{{Oesch}
  et~al.}{2015}]{Oesch2015}
{Oesch} P.~A.,  et~al., 2015, \mn@doi [\apj] {10.1088/2041-8205/804/2/L30},
  \href {https://ui.adsabs.harvard.edu/#abs/2015ApJ...804L..30O} {804, L30}

\bibitem[\protect\citeauthoryear{{Oesch}, {Bouwens}, {Illingworth}, {Labb{\'e}}
   \& {Stefanon}}{{Oesch} et~al.}{2018}]{Oesch2018}
{Oesch} P.~A.,  {Bouwens} R.~J.,  {Illingworth} G.~D.,  {Labb{\'e}} I.,
  {Stefanon} M.,  2018, \mn@doi [\apj] {10.3847/1538-4357/aab03f}, \href
  {https://ui.adsabs.harvard.edu/#abs/2018ApJ...855..105O} {855, 105}

\bibitem[\protect\citeauthoryear{{Oke} \& {Gunn}}{{Oke} \&
  {Gunn}}{1983}]{Oke1983}
{Oke} J.~B.,  {Gunn} J.~E.,  1983, \mn@doi [\apj] {10.1086/160817}, \href
  {https://ui.adsabs.harvard.edu/#abs/1983ApJ...266..713O} {266, 713}

\bibitem[\protect\citeauthoryear{{Ono} et~al.}{{Ono} et~al.}{2012}]{Ono2012}
{Ono} Y.,  et~al., 2012, \mn@doi [\apj] {10.1088/0004-637X/744/2/83}, \href
  {http://adsabs.harvard.edu/abs/2012ApJ...744...83O} {744, 83}

\bibitem[\protect\citeauthoryear{{Osterbrock} \& {Ferland}}{{Osterbrock} \&
  {Ferland}}{2006}]{Osterbrock2006}
{Osterbrock} D.~E.,  {Ferland} G.~J.,  2006, {Astrophysics of gaseous nebulae
  and active galactic nuclei}

\bibitem[\protect\citeauthoryear{{Paardekooper}, {Khochfar}  \& {Dalla
  Vecchia}}{{Paardekooper} et~al.}{2015}]{Paardekooper2015}
{Paardekooper} J.-P.,  {Khochfar} S.,   {Dalla Vecchia} C.,  2015, \mn@doi
  [\mnras] {10.1093/mnras/stv1114}, \href
  {https://ui.adsabs.harvard.edu/#abs/2015MNRAS.451.2544P} {451, 2544}

\bibitem[\protect\citeauthoryear{{Pei}}{{Pei}}{1992}]{Pei1992}
{Pei} Y.~C.,  1992, \mn@doi [\apj] {10.1086/171637}, \href
  {https://ui.adsabs.harvard.edu/#abs/1992ApJ...395..130P} {395, 130}

\bibitem[\protect\citeauthoryear{{Penston} et~al.}{{Penston}
  et~al.}{1990}]{Penston1990}
{Penston} M.~V.,  et~al., 1990, \aap, \href
  {http://adsabs.harvard.edu/abs/1990A\%26A...236...53P} {236, 53}

\bibitem[\protect\citeauthoryear{{Pentericci} et~al.}{{Pentericci}
  et~al.}{2014}]{Pentericci2014}
{Pentericci} L.,  et~al., 2014, \mn@doi [\apj] {10.1088/0004-637X/793/2/113},
  \href {https://ui.adsabs.harvard.edu/#abs/2014ApJ...793..113P} {793, 113}

\bibitem[\protect\citeauthoryear{{Petrosian}, {Silk}  \& {Field}}{{Petrosian}
  et~al.}{1972}]{Petrosian1972}
{Petrosian} V.,  {Silk} J.,   {Field} G.~B.,  1972, \mn@doi [\apj]
  {10.1086/181054}, \href
  {https://ui.adsabs.harvard.edu/#abs/1972ApJ...177L..69P} {177, L69}

\bibitem[\protect\citeauthoryear{{Rasappu}, {Smit}, {Labb{\'e}}, {Bouwens},
  {Stark}, {Ellis}  \& {Oesch}}{{Rasappu} et~al.}{2016}]{Rasappu2016}
{Rasappu} N.,  {Smit} R.,  {Labb{\'e}} I.,  {Bouwens} R.~J.,  {Stark} D.~P.,
  {Ellis} R.~S.,   {Oesch} P.~A.,  2016, \mn@doi [\mnras]
  {10.1093/mnras/stw1484}, \href
  {https://ui.adsabs.harvard.edu/#abs/2016MNRAS.461.3886R} {461, 3886}

\bibitem[\protect\citeauthoryear{{Reddy} et~al.,}{{Reddy}
  et~al.}{2015}]{Reddy2015}
{Reddy} N.~A.,  et~al., 2015, \mn@doi [\apj] {10.1088/0004-637X/806/2/259},
  \href {http://adsabs.harvard.edu/abs/2015ApJ...806..259R} {806, 259}

\bibitem[\protect\citeauthoryear{{Roberts-Borsani} et~al.}{{Roberts-Borsani}
  et~al.}{2016}]{Roberts-Borsani2016}
{Roberts-Borsani} G.~W.,  et~al., 2016, \mn@doi [\apj]
  {10.3847/0004-637X/823/2/143}, \href
  {http://adsabs.harvard.edu/abs/2016ApJ...823..143R} {823, 143}

\bibitem[\protect\citeauthoryear{{Robertson}, {Ellis}, {Furlanetto}  \&
  {Dunlop}}{{Robertson} et~al.}{2015}]{Robertson2015}
{Robertson} B.~E.,  {Ellis} R.~S.,  {Furlanetto} S.~R.,   {Dunlop} J.~S.,
  2015, \mn@doi [\apjl] {10.1088/2041-8205/802/2/L19}, \href
  {http://adsabs.harvard.edu/abs/2015ApJ...802L..19R} {802, L19}

\bibitem[\protect\citeauthoryear{{Sanders} et~al.,}{{Sanders}
  et~al.}{2016}]{Sanders2016}
{Sanders} R.~L.,  et~al., 2016, \mn@doi [\apj] {10.3847/0004-637X/816/1/23},
  \href {http://adsabs.harvard.edu/abs/2016ApJ...816...23S} {816, 23}

\bibitem[\protect\citeauthoryear{{Sanders} et~al.}{{Sanders}
  et~al.}{2018}]{Sanders2018}
{Sanders} R.~L.,  et~al., 2018, \mn@doi [\apj] {10.3847/1538-4357/aabcbd},
  \href {https://ui.adsabs.harvard.edu/#abs/2018ApJ...858...99S} {858, 99}

\bibitem[\protect\citeauthoryear{{Schenker}, {Ellis}, {Konidaris}  \&
  {Stark}}{{Schenker} et~al.}{2014}]{Schenker2014}
{Schenker} M.~A.,  {Ellis} R.~S.,  {Konidaris} N.~P.,   {Stark} D.~P.,  2014,
  \mn@doi [\apj] {10.1088/0004-637X/795/1/20}, \href
  {http://adsabs.harvard.edu/abs/2014ApJ...795...20S} {795, 20}

\bibitem[\protect\citeauthoryear{{Schmidt}, {Weymann}  \& {Foltz}}{{Schmidt}
  et~al.}{1989}]{Schmidt1989}
{Schmidt} G.~D.,  {Weymann} R.~J.,   {Foltz} C.~B.,  1989, \mn@doi [\pasp]
  {10.1086/132495}, \href
  {https://ui.adsabs.harvard.edu/#abs/1989PASP..101..713S} {101, 713}

\bibitem[\protect\citeauthoryear{{Schmidt} et~al.}{{Schmidt}
  et~al.}{2017}]{Schmidt2017}
{Schmidt} K.~B.,  et~al., 2017, \mn@doi [\apj] {10.3847/1538-4357/aa68a3},
  \href {https://ui.adsabs.harvard.edu/#abs/2017ApJ...839...17S} {839, 17}

\bibitem[\protect\citeauthoryear{{Senchyna} et~al.,}{{Senchyna}
  et~al.}{2017}]{Senchyna2017}
{Senchyna} P.,  et~al., 2017, \mn@doi [\mnras] {10.1093/mnras/stx2059}, \href
  {http://adsabs.harvard.edu/abs/2017MNRAS.472.2608S} {472, 2608}

\bibitem[\protect\citeauthoryear{{Shapley}, {Steidel}, {Strom},
  {Bogosavljevi{\'c}}, {Reddy}, {Siana}, {Mostardi}  \& {Rudie}}{{Shapley}
  et~al.}{2016}]{Shapley2016}
{Shapley} A.~E.,  {Steidel} C.~C.,  {Strom} A.~L.,  {Bogosavljevi{\'c}} M.,
  {Reddy} N.~A.,  {Siana} B.,  {Mostardi} R.~E.,   {Rudie} G.~C.,  2016,
  \mn@doi [\apj] {10.3847/2041-8205/826/2/L24}, \href
  {https://ui.adsabs.harvard.edu/#abs/2016ApJ...826L..24S} {826, L24}

\bibitem[\protect\citeauthoryear{{Shapley} et~al.}{{Shapley}
  et~al.}{2017}]{Shapley2017}
{Shapley} A.~E.,  et~al., 2017, \mn@doi [\apj] {10.3847/2041-8213/aa8815},
  \href {https://ui.adsabs.harvard.edu/#abs/2017ApJ...846L..30S} {846, L30}

\bibitem[\protect\citeauthoryear{{Shivaei} et~al.}{{Shivaei}
  et~al.}{2018}]{Shivaei2018}
{Shivaei} I.,  et~al., 2018, \mn@doi [\apj] {10.3847/1538-4357/aaad62}, \href
  {https://ui.adsabs.harvard.edu/#abs/2018ApJ...855...42S} {855}

\bibitem[\protect\citeauthoryear{{Skelton} et~al.}{{Skelton}
  et~al.}{2014}]{Skelton2014}
{Skelton} R.~E.,  et~al., 2014, \mn@doi [\apjs] {10.1088/0067-0049/214/2/24},
  \href {http://adsabs.harvard.edu/abs/2014ApJS..214...24S} {214, 24}

\bibitem[\protect\citeauthoryear{{Smit} et~al.}{{Smit} et~al.}{2014}]{Smit2014}
{Smit} R.,  et~al., 2014, \mn@doi [\apj] {10.1088/0004-637X/784/1/58}, \href
  {http://adsabs.harvard.edu/abs/2014ApJ...784...58S} {784, 58}

\bibitem[\protect\citeauthoryear{{Smit} et~al.,}{{Smit}
  et~al.}{2015}]{Smit2015}
{Smit} R.,  et~al., 2015, \mn@doi [\apj] {10.1088/0004-637X/801/2/122}, \href
  {http://adsabs.harvard.edu/abs/2015ApJ...801..122S} {801, 122}

\bibitem[\protect\citeauthoryear{{Stanway}, {Eldridge}  \& {Becker}}{{Stanway}
  et~al.}{2016}]{Stanway2016}
{Stanway} E.~R.,  {Eldridge} J.~J.,   {Becker} G.~D.,  2016, \mn@doi [\mnras]
  {10.1093/mnras/stv2661}, \href
  {http://adsabs.harvard.edu/abs/2016MNRAS.456..485S} {456, 485}

\bibitem[\protect\citeauthoryear{{Stark}}{{Stark}}{2016}]{Stark2016}
{Stark} D.~P.,  2016, \mn@doi [\araa] {10.1146/annurev-astro-081915-023417},
  \href {http://adsabs.harvard.edu/abs/2016ARA\%26A..54..761S} {54, 761}

\bibitem[\protect\citeauthoryear{{Stark}, {Ellis}, {Chiu}, {Ouchi}  \&
  {Bunker}}{{Stark} et~al.}{2010}]{Stark2010}
{Stark} D.~P.,  {Ellis} R.~S.,  {Chiu} K.,  {Ouchi} M.,   {Bunker} A.,  2010,
  \mn@doi [\mnras] {10.1111/j.1365-2966.2010.17227.x}, \href
  {https://ui.adsabs.harvard.edu/#abs/2010MNRAS.408.1628S} {408, 1628}

\bibitem[\protect\citeauthoryear{{Stark} et~al.,}{{Stark}
  et~al.}{2015a}]{Stark2015a}
{Stark} D.~P.,  et~al., 2015a, \mn@doi [\mnras] {10.1093/mnras/stv688}, \href
  {http://adsabs.harvard.edu/abs/2015MNRAS.450.1846S} {450, 1846}

\bibitem[\protect\citeauthoryear{{Stark} et~al.,}{{Stark}
  et~al.}{2015b}]{Stark2015b}
{Stark} D.~P.,  et~al., 2015b, \mn@doi [\mnras] {10.1093/mnras/stv1907}, \href
  {http://adsabs.harvard.edu/abs/2015MNRAS.454.1393S} {454, 1393}

\bibitem[\protect\citeauthoryear{{Stark} et~al.}{{Stark}
  et~al.}{2017}]{Stark2017}
{Stark} D.~P.,  et~al., 2017, \mn@doi [\mnras] {10.1093/mnras/stw2233}, \href
  {http://adsabs.harvard.edu/abs/2017MNRAS.464..469S} {464, 469}

\bibitem[\protect\citeauthoryear{{Stasi{\'n}ska}, {Izotov}, {Morisset}  \&
  {Guseva}}{{Stasi{\'n}ska} et~al.}{2015}]{Stasinska2015}
{Stasi{\'n}ska} G.,  {Izotov} Y.,  {Morisset} C.,   {Guseva} N.,  2015, \mn@doi
  [\aap] {10.1051/0004-6361/201425389}, \href
  {https://ui.adsabs.harvard.edu/#abs/2015A&A...576A..83S} {576, A83}

\bibitem[\protect\citeauthoryear{{Steidel} et~al.}{{Steidel}
  et~al.}{2014}]{Steidel2014}
{Steidel} C.~C.,  et~al., 2014, \mn@doi [\apj] {10.1088/0004-637X/795/2/165},
  \href {https://ui.adsabs.harvard.edu/#abs/2014ApJ...795..165S} {795, 165}

\bibitem[\protect\citeauthoryear{{Steidel}, {Strom}, {Pettini}, {Rudie},
  {Reddy}  \& {Trainor}}{{Steidel} et~al.}{2016}]{Steidel2016}
{Steidel} C.~C.,  {Strom} A.~L.,  {Pettini} M.,  {Rudie} G.~C.,  {Reddy} N.~A.,
    {Trainor} R.~F.,  2016, \mn@doi [\apj] {10.3847/0004-637X/826/2/159}, \href
  {http://adsabs.harvard.edu/abs/2016ApJ...826..159S} {826, 159}

\bibitem[\protect\citeauthoryear{{Strom}, {Steidel}, {Rudie}, {Trainor},
  {Pettini}  \& {Reddy}}{{Strom} et~al.}{2017}]{Strom2017}
{Strom} A.~L.,  {Steidel} C.~C.,  {Rudie} G.~C.,  {Trainor} R.~F.,  {Pettini}
  M.,   {Reddy} N.~A.,  2017, \mn@doi [\apj] {10.3847/1538-4357/836/2/164},
  \href {http://adsabs.harvard.edu/abs/2017ApJ...836..164S} {836, 164}

\bibitem[\protect\citeauthoryear{{Tilvi} et~al.}{{Tilvi}
  et~al.}{2016}]{Tilvi2016}
{Tilvi} V.,  et~al., 2016, \mn@doi [\apj] {10.3847/2041-8205/827/1/L14}, \href
  {https://ui.adsabs.harvard.edu/#abs/2016ApJ...827L..14T} {827, L14}

\bibitem[\protect\citeauthoryear{{Ueda} et~al.}{{Ueda} et~al.}{2008}]{Ueda2008}
{Ueda} Y.,  et~al., 2008, \mn@doi [\apjs] {10.1086/591083}, \href
  {http://adsabs.harvard.edu/abs/2008ApJS..179..124U} {179, 124}

\bibitem[\protect\citeauthoryear{{Vanzella} et~al.}{{Vanzella}
  et~al.}{2016}]{Vanzella2016}
{Vanzella} E.,  et~al., 2016, \mn@doi [\apj] {10.3847/0004-637X/825/1/41},
  \href {https://ui.adsabs.harvard.edu/#abs/2016ApJ...825...41V} {825, 41}

\bibitem[\protect\citeauthoryear{{Weinberger}, {Kulkarni}, {Haehnelt},
  {Choudhury}  \& {Puchwein}}{{Weinberger} et~al.}{2018}]{Weinberger2018}
{Weinberger} L.~H.,  {Kulkarni} G.,  {Haehnelt} M.~G.,  {Choudhury} T.~R.,
  {Puchwein} E.,  2018, \mn@doi [\mnras] {10.1093/mnras/sty1563}, \href
  {https://ui.adsabs.harvard.edu/#abs/2018MNRAS.479.2564W} {479, 2564}

\bibitem[\protect\citeauthoryear{{Wilkins}, {Feng}, {Di-Matteo}, {Croft},
  {Stanway}, {Bouwens}  \& {Thomas}}{{Wilkins} et~al.}{2016}]{Wilkins2016}
{Wilkins} S.~M.,  {Feng} Y.,  {Di-Matteo} T.,  {Croft} R.,  {Stanway} E.~R.,
  {Bouwens} R.~J.,   {Thomas} P.,  2016, \mn@doi [\mnras]
  {10.1093/mnrasl/slw007}, \href
  {https://ui.adsabs.harvard.edu/\#abs/2016MNRAS.458L...6W} {458, L6}

\bibitem[\protect\citeauthoryear{{Wise}, {Demchenko}, {Halicek}, {Norman},
  {Turk}, {Abel}  \& {Smith}}{{Wise} et~al.}{2014}]{Wise2014}
{Wise} J.~H.,  {Demchenko} V.~G.,  {Halicek} M.~T.,  {Norman} M.~L.,  {Turk}
  M.~J.,  {Abel} T.,   {Smith} B.~D.,  2014, \mn@doi [\mnras]
  {10.1093/mnras/stu979}, \href
  {https://ui.adsabs.harvard.edu/#abs/2014MNRAS.442.2560W} {442, 2560}

\bibitem[\protect\citeauthoryear{{Xue} et~al.}{{Xue} et~al.}{2011}]{Xue2011}
{Xue} Y.~Q.,  et~al., 2011, \mn@doi [\apjs] {10.1088/0067-0049/195/1/10}, \href
  {http://adsabs.harvard.edu/abs/2011ApJS..195...10X} {195, 10}

\bibitem[\protect\citeauthoryear{{Xue}, {Luo}, {Brandt}, {Alexander}, {Bauer},
  {Lehmer}  \& {Yang}}{{Xue} et~al.}{2016}]{Xue2016}
{Xue} Y.~Q.,  {Luo} B.,  {Brandt} W.~N.,  {Alexander} D.~M.,  {Bauer} F.~E.,
  {Lehmer} B.~D.,   {Yang} G.,  2016, \mn@doi [\apjs]
  {10.3847/0067-0049/224/2/15}, \href
  {http://adsabs.harvard.edu/abs/2016ApJS..224...15X} {224, 15}

\bibitem[\protect\citeauthoryear{{Yang} et~al.,}{{Yang}
  et~al.}{2017}]{Yang2017}
{Yang} H.,  et~al., 2017, \mn@doi [\apj] {10.3847/1538-4357/aa7d4d}, \href
  {https://ui.adsabs.harvard.edu/#abs/2017ApJ...844..171Y} {844, 171}

\bibitem[\protect\citeauthoryear{{Zitrin} et~al.,}{{Zitrin}
  et~al.}{2015}]{Zitrin2015}
{Zitrin} A.,  et~al., 2015, \mn@doi [\apjl] {10.1088/2041-8205/810/1/L12},
  \href {http://adsabs.harvard.edu/abs/2015ApJ...810L..12Z} {810, L12}

\bibitem[\protect\citeauthoryear{{de Barros} et~al.}{{de Barros}
  et~al.}{2016}]{deBarros2016}
{de Barros} S.,  et~al., 2016, \mn@doi [\aap] {10.1051/0004-6361/201527046},
  \href {https://ui.adsabs.harvard.edu/#abs/2016A&A...585A..51D} {585, A51}

\bibitem[\protect\citeauthoryear{{van Dokkum} et~al.}{{van Dokkum}
  et~al.}{2011}]{vanDokkum2011}
{van Dokkum} P.~G.,  et~al., 2011, \mn@doi [\apjl]
  {10.1088/2041-8205/743/1/L15}, \href
  {http://adsabs.harvard.edu/abs/2011ApJ...743L..15V} {743, L15}

\bibitem[\protect\citeauthoryear{{van der Wel} et~al.}{{van der Wel}
  et~al.}{2011}]{vanderWel2011}
{van der Wel} A.,  et~al., 2011, \mn@doi [\apj] {10.1088/0004-637X/742/2/111},
  \href {http://adsabs.harvard.edu/abs/2011ApJ...742..111V} {742, 111}

\makeatother
\end{thebibliography}



\appendix


\bsp	
\label{lastpage}
\end{document}